\documentclass[a4paper,11pt]{article}
\pdfoutput=1 % if your are submitting a pdflatex (i.e. if you have
\usepackage{jheppub} % for details on the use of the package, please
\usepackage[T1]{fontenc} % if needed

\def\dd{{\rm d}}
\def\rmd{{\rm d}}
\def\rme{{\rm e}}

\def\pT{p_{\scriptscriptstyle T}}
\def\k{\boldsymbol{k}}
\def\q{\boldsymbol{q}}

\def\x{\boldsymbol{x}}
\def\y{\boldsymbol{y}}
\def\r{\boldsymbol{r}}

\usepackage{xcolor}

\usepackage{bm}

\title{Quenching effects in the cumulative jet spectrum}

\author[a,b]{Adam Takacs}
\author[a]{and Konrad Tywoniuk}
\affiliation[a]{Department of Physics and Technology, University of Bergen,
5007 Bergen, Norway}
\affiliation[b]{Department of Astronomy and Theoretical Physics, Lund University, S-223 62 Lund, Sweden}
\emailAdd{adam.takacs@uib.no}
\emailAdd{konrad.tywoniuk@uib.no}

\abstract{
The steeply falling jet spectrum induces bias on the medium modifications of jet observables in heavy-ion collisions. To explore this bias, we develop a novel analytic framework to study the quenched jet spectrum, and its cumulative. We include many energy-loss-related effects, such as soft and hard medium induced emissions, broadening, elastic scattering, jet fragmentation, cone size, coherence effects, etc. We show that, different jet spectrum-based observables are connected, e.g., the nuclear modification, spectrum shift, and the quantile procedure. We present the first predictions for the nuclear modification factor and the quantile procedure with cone size dependence. As an example, we compare dijet and boson+jet events to unfold the spectrum bias effects. We improve quark-, and gluon-jet classification using arguments based on the cumulative. Besides pointing out its flexibility, we apply our framework with other energy loss models such as the hybrid weak-, strong-coupling.
}

\begin{document}
\maketitle
\flushbottom

%%%%%%%%%%%%%%%%%%%%%%%%%%%%%%%%%%%%%%%%%%%%%%%%%%%%%%%
%%%%%%%%%%%%%%%%%%%%%%%%%%%%%%%%%%%%%%%%%%%%%%%%%%%%%%%
\section{Introduction}
%%%%%%%%%%%%%%%%%%%%%%%%%%%%%%%%%%%%%%%%%%%%%%%%%%%%%%%
%%%%%%%%%%%%%%%%%%%%%%%%%%%%%%%%%%%%%%%%%%%%%%%%%%%%%%%

At high-energy particle colliders, such as LHC at CERN, the exchanged momentum is large enough to resolve and scatter partonic constituents of the matter. These violent scatterings deviate the partonic constituents of nucleons and allow for intense bremsstrahlung radiation that ultimately result in collimated bunches of hadronic particles and energy. These so-called jets open a new perspective on the understanding of strong interactions at intermediate scales between the scale of the hard partonic scattering and the hadronization scale. Jets can be calculated with high-precision within perturbative QCD both gin electron-positron and proton-proton collisions  \cite{Larkoski:2017jix,Marzani:2019hun,Dasgupta:2020fwr}.

For accelerated heavy nuclei, processes involving large momentum exchanges between the incoming partons have additionally the potential to be well-calibrated probes of the hot and dense quark-gluon plasma (QGP) that is formed in the aftermath of such violent collisions. In this context, jets are particularly interesting since their typical formation time-scales overlap with the time-scales governing the creation and evolution of the QGP, suggesting potentially substantial jet-medium interactions. 
While the strong separation of the medium scale and the jet scale motivates a perturbative description of hard jet-medium interactions, many aspects of this processes are in the realm of non-perturbative physics and have to be modeled on the phenomenological level. A satisfactory description of jet production in heavy-ion collisions would therefore allow us to separate perturbative from non-perturbative phenomena. Besides, a well-controlled scale separation is an indispensable insight when studying, e.g., jet-medium coupling, thermalization, or medium modification of hadronization.

Jet studies have a rich history and wide selection of observables have been discussed involving fully reconstructed jets or their substructure, for recent reviews see~\cite{Larkoski:2017jix, Asquith:2018igt, Marzani:2019hun}. A fundamental observable is the momentum spectrum of jets for different reconstruction parameters $R$ \cite{Dasgupta:2014yra, Dasgupta:2016bnd}. The nuclear modification factor $R_{\rm AA}$, compares the spectrum in heavy-ion collisions (AA) to proton-proton (pp) at the same reconstructed jet $p_T$. However, jets that interact with a surrounding medium lose energy and end up with smaller $p_T$. Therefore---and it is not emphasized enough---the jet selection for $R_{\rm AA}$ compares two jet populations originated at different $p_T$. The equal $p_T$ selection induces a bias in the observables because the probability of creating a jet is steeply falling with $p_T$~\cite{Rajagopal:2016uip, Casalderrey-Solana:2018wrw}. We explore the bias in this paper. 

Recently, there has been efforts toward mitigating such bias effects by investigating novel observables or by using machine learning techniques~\cite{Du:2020pmp}. One alternative, that we will investigate in detail in the current work, is to introduce a quantile procedure~\cite{Brewer:2018dfs} to reconstruct a $p_T$ that is closer to the initial jet $p_T$ before quenching sets in. In contrast to $R_{\rm AA}$, the quantile procedure uses the tail-cumulative of the jet spectrum and momentum ratio to reduce the bias coming from the steepness of the spectrum. We demonstrate the properties of the quantile procedure within a versatile framework to incorporate quenching effects and explain its robustness for the first time. 

New measurements have been suggested, e.g., boson+jet, to have a better handle on the momentum mitigation in heavy-ions~\cite{Chatrchyan:2012gt,Sirunyan:2017jic,Aaboud:2018anc,Aaboud:2019oac}. Bosons suffer less medium effect, and their momenta are related to the initiator of the recoiling jet. The jet spectrum in boson+jet events is slightly different from inclusive QCD jets, and we will use it to illustrate the bias on the quenching. Moreover, quark- and gluon-jet contribution in the inclusive and boson+jet samples differ. This can be used for quark-gluon jet discrimination in a model-independent fashion~\cite{Brewer:2020och}. Using arguments on the cumulative, we improve the statistics of the classification task. Parallel with the works mentioned above, some numerical studies also appeared using Bayesian and machine learning techniques to extract the energy loss properties from data~\cite{He:2018gks}. 

During the evolution of jets inside a QGP, their constituents scatter elastically and inelastically on the medium. The scatterings redistribute energy to larger angles out of the jet cone, resulting in energy loss. The inelastic (or radiative) part describes the emissions induced by the medium (medium-induced emissions, or MIE for short). The MIE has well-known limits: (i) in the multiple soft scattering limit, the scattering centers act coherently, resulting in suppression of emissions (QCD analog of the Landau--Pomeranchuk--Migdal effect), captured by the BDMPSZ formula, which describes the induced emission of soft gluons~\cite{Baier:1996sk, Zakharov:1996fv, Baier:1998kq}. (ii) the single hard scattering limit is captured by the GLV formula, which describes emissions of harder gluons~\cite{Gyulassy:2000er}. Beyond the analytic limits, MIE is also amenable to direct numerical methods, such as in Refs.~\cite{CaronHuot:2010bp, Feal:2018sml, Andres:2020vxs}. Recently, there has also been a progression in the better understanding of the two regimes and the scales involved~\cite{Mehtar-Tani:2019tvy, Mehtar-Tani:2019ygg, Barata:2020sav}. In our work, we adopt the latter strategy to explore the impact of the MIE spectrum. The MIE can be resummed accounting for multiple induced emissions. The resulting formalism is the quenching weight~\cite{Baier:2001yt, Salgado:2003gb}. It is easy to generalize for all jet constituents, including coherence effects~\cite{Mehtar-Tani:2017web}, spectrum shapes, and elastic energy loss. Within this framework, we show the appearance of the spectrum bias.

The paper is organized in the following; in Sec.~\ref{sec:Quenching} we define our novel framework to calculate the medium jet spectrum. In Sec.~\ref{sec:Quantile}, for the first time, we show the properties of the cumulative spectrum, and we apply the quantile procedure. In Sec.~\ref{sec:Numerical_results} we give predictions for measurements, considering cone size dependence for the single-inclusive jet sample produced in dijet and boson+jet events (we focus concretely on Z+jet). We show how to use cumulative arguments to improve quark-gluon discrimination. Besides in App.~\ref{sec:Other_models}, we use the quenching weight formalism for elastic scatterings and in the hybrid weak-, strong-coupling model.

%%%%%%%%%%%%%%%%%%%%%%%%%%%%%%%%%%%%%%%%%%%%%%%%%%%%%%%
%%%%%%%%%%%%%%%%%%%%%%%%%%%%%%%%%%%%%%%%%%%%%%%%%%%%%%%
\section{Quenching effects in the spectrum}\label{sec:Quenching}
%%%%%%%%%%%%%%%%%%%%%%%%%%%%%%%%%%%%%%%%%%%%%%%%%%%%%%%
%%%%%%%%%%%%%%%%%%%%%%%%%%%%%%%%%%%%%%%%%%%%%%%%%%%%%%%

The main observable considered in this work is the single-inclusive spectrum of reconstructed jets in heavy-ion collisions. In this work, we will both consider jets produced in conventional QCD processes, that is dijet events, and jets produced in conjunction with a photon or weak boson, so-called boson-jet events. In the context of high-energy collisions, it is natural to assume a factorization of the partonic hard cross-section from the subsequent medium processes. This can be justified by invoking the large separation of momentum scales involved in jet production; typically the hard scattering $Q_{\rm hard}\sim 10^3$ GeV, is much bigger than the jet scale $Q_{\rm jet} \sim p_T R\sim 10^2$ GeV, where $p_T$ is the reconstructed transverse momentum of the jet and $R$ is jet cone parameter. These scales are much bigger than the typical medium scale, for instance, the temperature of the medium $T \sim 0.5$ GeV. Hence, one can write the medium modification of the vacuum jet spectrum due to energy loss~\cite{Baier:2001yt},
\begin{equation}\label{eq:medium_spectrum}
    \frac{\rmd\sigma^{\rm med}_R}{\rmd p_T}(p_T)=\int^\infty_0\rmd\varepsilon\, \mathcal{P}_>(\varepsilon) \left.\frac{\rmd\sigma^{\rm vac}_R}{\rmd p'_T}\right\vert_{p'_T=p_T+\varepsilon}\,.
\end{equation}
The $\mathcal P_>(\varepsilon)$ describes the probability of a vacuum jet to distribute (or lose) energy out of the jet cone. Above, $\rmd \sigma^\text{vac}_R/\rmd p_T$ refers to the partonic cross-section to produce a jet with $R$ in the collinear factorization and can be calculated up to high precision~\cite{Dasgupta:2014yra,Dasgupta:2016bnd}. In this work, instead, we extract the spectrum using a Monte Carlo event generator, see Sec.~\ref{sec:Numerical_results} for further details.

The vacuum spectrum is well approximated by a power-law, $\rmd \sigma^{\rm vac}_R/\rmd p_T \propto 1/p_T^n$, and is steeply falling, i.e. $n \gg 1$. Therefore, we approximate $\rmd\sigma_{R}^{\rm vac}(p_T+\varepsilon)/\rmd p_T = A\,(p_T+\varepsilon)^{-n}\approx \rme^{-n \varepsilon/p_T}\rmd\sigma_{\rm vac}(p_T)/\rmd p_T$, where $A$ is a constant and $n\equiv n(p_T,R)$ is the power index of the spectrum. In the last step, we additionally assumed that $\varepsilon \ll p_T$. The cone-size dependent nuclear modification factor defined as
\begin{equation}\label{eq:Q_def}
    R_{\rm med}(p_T,R)\equiv \left.\frac{\rmd\sigma_R^{\rm med}}{\rmd p_T}\middle/\frac{\rmd\sigma_R^{\rm vac}}{\rmd p_T} \right.\,,
\end{equation}
is consequently related to the quenching factor $\mathcal{Q}(\nu\equiv n/p_T)$, which is the Laplace transform of $\mathcal P_>(\varepsilon)$, i.e.
\begin{equation}\label{eq:Q_def2}
    R_{\rm med}(p_T,R)\approx\int^\infty_0\rmd\varepsilon\,\mathcal{P}_>(\varepsilon)\,\rme^{-\frac{n\varepsilon}{p_T}} \equiv\mathcal{Q}_>(\nu) \,.
\end{equation} 
This approximation is precise within at most a few percent for realistic parameters in a wide kinematic range $10 < p_T < 1000$ GeV that we consider here. Therefore, in what follows, we simply identify $R_{\rm med}=\mathcal Q_>$. For further details on such corrections, see Appendix~\ref{sec:Corrections_LaplaceAndPower}. In order to go from $R_{\rm med}$ to the experimentally measured $R_{\rm AA}$, one needs to include a $p_T$-dependent quark and gluon jet production. Other differences are mostly due to geometry (for a review see Ref.~\cite{Miller:2007ri}), and cold nuclear effects (nPDF), that we include in Sec.~\ref{sec:Numerical_results}.

The energy loss distribution $\mathcal{P}_>(\varepsilon)$---and therefore the quenching factor $\mathcal{Q}_>(\nu)$---depends on vacuum jet properties (like the jet $p_T$ and the cone size $R$) and the properties of the medium (e.g., medium length $L$, and the jet transport coefficient $\hat q$). Its normalization condition, $\int_0^\infty \rmd \varepsilon\, \mathcal P_>(\varepsilon)=1$, translates to $\mathcal{Q}_>(0)=1$. The assumptions underlying Eq.~\eqref{eq:Q_def} are quite robust for a wide range of applications. Therefore, one is flexible in defining the precise nature of the energy loss distribution $\mathcal{P}_>(\varepsilon)$. The introduction of a probability distribution to describe effects of quenching in heavy-ion collisions has a long history in the analysis of single-inclusive hadron~\cite{Baier:2001yt,Salgado:2003gb,Arleo:2017ntr} and jet spectra~\cite{Mehtar-Tani:2017web,Spousta:2015fca,Qiu:2019sfj,Mehtar-Tani:2021fud}. 

We will derive the quenching factor $\mathcal Q_>$ of the jet in several steps. First, we consider the induced radiation spectrum of a single color-charge propagating through the medium and how to account for their multiple emissions. We thus arrive at the quenching factor for a single parton, $\mathcal{Q}_>^{(0)}(\nu)$. Next, we consider the effect of jet fragmentation which leads to multiple vacuum-like emissions on short time-scales inside the jets. Partons from these emissions contribute to the quenching of the full jet. This is accounted for by the so-called collimator function that provides a fully resummed quenching factor $\mathcal{Q}_>(\nu)$. The corresponding jet quenching probability distribution can then be found via an inverse Laplace transform, but we will not pursue this further in this work.

In our numerical results in Sec.~\ref{sec:Numerical_results}, we will also include elastic energy loss, since it potentially can contribute to the $\sim 10-30\%$ level to the final jet suppression factor. As we said, the formulation above is quite general and allows to separately formulate \texttt{a)} the mechanism of quenching, and \texttt{b)} the phase space where the jet is affected. In App.~\ref{sec:Other_models}, therefore, we show how to formulate other energy loss models in terms of quenching weights.

%%%%%%%%%%%%%%%%%%%%%%%%%%%%%%%%%%%%%%%%%%%%%%%%%%%%%%%
\subsection{Constructing the quenching weight}
%%%%%%%%%%%%%%%%%%%%%%%%%%%%%%%%%%%%%%%%%%%%%%%%%%%%%%%

We construct the radiative energy loss distribution $\mathcal P_>(\varepsilon)$ (or the quenching factor $\mathcal Q_>(p_T)$) of a jet starting from a single parton, that we denote $\mathcal{P}_>^{(0)}(\varepsilon)$. For the radiation of a single medium-induced gluon, this energy loss probability is simply given by
\begin{equation}\label{eq:eloss-prob-one}
    \mathcal{P}_>^{(0)}(\varepsilon) \approx \left.\frac{\rmd I_>}{\rmd \omega}\right|_{\omega = \varepsilon} + \delta(\varepsilon) \left[ 1 - \int_0^\infty \rmd \omega\, \frac{\rmd I_>}{\rmd \omega}\right] \,,
\end{equation}
where $\rmd I_>/\rmd \omega$ is the spectrum of medium-induced gluon radiation spectrum that emerges at angles larger than the jet cone, or
\begin{equation}\label{eq:out_of_cone_def}
    \frac{\rmd I_>}{\rmd\omega}=\int^\infty_{(\omega R)^2}\rmd\bm k^2 \frac{\rmd I}{\rmd \omega \, \rmd \k^2}\,.
\end{equation}
This is an important difference with respect to the more common use of the quenching weights applied to single-hadron spectra in Refs.~\cite{Baier:2001yt,Salgado:2003gb}.
The first term in Eq.~\eqref{eq:eloss-prob-one} describes an emission, while the second term is a virtual correction and provides the normalization. 
Our starting point in Sec.~\ref{sec:spectra} is to discuss the specific details of this induced-emission spectrum. Accounting for multiple such emissions in course of the medium propagation allows us to derive the single parton quenching weight in Sec.~\ref{sec:single-parton-quenching}.

Then, having derived how \emph{one parton} contributes to the energy loss of the whole jet, we next turn to the calculation of how \emph{multiple partons} in the jet, resolved by the medium during their fragmentation process, add up to the total quenching effect. This will be described in Sec.~\ref{sec:multi-parton-quenching}.

%%%%%%%%%%%%%%%%%%%%%%%%%%%%%%%%%%%%%%%%%%%%%%%%%%%%%%%%%%%%%%%%%%%
%%%%%%%%%%%%%%%%%%%%%%%%%%%%%%%%%%%%%%%%%%%%%%%%%%%%%%%%%%%%%%%%%%%
\subsection{The medium induced gluon spectrum at finite cone}
\label{sec:spectra}
%%%%%%%%%%%%%%%%%%%%%%%%%%%%%%%%%%%%%%%%%%%%%%%%%%%%%%%%%%%%%%%%%%%
%%%%%%%%%%%%%%%%%%%%%%%%%%%%%%%%%%%%%%%%%%%%%%%%%%%%%%%%%%%%%%%%%%%

The medium-induced spectrum from multiple scattering in a QCD medium was derived independently by Zakharov \cite{Zakharov:1996fv,Zakharov:1997uu} and Baier-Dokshitzer-Mueller-Peigne-Schiff (BDMPS) \cite{Baier:1996sk,Baier:1996kr,Baier:1998kq}, see also Refs.~\cite{Blaizot:2012fh,Apolinario:2014csa}. In the limit of soft gluon emission with $\omega$ energy and $\bm k$ transverse momentum, this spectrum can be written as
\begin{align}
     \omega \frac{\rmd I}{\rmd \omega \rmd^2 \k} &= \frac{\alpha_s C_R}{(2\pi)^2\omega^2} 2 \text{Re} \int_0^\infty \rmd t_1 \int_{t_1}^\infty \rmd t_2 \int \rmd^2 \x \, \rme^{- \int_{t_2}^\infty \rmd s\, v(\x,s)} \nonumber\\
     &\times \partial_{\x} \cdot \partial_{\y} \mathcal{K}(\x,t_2; \y,t_1) \big\vert_{\y=0} \,,
\end{align}
where $C_R=C_F$ if the radiator is a quark ($C_A$ for gluon), the path integral is
\begin{equation}
\label{eq:path-integral}
    \mathcal{K}(t_2,\x; t_1,\y) = \int_{\r(t_1)=\y}^{\r(t_2)=\x} \mathcal{D} \r \, \exp \left\{  \int_{t_1}^{t_2} \rmd s \, \left[i\frac{\omega}{2}\dot{\r}^2 - v(\r,s)  \right] \right\} \,,
\end{equation}
and
\begin{equation}
    v(\x,t) = N_c \int \frac{\rmd^2 \q}{(2\pi)^2} \, \frac{\rmd^2 \sigma_\text{el}}{\rmd^2 \q} \left(1 - \rme^{i \x \cdot \q} \right) \,,
\end{equation}
is related to the elastic scattering potential in medium. The leading logarithmic behavior of the scattering potential reads for any hard Coulomb tailed elastic potential,
\begin{equation}\label{eq:IOE_LLpotential}
    v(\x,t)_{\rm \tiny LL} = \frac{1}{4} \hat q_0(t) \x^2 \log\left(\frac{1}{\x^2 \mu_\ast^2}\right) + \mathcal{O}\left(\x^4 \mu_\ast^2\right) \,, 
\end{equation}
where $\hat q_0(t) $ is a local transport coefficient and $\mu_\ast$ is related to the infrared scale that regularizes the in-medium scattering potential $\rmd^2 \sigma_\text{el}/\rmd^2 \q$. For example, the Gyulassy-Wang scattering potential \cite{Gyulassy:1993hr}, describes scattering off a plasma made up of Yukawa-screened quasi-particles, reads
\begin{equation}\label{eq:gw-potential}
    \left.\frac{\rmd^2 \sigma_\text{el}}{\rmd^2 \q} \right|_\text{GW} = \frac{g^4 n(t)}{(\q^2 + \mu^2)^2} \,,
\end{equation}
where $\mu$ is a screening mass, and $n(t)$ is the density of scattering centers in the medium. In this case $\hat q_0(t) = 4\pi \alpha_s^2 N_c n(t)$, and $\mu_\ast^2 = \mu^2 \rme^{-1 + 2 \gamma_E}/4$.

The path integral in Eq.~\eqref{eq:path-integral} can be solved numerically, see e.g. in Ref.~\cite{CaronHuot:2010bp,Feal:2018sml,Andres:2020vxs}, but analytic solutions are available in limiting cases. Here, two of the most frequent schemes are discussed. In the ``harmonic oscillator'' (HO) approximation, valid when the logarithm is slowly varying around a constant scale, i.e. $\mu_\ast^{2} \ll 1/\x^2 \sim Q^2$, one can absorb the logarithm in the definition of the transport coefficient so that $v(\x) \approx \hat q \x^2/4$. This corresponds to a purely Gaussian momentum broadening in transverse momentum given by $\langle \k^2 \rangle = \hat q t$ during the propagation in the medium. In this case, Eq.~\eqref{eq:path-integral} describes a harmonic oscillator with imaginary frequency $\Omega^2 = \hat q/(2 i \omega)$ in transverse to the propagation, and whose solution is well known. This approximation, however, fails to describe the hard tail of medium emissions. On the other hand, one can also truncate the resummation of medium scatterings at a fixed order $N$, giving rise to the so-called ``opacity expansion''~\cite{Gyulassy:2000er,Wiedemann:2000za}. The $N\!=\!1$ approximation amounts to considering a single, incoherent scattering with the medium during the propagation.

A discussion of the regions of validity of these approaches was recently addressed in Refs.~\cite{Mehtar-Tani:2019tvy,Mehtar-Tani:2019ygg,Barata:2020sav,Andres:2020kfg}. In the soft scattering regime, the formation time of emissions $t_{\rm f} = 2\omega/\k^2$ becomes modified due to Gaussian broadening, i.e. $t_{\rm f} \sim \sqrt{2\omega/\hat q}$. For emissions with large formation times $t_{\rm f} \sim L$ correspond to $\omega \sim \hat q L^2/2\equiv\omega_c$. It follows that emissions with $\omega > \omega_c$ cannot be produced by soft collisions and Gaussian broadening, and are dominated by a single, hard scattering with the medium constituents. Similarly, at short formation times of the order of the medium mean free path $t_{\rm} \sim \lambda\ll L$, or $\omega \sim \hat q \lambda^2/2 = \omega_{\rm \tiny BH}\ll\omega_c$, the spectrum is again dominated by single scattering~\cite{Wiedemann:2000za,Andres:2020kfg}. Since this latter regime gives a small contribution to energy loss, we will not discuss it further here.

Recently, the contribution of hard emissions has been shown to matter for precision comparisons with high-$p_T$ single-hadron spectra at RHIC and LHC \cite{Feal:2019xfl}. A systematic procedure to calculate the spectrum for a large range of relevant emission energies $\omega > \omega_{\rm \tiny BH}$
was developed in the so-called ``improved opacity expansion'' (IOE) \cite{Mehtar-Tani:2019tvy,Mehtar-Tani:2019ygg,Barata:2020sav}. This framework rewrites the leading-log scattering potential form Eq.~\eqref{eq:IOE_LLpotential} as
\begin{equation}\label{eq:LL-potential}
    v_{\rm \tiny LL}(\x,t) = \frac{1}{4} \hat q_0 \x^2 \left[ \log\left(\frac{Q_c^2}{\mu_\ast^2}\right) + \log\left(\frac{1}{Q_c^2 \x^2}\right) \right] = v_{\rm \tiny HO}(\x,t) + \delta v(\x,t) \,,
\end{equation}
where $Q_c$ is a separation scale of the harmonic potential. In the limit of $Q_c \gg \mu_\ast$, one can then expand the solution of the path integral in Eq.~\eqref{eq:path-integral} around the HO solution with an effective $\hat q(Q_c^2)$ and treat hard scatterings with the medium, given by $\delta v(\x,s)$, as higher-order perturbations. This approach is systematically improvable and, up to next-to-next-to-leading order in this expansion, the effective $\hat q$ parameter is given by \cite{Barata:2020sav}
\begin{equation}\label{eq:qhat-nnlo}
    \hat q(Q^2_c)=\hat q_0\ln\left(\frac{Q^2_c}{\mu_*^2}\right)\left[1+\frac{1.013}{\ln\left(\frac{Q^2_c}{\mu_*^2}\right)}+\frac{0.318}{\ln^2\left(\frac{Q^2_c}{\mu_*^2}\right)}\right]\,,
\end{equation}
where the scale $Q_c^2$ is itself found through an implicit equation,  $Q_c^4=\hat q_0\omega\ln(Q_c^2/\mu_*^2)$. This equation has solution, if $\omega\geq \omega_{\rm min}\equiv2e\mu^4_\star/\hat q_0$, when $Q^2_c=\mu^2_\star\exp[-W_{-1}(2\mu^4_\star/(w\hat q_0))/2]$, where $W_i(x)$ is the Lambert function on the $i^{\rm th}$ branch. Since jet quenching is not sensitive to the details of very soft gluon emissions, in our numerical results we freeze the logarithms at 1, i.e. $\ln Q_c^2/\mu_\ast^2 \geq 1$.

The IOE has so far only been developed for the spectrum integrated over transverse momenta, i.e.\ $\rmd I/\rmd \omega$, and for the momentum broadening of a single particle in the medium \cite{Barata:2020rdn}. Since the contributions to jet energy loss rely on out-of-cone emissions, cf. Eq.~\eqref{eq:out_of_cone_def}, we instead have to consider the matching of the partially integrated spectrum $\rmd I_>/\rmd \omega$ between the multiple-scattering HO and single-scattering $N=1$ regimes. We use the quenching parameter $\hat q$ from Eq.~\eqref{eq:qhat-nnlo} which correctly connects the $\hat q_0$ parameter from the elastic potential with the multiple scattering formalism. We propose a simple interpolation scheme that relies on a single matching scale $\omega_\star$, that is related to the broadening of soft gluons and will be defined below. At small gluon energies, below the matching scale, the spectrum is described by the HO approach with the effective $\hat q$. Above the matching scale, the spectrum is given by the $N=1$ spectrum.

\begin{figure}
    \centering
    \includegraphics[width=0.49\textwidth]{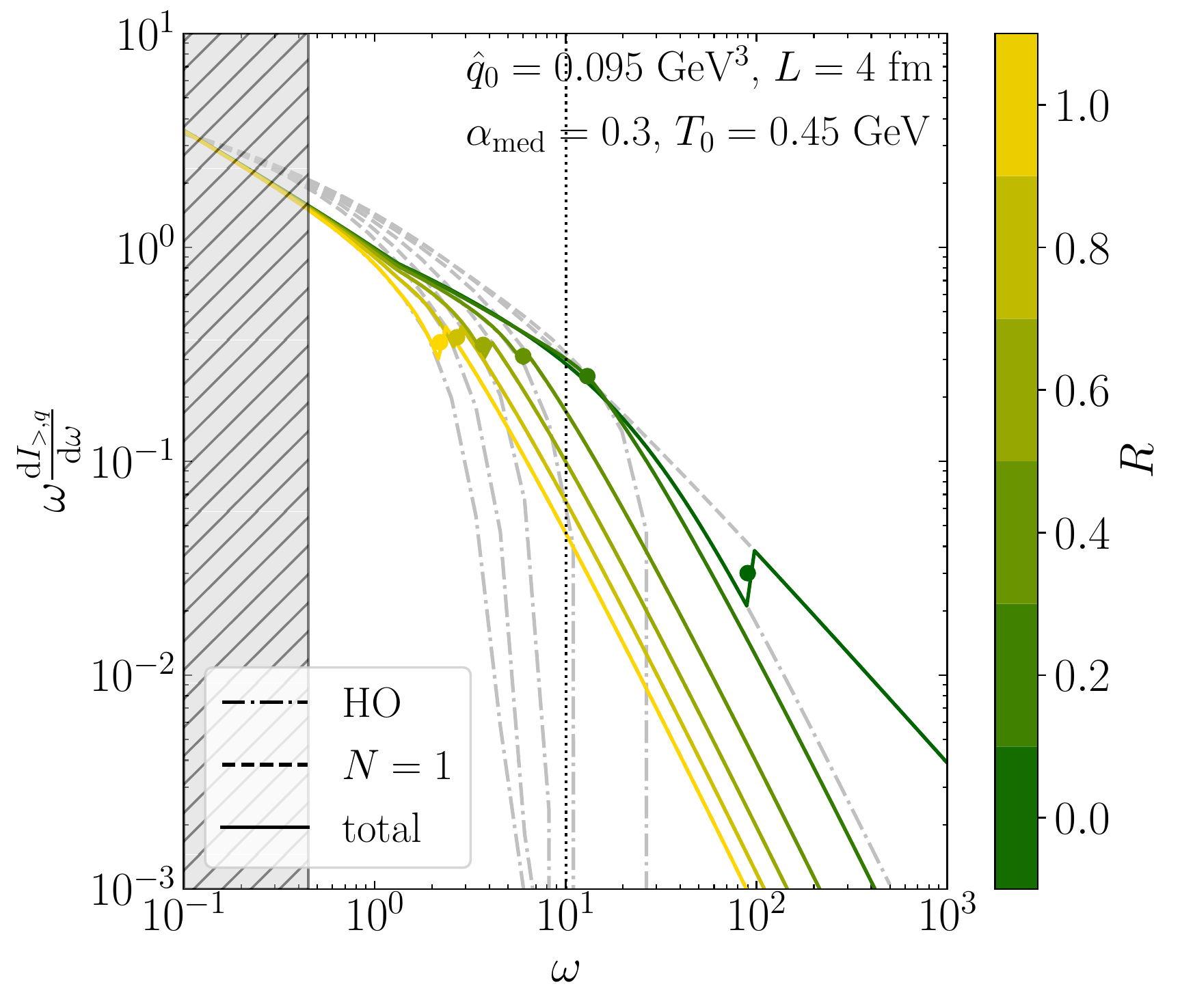}
    \includegraphics[width=0.49\textwidth]{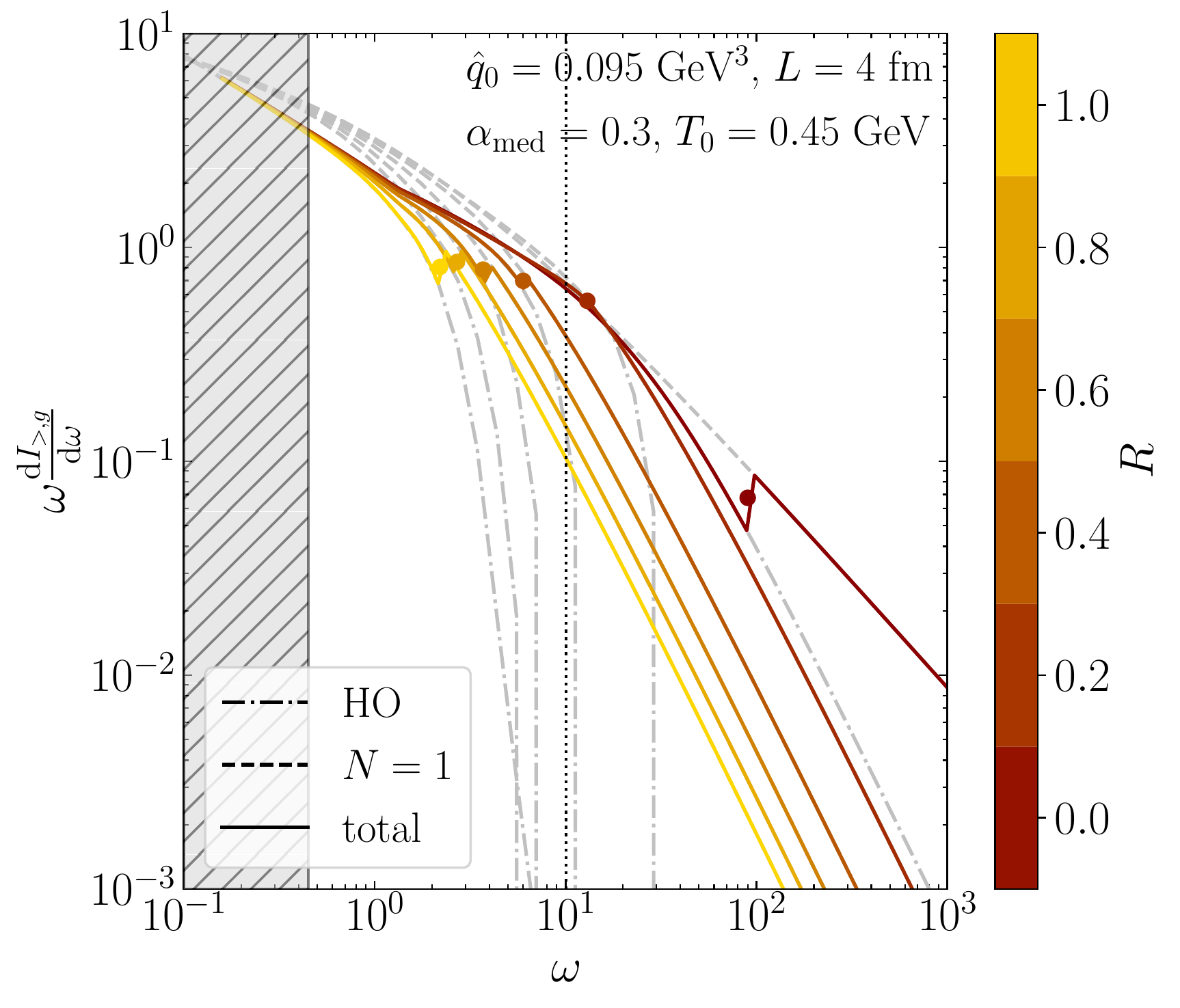}
    \caption{The out of cone emission spectrum, matched at NNLO for quarks (left) and gluons (right) from Eq.~\eqref{eq:Matched_spectrum} for different cone sizes. The dash-dotted (dashed) grey lines are the HO ($N=1$) spectra, the bullets show the matching points $\omega_\ast$. The dotted line shows the region $\omega_s$ below in which emitted gluons thermalize during the broadening. The hatched band shows where $\omega<T_0$, considered and neglected as background.}
    \label{fig:dIdwo}
\end{figure}

We will treat the medium as a ``brick'' of constant $\hat q_0$ and fixed length $L$. In the absence of the cone constraint, a natural matching scale is $\omega_\star\sim\omega_c \equiv \hat q L^2/2$. For the out-of-cone spectrum, the effect of broadening after emission cannot be neglected. For Gaussian broadening that presents in HO, a particle emitted at initial time accumulates $\langle \k^2 \rangle \sim \hat qL$ after propagating through the medium. This corresponds to an angle $\theta \sim \sqrt{\hat qL}/\omega$ in the small-angle approximation. Demanding that this angle is larger than the jet cone $R$ for energy loss, a cut-off in energy arises $\omega < \omega_R \equiv \sqrt{\hat q L}/R$,  above which the HO spectrum falls rapidly. For more details, see App.~\ref{sec:qf-scaling}. It turns out that a relatively smooth matching between the HO and $N=1$ regimes is achieved by choosing
\begin{equation}\label{eq:omega_s}
    \omega_\star = \min\big(\omega_c, \omega_R \big) \,,
\end{equation}
where $\omega_c $ and $\omega_R$ are defined with the effective $\hat q$ parameter in Eq.~\eqref{eq:qhat-nnlo}. For our final results, see the left and right panels in Fig.~\ref{fig:dIdwo}. The postulated matching works extremely well, up to some negligible discontinuities in the spectrum.

%\paragraph{HO spectrum details.}
The regime of soft gluon emissions, $\omega < \omega_\star$, is dominated by multiple scattering where we can employ the HO approximation. The spectrum in this approximation is given by
\begin{equation}
    \omega \frac{\rmd I^{\rm HO}}{\rmd \omega \, \rmd^2 \k} =\frac{\bar \alpha}{\pi \omega}\, \text{Im} \Big[ \mathcal{R}^\text{in-in} + \mathcal{R}^\text{in-out} \Big] \,,
\end{equation}
where $\bar \alpha = \alpha_s C_R/\pi$ and the two factors read \cite{MehtarTani:2012cy}
\begin{align}
    \mathcal{R}^\text{in-in} &= \int_0^L \rmd t\, \frac{(1+i)\sqrt{\omega \hat q} \cot (\Omega t)}{\hat q(L-t) -(1+i)\sqrt{\omega \hat q} \cot (\Omega t)}\exp \left[-\frac{\k^2}{\hat q(L-t) - (1+i)\sqrt{\omega \hat q} \cot (\Omega t)} \right] \,,\\
    \mathcal{R}^\text{in-out} &= \int_0^L \rmd t \, \frac{1}{ \cos^2 (\Omega t)} \exp \left[-i \frac{\k^2}{2\omega \Omega} \tan (\Omega t)\right] \,,
\end{align}
where $\Omega = (1-i) \sqrt{\hat q/\omega}/2$. The integrated spectrum in Eq.~\eqref{eq:out_of_cone_def} reads then
\begin{equation}\label{eq:dIout-HO}
    \omega \frac{\rmd I^{\rm \tiny HO}_>}{\rmd \omega} = \frac{\bar \alpha}{\omega}\, \text{Im} \Big[ \mathcal{R}_>^\text{in-in} + \mathcal{R}_>^\text{in-out} \Big] \,, 
\end{equation}
where now
\begin{align}
    \mathcal{R}_>^\text{in-in} &= \int_0^L \rmd t\, (1+i)\sqrt{\omega \hat q} \cot (\Omega t) \exp \left[-\frac{(\omega R)^2}{\hat q(L-t) - (1+i)\sqrt{\omega \hat q} \cot (\Omega t)} \right] \,,\\
    \mathcal{R}_>^\text{in-out} &= -\int_0^L \rmd t \, \frac{(1+i) \sqrt{\hat q \omega}}{ \cos (\Omega t)\, \sin(\Omega t)} \exp \left[-i \frac{(\omega R)^2}{2\omega \Omega} \tan (\Omega t)\right] \,,
\end{align}
We find that in the $R\to0$ limit, Eq.~\eqref{eq:dIout-HO} yields 
\begin{align}\label{eq:BDMPSZ_spectrum}
    \omega \frac{\rmd I^{\rm \tiny HO}}{\rmd\omega} & = 2 \bar \alpha\, \ln \left\vert \cos (1-i)\sqrt{\frac{\omega_c}{2\omega}}\right\vert\,.
\end{align}
which is the celebrated BDMPS-Z spectrum~\cite{Zakharov:1996fv,Baier:1998kq}.

%\paragraph{$N=1$ spectrum details.}
As discussed above, at $\omega> \omega_\star$, the HO spectrum has to be corrected with the single hard gluon emission spectrum ($N=1$)~\cite{Mehtar-Tani:2019tvy,Mehtar-Tani:2019ygg,Barata:2020sav}, for which
\begin{equation}
    \omega \frac{\rmd I^{N=1}}{\rmd \omega \rmd^2 \k} = 8 \pi \bar \alpha N_c \int_0^L \rmd t \int \frac{\rmd^2 \q}{(2\pi)^2}\, \frac{\rmd^2 \sigma_\text{el}}{\rmd^2 \q} \, \frac{\k \cdot \q}{\k^2 (\k-\q)^2} \left[1- \cos \frac{(\k-\q)^2}{2\omega}t \right] \,.
\end{equation}
Using Eq.~\eqref{eq:out_of_cone_def} and the Gyulassy-Wang potential from Eq.~\eqref{eq:gw-potential}, we immediately find that the integrated spectrum reads
\begin{equation}\label{eq:GLV_spectrum_out}
    \omega \frac{\rmd I^{N=1}_>}{\rmd \omega} = \bar \alpha \frac{ \hat q_0 L^2}{\omega} \int_0^\infty \rmd u\, \frac{u - \sin u}{u^2} \frac{1}{[(\zeta + u + y)^2 - 4 \zeta u]^{1/2}} \,,
\end{equation}
where $\zeta \equiv \omega R^2 L/2$ and $y=\mu^2 L/(2\omega)$. Again, for $R \to 0$, we recover the familiar form of the integrated $N=1$ spectrum~\cite{Gyulassy:2000er,Salgado:2003gb},
\begin{equation}
    \omega \frac{\rmd I^{N=1}}{\rmd \omega} = \bar \alpha\frac{\hat q_0 L^2}{\omega} \int_0^\infty \rmd u\, \frac{u - \sin u}{u^2}\frac{1}{u+y} \,,
\end{equation}
as expected.

The full spectrum is therefore postulated to be well approximated by the following interpolation,
\begin{equation}\label{eq:Matched_spectrum}
    \frac{\rmd I_>}{\rmd \omega} = \Theta\big(\omega_\star - \omega\big) \frac{\rmd I^{\rm \tiny HO}_>}{\rmd \omega} + \Theta\big(\omega - \omega_\star\big)\frac{\rmd I^{N=1}_>}{\rmd \omega} \,,
\end{equation}
where all parameters in the HO spectrum and in the matching scales contain the effective $\hat q$, given in Eq.~\eqref{eq:qhat-nnlo}. Note that the spectrum $\rmd I_>/\rmd \omega$ depends on the initial energy only in the combination $\omega =x(1-x)E \approx x E$.

\begin{table}[b!]
    \centering
    \begin{tabular}{c|c}
        Medium parameter & Value \\
        \hline 
        $\alpha_s$ & 0.3 \\
        $T_0$ & 0.45 GeV \\
        $\hat q_0$ & 0.095 GeV$^3$ \\
        $L$ & 4 fm \\
    \end{tabular}
    \caption{Choice of medium parameters corresponding to  0--10\% central PbPb collisions with $\sqrt{s_{NN}} =5.02$ TeV.}
    \label{tab:medium-parameters}
\end{table}
At this point, we pause to discuss the choice of medium parameters. We choose the medium coupling to be fixed at $\alpha_s =\alpha_{\rm med} = 0.3$, corresponding to $g_{\rm med} \approx 1.94$. The IR cutoff scale is $\mu\equiv\sqrt{2/3}g_{\rm med}T_0\approx0.71$ GeV, and $\omega_{\rm min}\approx1.5$ GeV. The remaining parameters are chosen to reflect the conditions in 0--10\% central PbPb events at $\sqrt{s_{NN}} =5.02$ TeV, see Table~\ref{tab:medium-parameters}. As demonstrated below, this parameter set gives a good description of the experimentally measured jet suppression factor, see Sec.~\ref{sec:Numerical_results}. Since our work does not deal with the precise description of experimental data, we have not attempted to fix these parameters from a model of the underlying medium nor fitted them to experimental data, which was done in \cite{Mehtar-Tani:2021fud}. This choice finally leads to a matching scale $Q_c^2 =(1-100)\mu^2_\star$ and $\hat q = (1-5)\hat q_0$, depending on $\omega$, in the IOE-matched spectrum Eq.~\eqref{eq:Matched_spectrum}. 

The MIE spectrum is shown in Fig.~\ref{fig:dIdwo} for quarks (left) and for gluons (right). The matching points $\omega_\ast$ from Eq.~\eqref{eq:omega_s} are shown with bullets and below (above) the spectrum is the HO ($N=1$) spectrum. The matching works very well capturing the cone size dependence, however it is not perfectly smooth (see $R=0$). It is good enough to study the integral of this spectrum, presented in the quenching factor. With different colors the cone size dependence points out, it is less probable to lose energy by opening the cone (i.e., recapturing emissions). The difference in the quark and gluon spectrum is the color factor $C_A/C_F=9/4$, and thus gluons lose more energy. The dotted line in Fig.~\ref{fig:dIdwo}, represents the energy scale which below secondary branching start to dominate \cite{Blaizot:2012fh}, corresponding to $\omega \sim \alpha_s^2 \hat q L^2$, see Eq.~\eqref{eq:omegas} (for more details, see the next subsection). Finally, the grey band in Fig.~\ref{fig:dIdwo} corresponds to emissions with $\omega \sim \omega_{\rm BH}$, which are given by the Bethe--Heitler spectrum~\cite{Wiedemann:2000za,Andres:2020kfg}. In what follows, we will neglect such emissions since these emissions do not contribute significantly to jet energy loss at high-$p_T$ \cite{Baier:2001yt,Salgado:2003gb}. 

%%%%%%%%%%%%%%%%%%%%%%%%%%%%%%%%%%%%%%%%%%%%%%%%%%%%%%%
\subsection{Single-parton contribution to out-of-cone energy loss}
\label{sec:single-parton-quenching}
%%%%%%%%%%%%%%%%%%%%%%%%%%%%%%%%%%%%%%%%%%%%%%%%%%%%%%%

When the number of medium-induced gluon emissions becomes large, $\int_\omega^\infty \rmd \omega \, \rmd I\big/\rmd \omega >1$, one needs to go beyond \eqref{eq:eloss-prob-one} and account for multiple emissions to the energy loss distribution. Assuming independent emissions, we can treat it as a Poisson process. This allows to define a probability distribution of radiating energy $\varepsilon$ off a single parton at angles larger than the jet cone. This probability distribution reads
\begin{align}\label{eq:eloss-prob-multi}
    \mathcal{P}_>^{(0)}(\varepsilon)&=\sum^\infty_{n=0}\frac{1}{n!}\left[\prod^n_{j=1}\int\rmd\omega_j\frac{\rmd I_>}{\rmd\omega_j}\right]\delta\left(\varepsilon-\sum^n_{j=1}\omega_j\right)\exp\left[-\int\rmd\omega\,\frac{\rmd I_>}{\rmd\omega}\right],
\end{align} 
where the parton radiates $n$ soft gluons with $\omega_j$ energies summing up to $\varepsilon$, and there is a Sudakov exponential factor to resum virtual contributions~\cite{Baier:2001yt}. In Eq.~\eqref{eq:eloss-prob-multi}, we only account for the primary emissions off the leading particle and neglect any secondary splittings. Performing the Laplace transform, the quenching factor of a single parton emitting multiple gluons can be calculated using
\begin{equation}\label{eq:Quenching_factor_singleparton_nothermal}
    \mathcal Q^{(0)}_>(\nu)\equiv\int^\infty_0\rmd\varepsilon\,\mathcal P^{(0)}_>(\varepsilon)\,\rme^{-\nu \varepsilon}=\exp\left[-\int^\infty_0\dd\omega\frac{\dd I_>}{\dd\omega}\left(1-{\rm e}^{-\nu\omega }\right)\right],
\end{equation}
where $\nu=n/p_T$. 

It is here worth emphasizing the role of the hard emissions described by the $N=1$ spectrum. Neglecting for the moment broadening, i.e. setting $R=0$ in Eq.~\eqref{eq:GLV_spectrum_out}, the spectrum at large $\omega > \omega_c \gg \mu^2 L/2$ is simply
\begin{equation}
    \left.\frac{\rmd I^{N=1}}{\rmd \omega} \right|_{\omega > \omega_c} = \frac{\bar \alpha \pi}{4} \frac{\hat q_0 L^2}{\omega^2} \,.
\end{equation}
The resulting single-parton quenching factor for this regime behaves as
\begin{align}\label{eq:qf-single-hard}
    \mathcal Q^{(0),N=1}_>(\nu) &= \exp \left[- \frac{\bar \alpha \pi}{2} \frac{\hat q_0}{\hat q} \left( 1- \rme^{-\nu \omega_c} + \nu \omega_c \Gamma(0,\nu\omega_c) \right) \right]  \,,
\end{align}
where $\Gamma(s,x) = \int_x^\infty \rmd t\, t^{s-1}\rme^{-t}$ is the upper incomplete gamma function. At low $p_T$, i.e. $p_T \ll n\omega_c$, the quenching becomes at most $\mathcal Q^{(0),N=1}(\nu)|_{p_T \ll n \omega_c} \approx 1 - \alpha_s C_R \hat q_0/(2\hat q)$ constant factor. Hence, we conclude that the impact of hard radiation at LHC, where we expect $\hat q > \hat q_0$, is relatively small. However, at high-$p_T$, i.e. $p_T \gg n \omega_c$, where the leading behavior is $Q^{(0)}(\nu) \approx Q^{(0),N=1}(\nu)|_{p_T \gg n \omega_c} \approx 1 - \frac{\alpha_s}{4}C_R\,\hat q_0L^2 \nu (1-\gamma_E - \ln \nu \omega_c)$, it becomes more important. Also, since hard radiation takes place at small angles, the effect is even smaller at $R>0$. 

Equation~\eqref{eq:Quenching_factor_singleparton_nothermal} also includes contributions from small energy gluons, and an important modification should be included to improve their description in the medium. As mentioned above, gluons emitted with energy $\omega_{\rm BH}<\omega<\omega_s$, where
\begin{equation}\label{eq:omegas}
\omega_s \equiv \pi \left(\frac{N_c \alpha_s}{\pi} \right)^2 \hat q_0L^2 \,,
\end{equation}
will thermalize quasi-instantaneously in the plasma via multiple branching~\cite{Blaizot:2012fh}. Their energy will basically be redistributed randomly over a cone with characteristic opening angle $R_{\rm rec}\sim \pi/2$. Hence, instead of losing energy $\omega$ out of the cone, the jet loses $\omega[1-(R/R_{\rm rec})^2]$, where the second power comes from the area proportionality. This process describes the thermalization of soft jet particles. Moreover, if $R_{\rm rec}(\omega,\eta,\phi)$, where $(\eta,\phi)$ describes the jet direction with respect to the reaction plane, one could use it to describe back-propagation of the thermalized energy to the cone, and thus medium response. The dashed line in Fig.~\ref{fig:dIdwo} shows the location of $\omega_s\approx8.5$ GeV for our parameters. Depending on the medium and jet parameters, this scale can be below or above the matching scale $\omega_\star$. In Ref.~\cite{Mehtar-Tani:2021fud}, the importance of $R_{\rm rec}$ was studied and small dependence in the result was observed and thus we used $R_{\rm rec}=\pi/2$.

Finally, after neglecting emissions below the Bethe-Heitler energy (we assume that it is given by the plasma temperature $\omega_{\rm BH}=T_0$), our final form for the quenching factor of a single parton inside the jet is therefore
\begin{equation}\label{eq:Quenching_factor_singleparton}
    \mathcal Q^{(0)}_>(\nu)=\exp\left[-\int^\infty_{T_0}\dd\omega\frac{\dd I_>}{\dd\omega}\left(1-{\rm e}^{-\nu\omega\left(1- \Theta(\omega_s-\omega)\frac{R^2}{R^2_{\rm rec}}\right)}\right)\right]\,,
\end{equation}
where the cone size dependence is implicit in the integration limits of the out-of-cone spectrum. The single parton quenching factor Eq.~\eqref{eq:Quenching_factor_singleparton} is shown in Fig.~\ref{fig:RAA} with dashed lines for quark and gluon initiators. By opening the cone, the emitted energy gets gradually recovered, and thus the quenching factor becomes closer to 1. The difference in between quark and gluon initiators is $\mathcal Q^{(0)}_{>,g}=(\mathcal Q^{(0)}_{>,q})^{C_A/C_F}$, resulting more quenching for gluon. We expect our description to be less valid at smaller energies.

%%%%%%%%%%%%%%%%%%%%%%%%%%%%%%%%%%%%%%%%%%%%%%%%%%%%%%%
\subsection{Quenching the whole jet}
\label{sec:multi-parton-quenching}
%%%%%%%%%%%%%%%%%%%%%%%%%%%%%%%%%%%%%%%%%%%%%%%%%%%%%%%

Having derived how \emph{one parton} contributes to the energy loss of the whole jet, we next turn to the calculation of how \emph{multiple partons} in the jet, resolved by the medium during their fragmentation process, add up to the total quenching effect.

Due to the large phase space for radiation between the jet scale $\sim p_T R$ and the hadronization scale $\sim \Lambda_{\rm QCD}$, the jet forms through multiple emissions. It can be estimated, from formation time arguments, that many of these emissions occur while the parton(s) are interacting with the surrounding medium~\cite{Mehtar-Tani:2017web}. In the limit of complete decoherence, the constituents are expected to lose energy independently, following an incoherent superposition of single-particle quenching factors in Eq.~\eqref{eq:Quenching_factor_singleparton}. However, one has account for coherence effects leading to a finite resolution power of the medium. As long as two partons are closer to each other than the medium resolution length, the medium cannot resolve them individually. The two partons are affected coherently (as a whole color charge) by medium interactions and, in particular, by induced energy loss~\cite{Mehtar-Tani:2017ypq}. The relevant time-scale can be estimated by comparing the size of a dipole, that in the small-angle approximation roughly scales as $r_\perp \sim \theta t$ (where $\theta$ is the angle of the dipole and $t$ is the propagation time), to the resolution length of the medium, that scales as $\lambda_\perp \sim 1/\langle k_\perp^2\rangle^{1/2} \sim 1/\sqrt{\hat q t}$, where we assume Gaussian transverse-momentum broadening. The two transverse sizes are equal at the decoherence time $t_{\rm d} \sim (\hat q \theta^2)^{-1/3}$.

This condition can be translated to emission times: emissions with  formation times smaller than the medium decoherence time $t_{\rm f}<t_{\rm d}$ are vacuum-like. In other words, they are generated according to the probability distribution to split in the vacuum. The core constituents should ultimately be resolved while they are still in the medium, i.e. $t_{\rm d} < L$. Therefore, jet constituents produced in the phase space delimited by $t_{\rm f} \ll t_{\rm d} \ll L$, will be resolved by the medium and are affected by quenching. The rest of the phase space stays unaffected.

These two effects (vacuum fragmentation and medium resolution) are captured by the collimator function $\mathcal{C}(p_T,R)$~\cite{Mehtar-Tani:2017web}, which is a function of the jet and medium scales. It takes into account the additional energy loss of resolved vacuum-like emissions in the medium (see also Ref.~\cite{Blok:2019uny} for an application to heavy-quark jets). The total quenching of the jet is therefore given as a product of the quenching of the total charge of the jet and the collimator, that is
\begin{equation}\label{eq:Quenching_factor_jet}
    \mathcal{Q}_{>,i}(p_T,R)= \mathcal{Q}^{(0)}_{>,i}\left(\frac{n}{p_T} \right)\,\mathcal{C}_i(p_T,R)\,,
\end{equation}
where $i= q,g$ indicates the dependence on the color charge. This is what we refer to as the fully resummed quenching factor of a jet. The functions $\mathcal C_i$ obey a set of coupled, non-linear evolution equations, see in Ref.~\cite{Mehtar-Tani:2017web}. Here, we use its linear approximation, where the quark and gluon solutions decouple. This allows to write the solution explicitly as
\begin{equation}\label{eq:Collimator_lin}
    \mathcal C_i(p_T,R)=\exp\left[-\int^R_0\frac{\rmd\theta}{\theta}\int_0^1\rmd z\,\frac{\alpha_s(k_\perp)}{\pi}P_{gi}(z)\Theta_{\rm res}\left(\mathcal Q^{(0)}_{>,g}(n/p_T)-1\right)\right]\,,
\end{equation}
where $\alpha_s$ is the 1-loop running coupling, the relative transverse momentum is $k_\perp=z(1-z)p_T\theta$, and $P_{gi}(z)$ is the Altarelli--Parisi LO splitting function. 
The finiteness of the integrals is ensured by the phase space measure $\Theta_{\rm res}=\Theta(L-t_{\rm d})\Theta(t_{\rm d}-t_{\rm f})$, with the corresponding times $t_{\rm f}=2z(1-z)p_T/k_\perp^2$ and $t_{\rm d}=[12/(\hat q_0\theta^2)]^{1/3}$. At large $p_T$, $\pT \gtrsim \hat q_0 L^2$, this implies that the angular integral is directly regulated by $\theta > \theta_c$, where $\theta_c=[12/(\hat q_0L^3)]^{1/2}$. In the opposite case, $\pT \lesssim \hat q_0 L^2 $, the angular integral is regulated by $\theta \gtrsim (\hat q/p_T^3)^{1/4}$. Finally, if $\theta_c > R$ the jet is completely coherent and $\mathcal C(\pT,R) = 1$. 

This linearized version of the collimator function is analytically calculable, which is a big advantage in comparison to the full, non-linear version. We also tested against the full non-linear solution, which resulted in small, <10\%, deviations even for big $R \sim 1$ cones.

\begin{figure}
    \centering
    \includegraphics[width=0.49\textwidth]{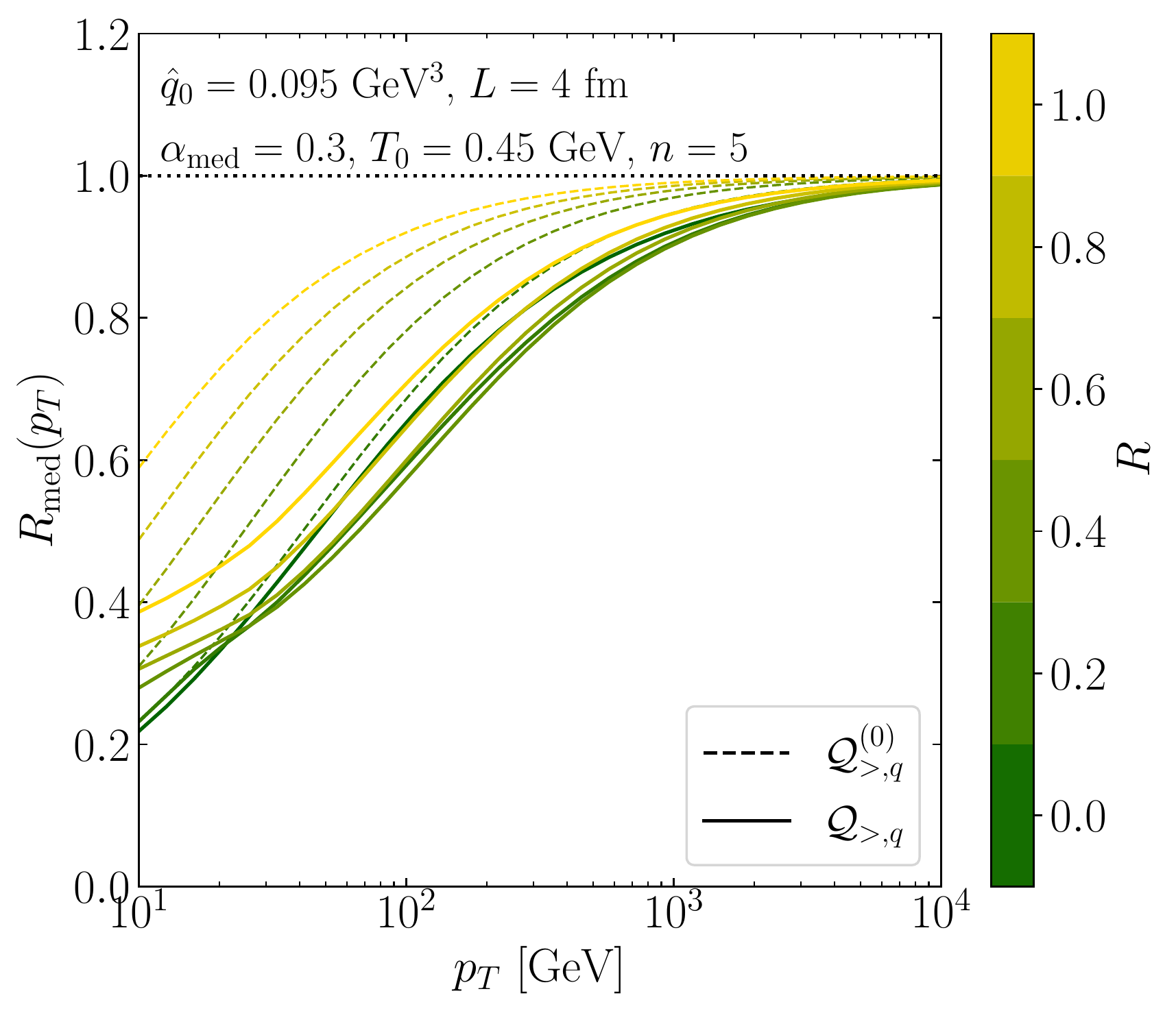}
    \includegraphics[width=0.49\textwidth]{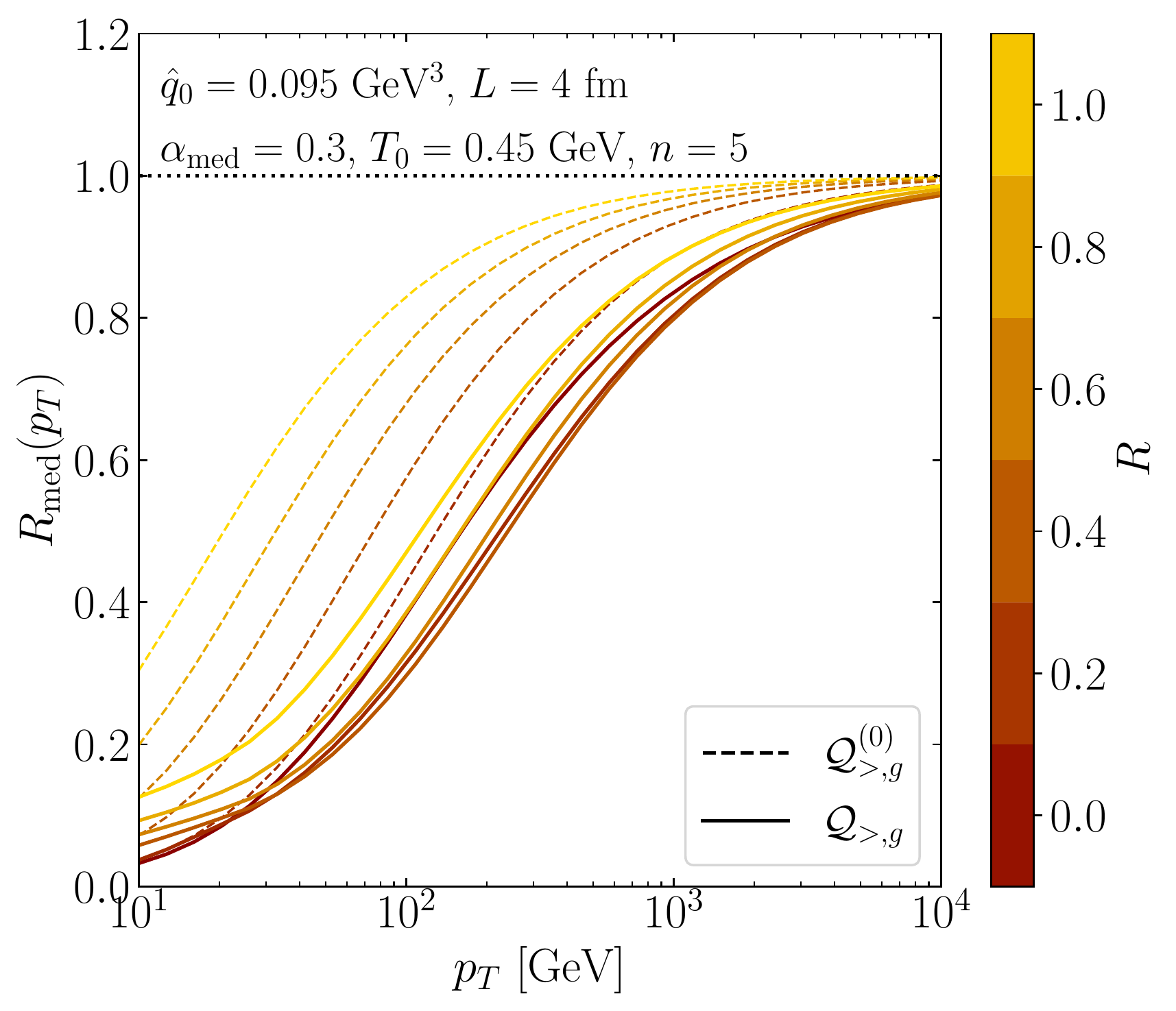}
    \caption{The nuclear modification factor using the matched radiation spectrum for single parton quenching form Eq.~\eqref{eq:Quenching_factor_singleparton} (dashed line), and quenching the whole jet with coherence effects from Eq.~\eqref{eq:Quenching_factor_jet} (solid line). Quark jets on the left and gluon jets on the right panel with different cone sizes.}
    \label{fig:RAA}
\end{figure}

The resummed quenching weight $Q_{>,i}(p_T,R)$ is shown in Fig.~\ref{fig:RAA} with solid lines for quarks (left) and gluons (right) for different cone sizes. It results in more quenching (with the same medium parameters) compared to the single-parton quenching factor because there are more jet constituents that contribute to the total energy loss. The $R$-dependence is a result of two competing effects: a) the recapture of medium-induced gluons by opening the cone, and b) the opening of phase space for vacuum-like emissions that source additional energy loss. The combination of these effects balances out, leading to a very mild cone size dependence. For a full discussion of the uncertainties related to the choice of medium scales and parameters involved in the quenching, see also Ref.~\cite{Mehtar-Tani:2021fud}.

We would like to emphasize the flexibility of the collimator function $\mathcal C_i$, which is independent of the particulars of the model of energy loss employed in the previous sections. One could start with other models for single-parton quenching $\mathcal Q^{(0)}$ and the resolved phase space $\Theta_{\rm res}$, and then use the collimator to describe the quenching of the multiple resolved sources inside the jet. 
As a concrete example, we provide an alternative calculation in the context of the hybrid weak- and strong-coupling model \cite{Casalderrey-Solana:2014bpa} in App.~\ref{sec:Other_models}.

%%%%%%%%%%%%%%%%%%%%%%%%%%%%%%%%%%%%%%%%%%%%%%%%%%%%%%%
%%%%%%%%%%%%%%%%%%%%%%%%%%%%%%%%%%%%%%%%%%%%%%%%%%%%%%%
\section{Quenching effects in the cumulative spectrum}\label{sec:Quantile}
%%%%%%%%%%%%%%%%%%%%%%%%%%%%%%%%%%%%%%%%%%%%%%%%%%%%%%%
%%%%%%%%%%%%%%%%%%%%%%%%%%%%%%%%%%%%%%%%%%%%%%%%%%%%%%%

The jet suppression factor, defined in Eq.~\eqref{eq:Q_def}, compares the jet spectra in heavy-ion collisions (medium), to that in proton-proton collisions (vacuum) at the same final $p_T$. 
In this section, we turn to the discussion of other observables that could be constructed from the inclusive jet spectra measured in these two systems. One alternative is to compare the cumulative of the jet spectra (i.e., integrated above a fixed $p_T$ cut). The cumulative is less affected by the initial shape of the hard spectrum and has better statistical uncertainties. Another approach is to estimate the $p_T$-shift necessary to match the pp and AA spectra. This is closely related to the typical amount of energy lost by a jet. Both of these procedures are straightforwardly related to the quenching factors entering the jet spectrum and will be discussed in further detail in Sec.~\ref{sec:Other_Quantiles}. Now, we turn to an observable that combines the strength of the two examples given above, namely the quantile procedure. 

%%%%%%%%%%%%%%%%%%%%%%%%%%%%%%%%%%%%%%%%%%%%%%%%%%%%%%%
\subsection{The quantile ratio}
\label{sec:quantile_ratio}
%%%%%%%%%%%%%%%%%%%%%%%%%%%%%%%%%%%%%%%%%%%%%%%%%%%%%%%

The quantile procedure was introduced in Ref.~\cite{Brewer:2018dfs} and aims to unfold the average momentum shift $\langle\varepsilon\rangle$ between vacuum and quenched jets. First, let us introduce the tail cumulative of the spectrum,
\begin{equation}\label{eq:cumulative_spectrum}
    \Sigma(p_T,R) \equiv \int^\infty_{p_T}\rmd p'_T\,\frac{\rmd \sigma_R}{\rmd p_T'} \,,
    \end{equation}
which is a probability after dividing with the full integral. The quantile procedure compares the medium and the vacuum spectrum at equal probabilities, $\Sigma^{\rm med}(p_T^{\rm q,med},R) = \Sigma^{\rm vac}(p_T^{\rm q,vac},R)$. In heavy-ion collisions, for a fixed $p_T^{\rm q, med}$, this condition allows identifying the corresponding $p_T^{\rm q, vac}$. Finally, the quantile momentum ratio is defined as 
\begin{align}\label{eq:Quantile_def}
    Q_{\rm med}(p^{\rm q,med}_T)&\equiv\left.\frac{p^{\rm q,med}_T}{p_T^{\rm q,vac}}\right|_{\Sigma}\,.
\end{align}
Therefore, $p_T^{\rm q,vac}$ is the momentum of vacuum jets above which vacuum and medium jets have equal probability to be produced. 

For a quick estimate, let us assume a steeply falling spectrum with a fixed power $n={\rm const}$, and neglect the $R$-dependence of the quenching. The tail cumulative cross-sections in vacuum and in medium (see Eq.~\eqref{eq:medium_spectrum}--\eqref{eq:Q_def}) are simply
\begin{align}
    \left. \Sigma^{\rm vac}\left(p^{\rm q,vac}_T\right)\right|_{n={\rm const}}&=\frac{1}{n-1}\left(p_T^{\rm q,vac}\right)^{1-n}\,,\\  \nonumber
    \left.\Sigma^{\rm med}\left(p^{\rm q,med}_T\right) \right|_{n={\rm const}} &= \int^\infty_{p_T^{\rm med}}\rmd p_T\, p_T^{-n}\mathcal{Q}(n/p_T)\,.
\end{align}
This results in the quantile momentum ratio
\begin{equation}\label{eq:Quantile_momentum_fixed_n}
    \left. Q_{\rm med}(p_T^{\rm q,med}) \right|_{n={\rm const}} = p_T^{\rm q,med}\left[(n-1)\int^\infty_{p^{\rm q,med}_T}\dd p_T\,p_T^{-n}\mathcal Q(n/p_T)\right]^{\frac{1}{n-1}}\,.
\end{equation}
To get the feeling for this quantity, it is instructive to consider a few simplified scenarios for the quenching factor $\mathcal Q(n/p_T)$. First, for a constant quenching factor $\mathcal{Q}(n/p_T)=\mathcal{Q}_0$, the quantile ratio is a trivial function of the quenching factor $Q_{\rm med}=\mathcal{Q}_0^{1/(n-1)}$. Next, we will consider the single-parton quenching factor obtained by using the soft limit ($\omega \ll \omega_c$ in Eq.~\eqref{eq:BDMPSZ_spectrum}) of the BDMPS-Z spectrum, which is derived in App.~\ref{sec:qf-scaling}. The interplay between the jet cone and the broadening introduces a characteristic energy scale $\omega_R = \sqrt{\hat q L}/R$ which defines two regimes that we discuss below:
\begin{itemize}
    \item  For $p_T \ll n \omega_R$, it scales parametrically as $\mathcal Q^{(0)}_>(p_T)= \exp[-\sqrt{ \omega_1 n/p_T}]$ where $\omega_1 = 8 \bar \alpha^2 \pi \omega_c$ is a characteristic energy scale of the medium. In this case the quantile ratio becomes 
    \begin{equation}
        Q_{\rm med}(p_T)\approx \exp\left[- \sqrt{\omega_1/((n-1)p_T)}\right]=\left[\mathcal Q^{(0)}_>(p_T)\right]^{\frac{1}{\sqrt{n(n-1)}}}\,.
    \end{equation}
    \item In the high-$p_T$ regime, $p_T \gg n \omega_R$, we instead have $\mathcal Q_>^{(0)} = \exp[- \omega_2 n/p_T]$ with $\omega_2 = 2 \bar \alpha \sqrt{2 \omega_c \omega_R}$. This $p_T$ dependence is similar to that of medium-induced single hard scattering, cf. Eq.~\eqref{eq:qf-single-hard}, and elastic drag, cf. Eq.~\eqref{eq:Quenching_factor_singleparton_elastic}. The quantile reads
    \begin{equation}
        Q_{\rm med}(p_T) = \left[\frac{n-1}{y^{n-1}} \,\gamma\left(n-1,y\right) \right]^{\frac{1}{n-1}}  \,,
    \end{equation}
    where $y=n \omega_2/p_T$ is the scaling variable and $\gamma(s,x)= \Gamma(s) - \Gamma(s,x)$ is the lower incomplete gamma function.
\end{itemize}
Both the quantile ratio $Q_{\rm med}$, and the jet suppression factor $R_{\rm med}$ depends identically on a dimensionless ratio of a medium scale over the jet transverse momentum. Strikingly, the main difference resides in the $n$ dependence. It turns out that the relation between quantile and quenching factor $Q_{\rm med} \simeq \mathcal Q_>^{1/(n-1)}$ holds approximately also for $p_T$-dependent quenching factors---at least for the case of fixed $n$. In particular, given that $\ln \mathcal Q^{(0)}_> \propto 1-n \omega_2/p_T$ at $p_T \gg n \omega_R$, we should expect a reduced sensitivity of the quantile to the details of the initial spectrum at high-$p_T$, i.e. $\ln Q_{\rm med} \propto 1-\omega_2/p_T$.

Here, we have mostly focused on the contribution from the out-of-cone, soft radiation spectrum to quenching. However, both quenching by hard emissions, see Eq.~\eqref{eq:qf-single-hard} and discussion below, and elastic energy loss, see App.~\ref{sec:Corrections_elastic}, behave in a similar fashion. For a single parton species, we should therefore expect to see a universal behavior, independent of the hard spectrum of the quenching at high-$p_T$.

\begin{figure}
    \centering
    \includegraphics[width=0.49\textwidth]{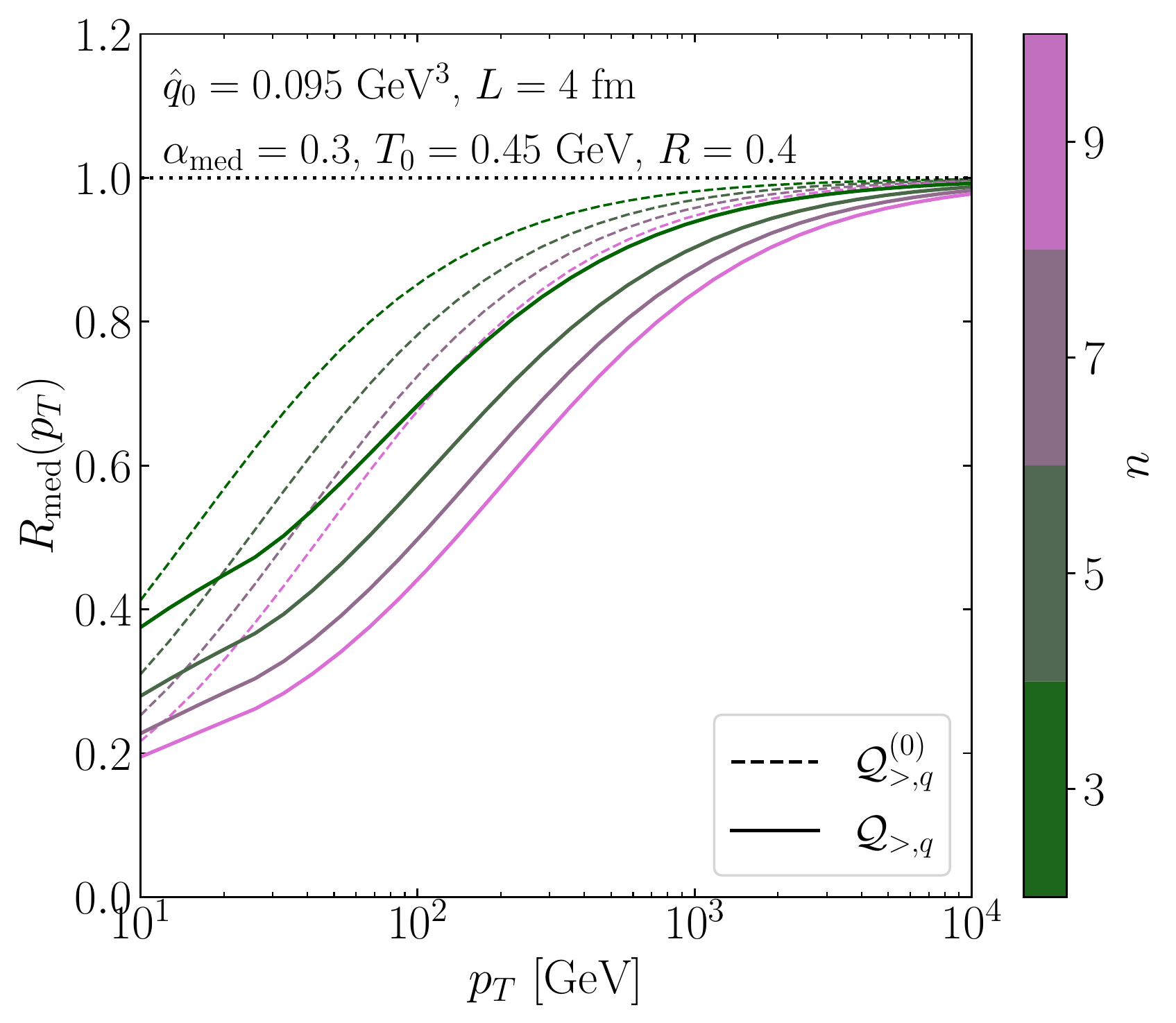}
    \includegraphics[width=0.49\textwidth]{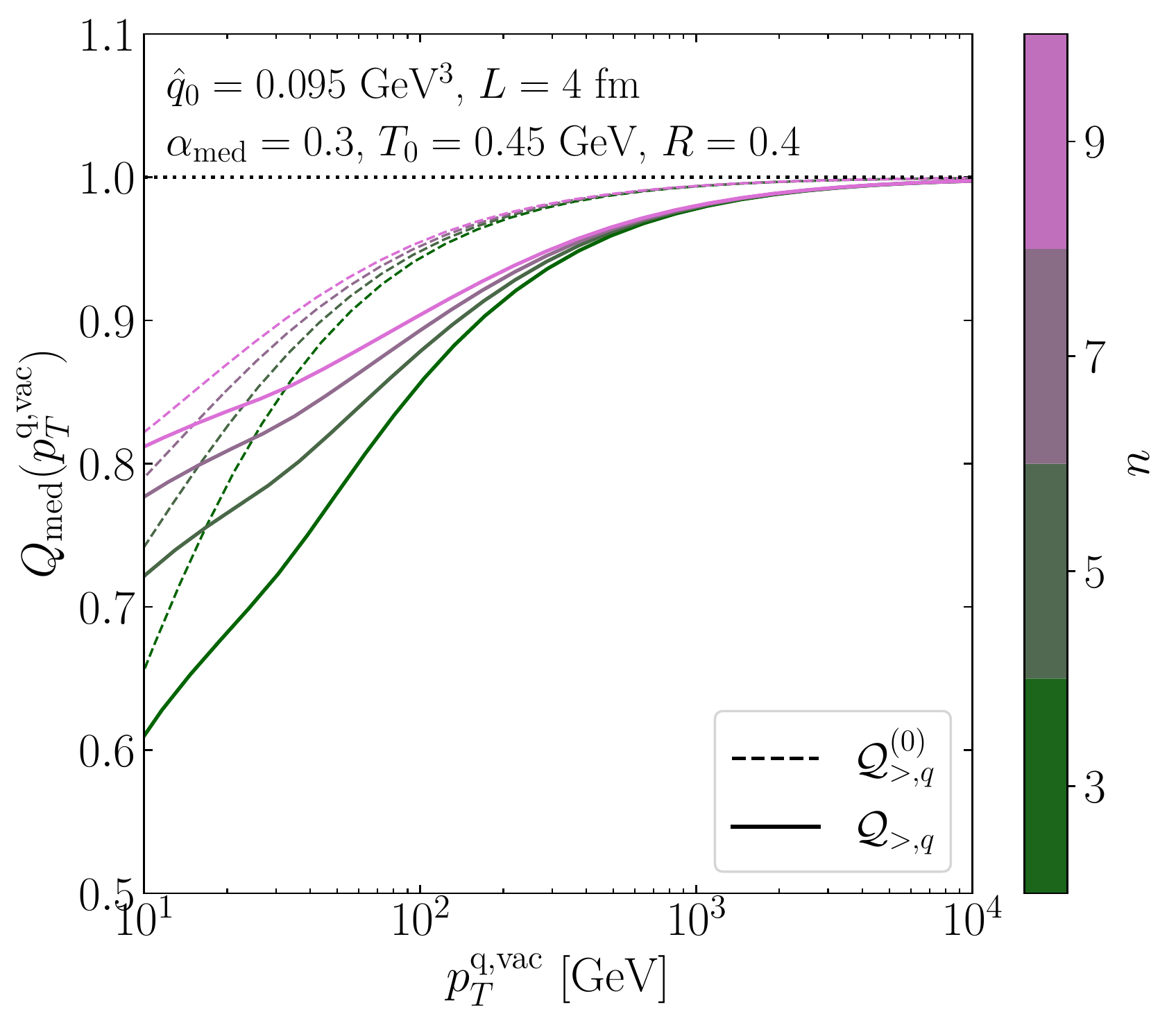}
    \caption{The quenching factor (left) and the quantile ratio (right) for quarks only, using single parton quenching (dashed lines) and quenching the whole jet (solid lines) for different spectrum power.}
    \label{fig:R_AA_QAA_n_dependence}
\end{figure}
We study the onset of the independence of the power-index of the hard spectrum $n$ in Fig.~\ref{fig:R_AA_QAA_n_dependence}. In these plots, we have computed the single-parton quenching factor $\mathcal Q^{(0)}_>$ for a \emph{single parton species}, in this case quarks, using as input the full radiative spectrum from Eq.~\eqref{eq:Matched_spectrum}, which encompass both soft and hard medium-induced emissions (dashed lines). We also plot the resummed quenching factor $\mathcal Q_>$, given in Eq.~\eqref{eq:Quenching_factor_jet}, (solid lines). On the left in Fig.~\ref{fig:R_AA_QAA_n_dependence}, we plot quenching factor $R_{\rm med}(p_T)$ for a wide range of constant $n$ values, $3 \leq n \leq 9$. Strikingly, the amount of quenching varies significantly with $n$ as a function of $p_T$, encapsulating the strong bias effects. The quantile ratio, on the other hand, plotted on the right in Fig.~\ref{fig:R_AA_QAA_n_dependence}, is remarkably resilient to the details of the hard spectrum. This holds both for the single-parton quenching factor, which was expected based on the discussion above, and the fully resummed $\mathcal Q_>$. 

This robustness to the details of the hard, partonic spectrum was observed but not derived analytically in the original paper~\cite{Brewer:2018dfs}. It is the result of the combined effect of using a momentum ratio, and using the cumulative distribution instead of the spectrum. Note, however, that we have observed scaling for quark and gluon contribution independently which only makes sense for pure samples of quark and gluon jets.\footnote{The scaling of the ``gluon quantile'' follows the same trends as for the quarks in Fig.~\ref{fig:R_AA_QAA_n_dependence}, the only difference being the Casimir scaling of the quenching factors, $\mathcal Q_{q,>} = (\mathcal Q_{g,>})^{4/9}$.} For realistic situations, e.g. dijet or boson-jet events, one has first to add up these contributions to the total spectrum (cf. Eq.~\eqref{eq:RAA_PYTHIA}), before computing the cumulative. In this case, the scaling features of the quantile ratio are not necessarily as transparent. We will discuss these issues in more detail in Sec.~\ref{sec:Numerical_results}.

\begin{figure}
    \centering
    \includegraphics[width=0.49\textwidth]{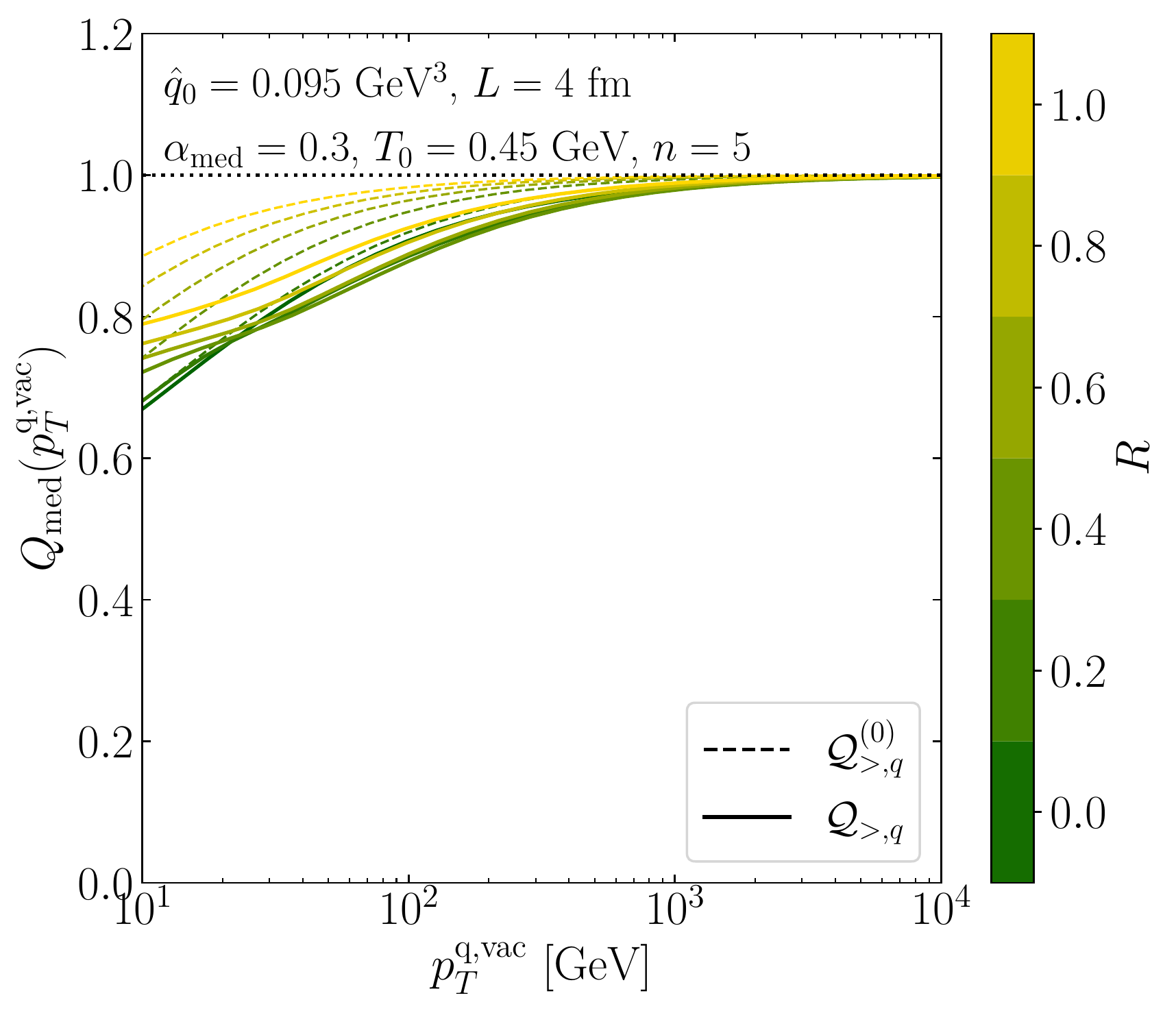}
    \includegraphics[width=0.49\textwidth]{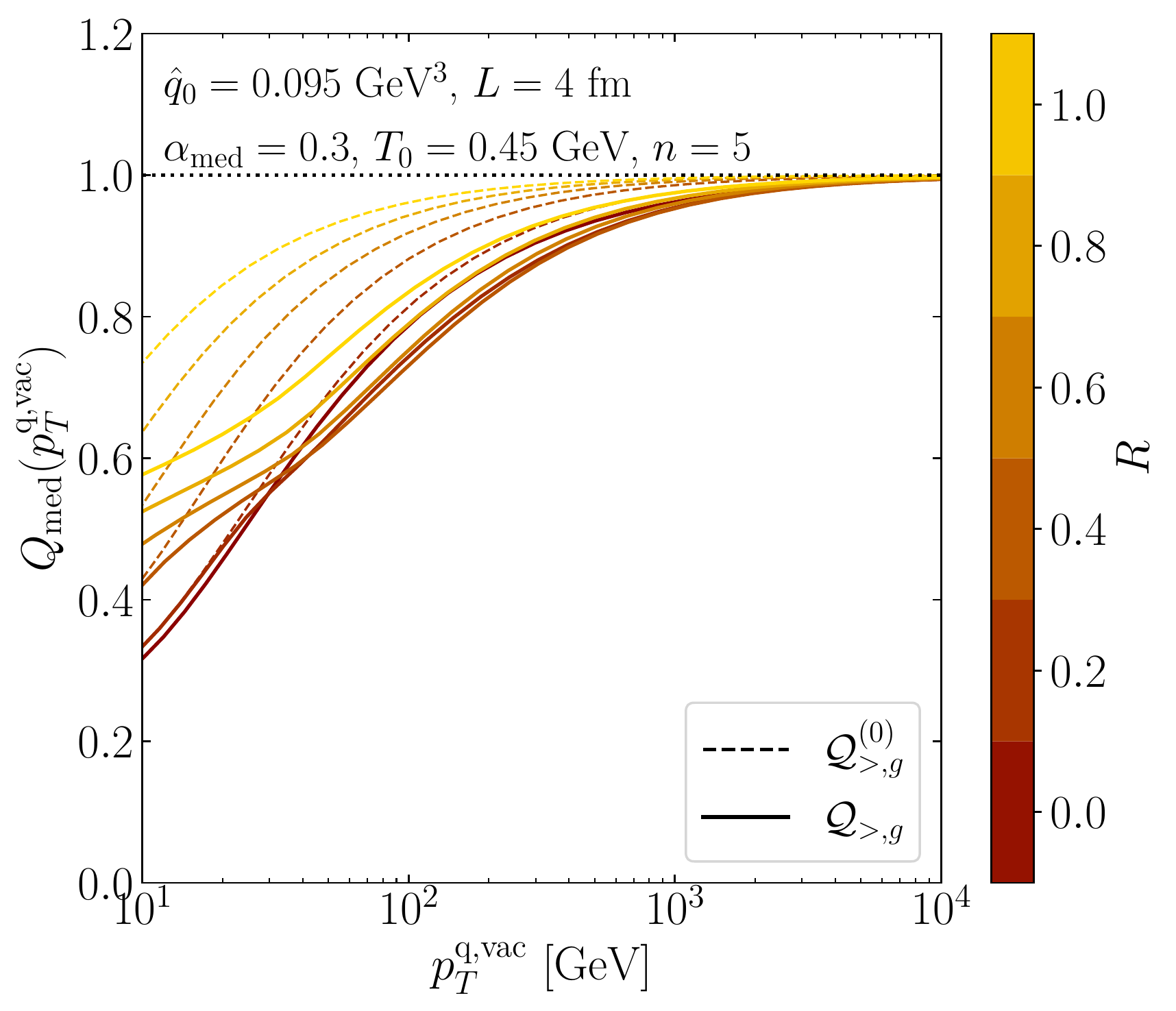}
    \caption{The quantile ratio using the Eq.~\eqref{eq:Quenching_factor_singleparton} for a single parton quenching (dashed line) and quenching the whole jet with coherence effects from Eq.~\eqref{eq:Quenching_factor_jet} (solid line). Quark jets on the left and gluon jets on the right panel for different cone sizes.}
    \label{fig:QAA}
\end{figure}

In order to make contact with the main objective of this paper, namely to provide predictions for the quantile in heavy-ion collisions at the LHC, we round off this section by studying the $R$-dependence of the quantile for pure quark and gluon jets in Fig.~\ref{fig:QAA} for fixed $n=5$ and  medium parameters given in Tab.~\ref{tab:medium-parameters}. As before,  the dashed lines correspond to using Eq.~\eqref{eq:Quenching_factor_singleparton} for the quenching factor, which assumes that the whole jet is quenched as a single parton, while the solid lines employ the resummed quenching factor Eq.~\eqref{eq:Quenching_factor_jet}. The single parton quenching trivially results in less modification and thus smaller momentum shift and quantile ratio for the same medium parameters. Generally, the qualitative features follows the naive expectation $Q_{\rm med} \simeq \mathcal Q_>^{1/(n-1)}$, cf. Fig.~\ref{fig:RAA}. The $R$-dependence is analogous to our previous discussion; less quenching results in a quenching factor closer to 1, and thus a smaller difference between the quantile momenta.

%%%%%%%%%%%%%%%%%%%%%%%%%%%%%%%%%%%%%%%%%%%%%%%%%%%%%%%
\subsection{Statistical advantage of the cumulative spectrum}
\label{sec:tail-cumulative}
%%%%%%%%%%%%%%%%%%%%%%%%%%%%%%%%%%%%%%%%%%%%%%%%%%%%%%%

Using the tail-cumulative distribution has certain advantages in statistical analysis. For a steeply falling distribution $f(p_T)\sim p_T^{-n}$ (like the hadronic or jet spectrum), there are fewer and fewer entries in the higher $p_T$ bins. This results in a rapid increase of the relative statistical uncertainty with $p_T$ such as $\Delta f/f(p_T)\sim p_T^{n/2}(N\Delta p_T)^{-1/2}$, where $N$ is the total number of hits, and $\Delta p_T$ is the size of a bin. By using $\Sigma_f(p_T)\equiv\int^\infty_{p_T}\rmd x\,f(x)$ tail-cumulative distribution, the sum of the higher bins results in less uncertainty $\Delta \Sigma_f/\Sigma_f(p_T)\sim p_T^{-1/2}\Delta f/f(p_T)$. This can be further slowed down using $\Delta p_T(p_T)\sim p_T$ logarithmic binning, resulting in $\Delta f/f\sim p_T^{(n-1)/2}$ and $\Delta \Sigma_f/\Sigma_f\sim p_T^{(n-2)/2}$. It is also true that the cumulative is equivalent to the original distribution, therefore for a given set of statistical samples the tail-cumulative distribution could be advantageous.\footnote{Up to truncation in the domain or co-domain of the distribution.}

We would like to note, however, that one also has to consider systematic uncertainties. Some of them cancel in ratio observables, such as the nuclear modification factor $R_{\rm AA}$. This cancellation is less trivial in the cumulative case.

%%%%%%%%%%%%%%%%%%%%%%%%%%%%%%%%%%%%%%%%%%%%%%%%%%%%%%%
\subsection{Other types of observables based on the spectrum}\label{sec:Other_Quantiles}
%%%%%%%%%%%%%%%%%%%%%%%%%%%%%%%%%%%%%%%%%%%%%%%%%%%%%%%

So far we have discussed the nuclear modification factor and the quantile procedure. But other observables related to the jet spectrum could also be defined (see also in Refs.~\cite{Baier:2001yt,Adler:2006bw,Brewer:2018dfs}). We demonstrate the relation between these observables within the quenching weight formalism and show how these observables are related to the quenching factor $\mathcal Q(n/p_T)$ (or $R_{\rm med}$) and the quantile ratio $Q_{\rm med}$ below.
\begin{itemize}
    \item {\bf Pseudo-quantile~\cite{Brewer:2018dfs}}: is a version of the quantile procedure which matches directly the spectrum instead of the cumulative, $\rmd\sigma^{\rm med}/\rmd \tilde p_T^{\rm q,med} \equiv \rmd\sigma^{\rm vac}/\rmd \tilde p^{\rm q,vac}_T$. The condition relates the two momenta, i.e. $\tilde p_T^{\rm q,vac}(\tilde p_T^{\rm q,med})$. We obtain then,
    \begin{equation}
        \widetilde Q_{\rm med}\left(\tilde p^{\rm q,med}_T\right)\equiv\left.\frac{\tilde p^{\rm q,med}_T}{\tilde p^{\rm q,vac}_T}\right|_{\sigma}\approx \Big[ \mathcal Q_>\left(\tilde p^{\rm q,med}_T\right)\Big]^{-\frac{1}{n(\tilde p^{\rm med}_T)}}\,.
    \end{equation}
    Because of the momentum ratio, this observable has similar $n$ dependence to the quantile ratio, and therefore it is more robust against the initial spectrum. The statistical uncertainty, however, is similar to the $R_{\rm med}$ since bins are not summed. It is equivalent with the momentum shift parameter.
    \item {\bf Momentum shift~\cite{Baier:2001yt,Adler:2006bw}}: is another interpretation of the pseudo-quantile, defined by demanding
    \begin{equation}
        \frac{\rmd\sigma^{\rm med}(p_T)}{\rmd p_T}\equiv\frac{\rmd\sigma^{\rm vac}(p_T+S(p_T))}{\rmd p_T}\,.
    \end{equation}
    The trivial connection between the definitions is $ p^{\rm vac}_T = p_T+S(p_T)$, therefore
    \begin{equation}
        \widetilde Q_{\rm med}(p_T)=1+\frac{S(p_T)}{p_T}\,.
    \end{equation}
    Therefore, the pseudo-quantile and the momentum shift are equivalent. The spectrum shift can also be expressed with the quenching factor,
    \begin{equation}
        S(p_T)=p_T\left[\mathcal Q_>(p_T)^{-\frac{1}{n(p_T)}}-1\right]\,,
    \end{equation}
    expressing the connection between the pseudo-quantile $\widetilde Q_{\rm med}$ and the shift parameter $S(p_T)$.     
    
    A different definition of the momentum shift parameter was used by the PHENIX collaboration $S_{\rm loss}(p_T)$~\cite{Adler:2006bw}. Their definition, however, assumes $n={\rm const}$ and postulates $S_{\rm loss}(p_T)=S_0 p_T$. The $S(p_T)$ and $\widetilde Q(p_T)$ are more general. We found to be necessary to consider changing power $n(p_T)$ in the spectrum, see Sec.~\ref{sec:Numerical_results} and App.~\ref{sec:Corrections_LaplaceAndPower}.
    
    \item {\bf Cumulative-}$\bm{R_{\rm med}}$ (or pseudo-ratio~\cite{Brewer:2018dfs}): is similar to the $R_{\rm med}$ but uses the ratio of the cumulative spectrum instead,
    \begin{equation}\label{eq:Cumulative-RAA}
        \widetilde R_{\rm med}(p_T)\equiv \frac{\Sigma^{\rm med}(p_T)}{\Sigma^{\rm vac}(p_T)}\,,
    \end{equation}
    where the cumulative of the spectrum is defined in Eq.~\eqref{eq:cumulative_spectrum}, and where we have suppressed the $R$ dependence for now.
    The integral reduces bias effects from the initial spectrum and improves the statistics as we showed in Sec.~\ref{sec:tail-cumulative}.
    For a $p_T$-independent quenching weight, as in the single-parton $\mathcal{P}(\varepsilon)$ in the soft BDMPS-Z limit, the cumulative $\widetilde R_{\rm med}$ can be written as 
    \begin{equation}
        \widetilde R_{\rm med} = \int_0^\infty \rmd \varepsilon \, \mathcal{P}(\varepsilon) \frac{\Sigma^{\rm vac}(\pT+ \varepsilon)}{\Sigma^{\rm vac}(p_T)} \,.
    \end{equation}
    A deviation from this expectation indicates a $p_T$ dependence of the quenching weight.
    Furthermore, for a spectrum with constant $n$, where $\Sigma^{\rm vac}(\pT+ \varepsilon)/\Sigma^{\rm vac}(p_T) = (1+\varepsilon/p_T)^{1-n}$, we simply get that $\widetilde R_{\rm med} = \mathcal{Q}((n-1)/p_T)$. Finally, a trivial connection between the $\widetilde R_{\rm med}$ and the quantile ratio for constant $n$ is
    $\widetilde R_{\rm med}(p_T)\approx Q_{\rm med}(p_T)^{n-1}$. 
\end{itemize}

In conclusion, we demonstrated the relationship between the different observables one can construct from the jet spectrum. It turns out that the quantile procedure is particularly appealing due to the reduced sensitivity to the hard spectrum and because of the improvement of the statistical uncertainties.

%%%%%%%%%%%%%%%%%%%%%%%%%%%%%%%%%%%%%%%%%%%%%%%%%%%%%%%
%%%%%%%%%%%%%%%%%%%%%%%%%%%%%%%%%%%%%%%%%%%%%%%%%%%%%%%
\section{Predictions}
\label{sec:Numerical_results}
%%%%%%%%%%%%%%%%%%%%%%%%%%%%%%%%%%%%%%%%%%%%%%%%%%%%%%%
%%%%%%%%%%%%%%%%%%%%%%%%%%%%%%%%%%%%%%%%%%%%%%%%%%%%%%%

In the previous sections, we introduced the quenching weight formalism and showed the way one can construct the quenched jet spectrum and its cumulative. In this section, we extend our formalism to provide more reliable predictions for measurements in heavy-ion collisions.

%%%%%%%%%%%%%%%%%%%%%%%%%%%%%%%%%%%%%%%%%%%%%%%%%%%%%%%
\subsection{Including the realistic spectrum and elastic energy loss}
\label{sec:PYTHIA_spectrum}
%%%%%%%%%%%%%%%%%%%%%%%%%%%%%%%%%%%%%%%%%%%%%%%%%%%%%%%

A realistic calculation of quenching effects has to include the partonic cross-section for jet production. This is evaluated at the hard scale of the collision, $Q_{\rm hard} \sim p_T$. The jet spectrum at a given cone size is computed then by including a DGLAP evolution to the scale $Q_{\rm jet} \sim p_T R$~\cite{Dasgupta:2014yra,Kang:2016mcy,Dai:2016hzf}. Alternatively, the partonic cross-section that results in a jet with a given cone can be parameterized using a Monte Carlo event generator in which the partonic cross-section and the parton shower are matched. In our study, we generate events with Pythia8~\cite{Sjostrand:2014zea} to fit the $p_T$-dependence of the spectrum with reconstruction parameter $R$, for quark-, and gluon-initiated jets. The large-angle DGLAP radiation results in additional $R$-dependence of the $R_{\rm AA}$, through the recapture of vacuum radiation. The spectrum parameterization, therefore, includes the vacuum radiation recaptured by the cone, resulting in an additional $R$ dependence~\cite{Dasgupta:2007wa}, see App.~\ref{sec:Generate_spectrum} and Fig.~\ref{fig:Spectrum_PYTHIA_fit}, in particular. Currently, we restrict our study to inclusive jets in dijet samples, generated in pp, and 0--10\% central PbPb collision at $\sqrt{s_{NN}}= 5.02$ TeV, with $p_T=20-1000$ GeV and $|\eta|<2.8$, similar to the kinematics used by ATLAS~\cite{Aaboud:2018twu}.\footnote{We refer the dijet $R_{\rm AA}$ as single-inclusive because jets contribute independently.} The details of the event generation (excluding ISR, MPI and including the effect of nuclear PDFs in Pb, jet selection, and quark/gluon flavor assignment procedure) are described in App.~\ref{sec:Generate_spectrum}. We also describe the proposed functional form, following Ref.~\cite{Spousta:2015fca}, to fit the spectral indices $n_q(p_T,R)$ and $n_g(p_T,R)$, that automatically parameterizes the $(p_T,R)$ dependence of the quark-gluon fraction. 

The extracted spectra $\rmd\sigma^{\rm pp/AA}_i(p_T,R)/\rmd p_T$, ($i=q,g$) were included to calculate the nuclear modification factor $R_{\rm AA}$, defined as 
\begin{equation}
    R_{\rm AA}(p_T) = \frac{\rmd N^{\rm AA}(p_T,R)/\rmd p_T}{N_{\rm coll}\; \rmd \sigma^{\rm pp}(p_T,R)/\rmd p_T} \,,
\end{equation}
where $N_{\rm coll}$ gives the number of collisions in the nuclear overlap at a given impact parameter and we identify $\rmd N^{\rm AA}/N_{\rm coll} = \rmd \sigma^{\rm AA}$. The main difference between this ratio and the previously defined $R_{\rm med}$, defined in Eq.~\eqref{eq:Q_def}, is the addition of both quark and gluon jets with their respective quenching factors and the nPDFs. As a result, in our framework, we get
\begin{equation}\label{eq:RAA_PYTHIA}
    R_{\rm AA}(p_T,R)=\left.\left[\sum_{i=q,g}\mathcal Q_i\left(p_T,R;n^{\rm AA}_i(p_T,R)\right)\frac{\rmd\sigma^{\rm AA}_i}{\rmd p_T}(p_T,R)\right]\middle/\sum_{i=q,g}\frac{\rmd\sigma^{\rm pp}_i}{\rmd p_T}(p_T,R)\right. \,,
\end{equation}
where we have explicitly written out the dependence of the quenching factor on the spectral index $n_i^{\rm AA}(\pT,R)$. Similarly, the cumulative spectrum in vacuum and medium are given by
\begin{align}
    \Sigma^{\rm pp}(p^{\rm q}_T,R) &= \int^\infty_{p^{\rm q}_T}\rmd p_T\sum_{i=q,g}\frac{\rmd\sigma_i^{\rm pp}}{\rmd p_T}(p_T,R) \,,\\
    \Sigma^{\rm AA}(p^{\rm q}_T,R) &=\int^\infty_{p_T^{\rm q}}\rmd p_T \sum_{j=q,g}\mathcal Q_j\left(p_T,R,n^{\rm AA}_i(p_T,R)\right)\frac{\rmd\sigma_j^{\rm AA}}{\rmd p_T}(p_T,R) \,.
\end{align}
The quantile is finally defined as,
\begin{equation} \label{eq:QAA_PYTHIA}
Q_{\rm AA} = \frac{p_T^{\rm q,AA}}{p_T^{\rm q,pp}} \,,
\end{equation}
where the two momenta are determined from the condition $\Sigma^{\rm pp}(p_T^{\rm q,pp},R) = \Sigma^{\rm AA}(p_T^{\rm q,AA},R)$.

In Sec.~\ref{sec:Quenching}, we focused our discussion on the quenching effects emerging from medium-induced radiation and broadening. However, for realistic predictions we should also include quenching from elastic scattering. Elastic energy loss is described by the transport coefficient $\hat e$, which is related to $\hat q$ through Einstein's fluctuation-dissipation relation $\hat e_g = \hat q/(4T)$ for gluons and $\hat e_q = \hat e_gC_F/N_c $ for quarks \cite{Moore:2004tg,Tachibana:2017syd}. Here, $T=T_0$ is the local temperature of the plasma. We model the single-particle energy loss distribution simply as $\mathcal{P}(\varepsilon) = \delta(\varepsilon - \hat e L)$, where the flavor index is suppressed. Assuming that the energy lost in elastic processes thermalize instantaneously, we also build in the possibility to recover part of this energy through the phenomenological parameter $R_{\rm rec}$, see Eq.~\eqref{eq:Quenching_factor_singleparton}. This finally results in a single-parton quenching factor from elastic energy loss, given by
\begin{equation}\label{eq:Quenching_factor_singleparton_elastic}
    \mathcal Q_{\rm el}^{(0)}(\nu) = \exp\left[- \hat e L n\nu \left( 1 - \frac{R^2}{R_{\rm rec}^2} \right)\right] \,.
\end{equation}
The criteria for resolving the partons in the jet are based on geometry and are therefore assumed to be identical for elastic and radiative processes. Therefore, the complete single-particle quenching factor $\mathcal Q^{(0)}_{>}$, appearing in Eqs.~\eqref{eq:Quenching_factor_jet}--\eqref{eq:Collimator_lin}, should be replaced by
\begin{equation}
    \mathcal Q^{(0)}_{>}(p_T) = \mathcal Q^{(0)}_{>,{\rm rad}}(p_T)\, \mathcal Q^{(0)}_{>,{\rm el}}(p_T) \,,
\end{equation}
where $\mathcal Q^{(0)}_{>,{\rm rad}}(p_T)$ is given by Eq.~\eqref{eq:Quenching_factor_singleparton} and $\mathcal Q^{(0)}_{>,{\rm el}}(p_T)$ is given by Eq.~\eqref{eq:Quenching_factor_singleparton_elastic}. Including elastic effects has an important effect on the magnitude of the total quenching factor. For further detalils see App.~\ref{sec:Corrections_elastic} and Fig.~\ref{fig:Corrections}.

%%%%%%%%%%%%%%%%%%%%%%%%%%%%%%%%%%%%%%%%%%%%%%%%%%%%%%%
\subsection{Numerical results for dijet events}
\label{sec:dijet-RAA}
%%%%%%%%%%%%%%%%%%%%%%%%%%%%%%%%%%%%%%%%%%%%%%%%%%%%%%%

\begin{figure}[t!]
    \centering
    \includegraphics[width=0.49\textwidth]{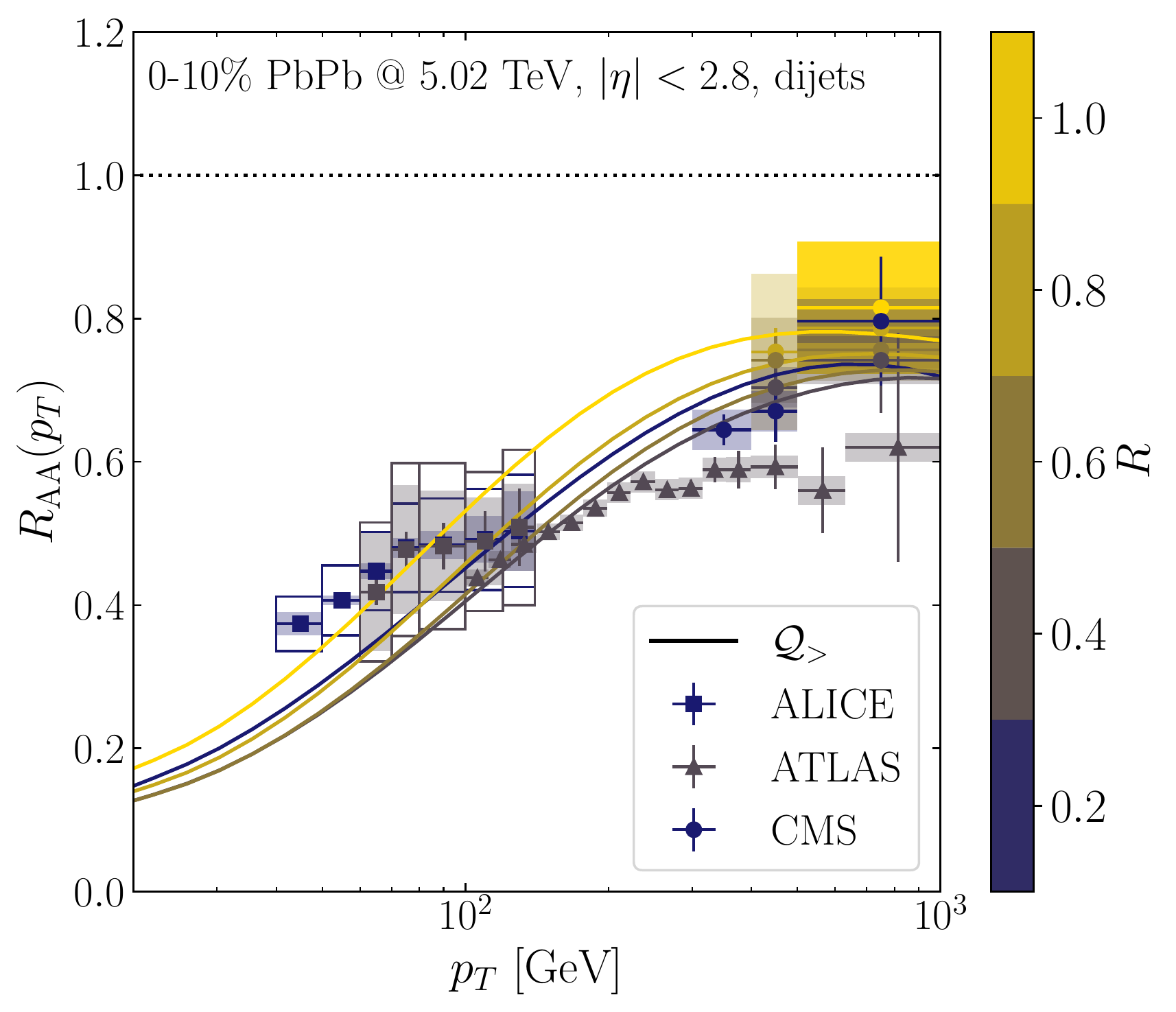}
    \includegraphics[width=0.49\textwidth]{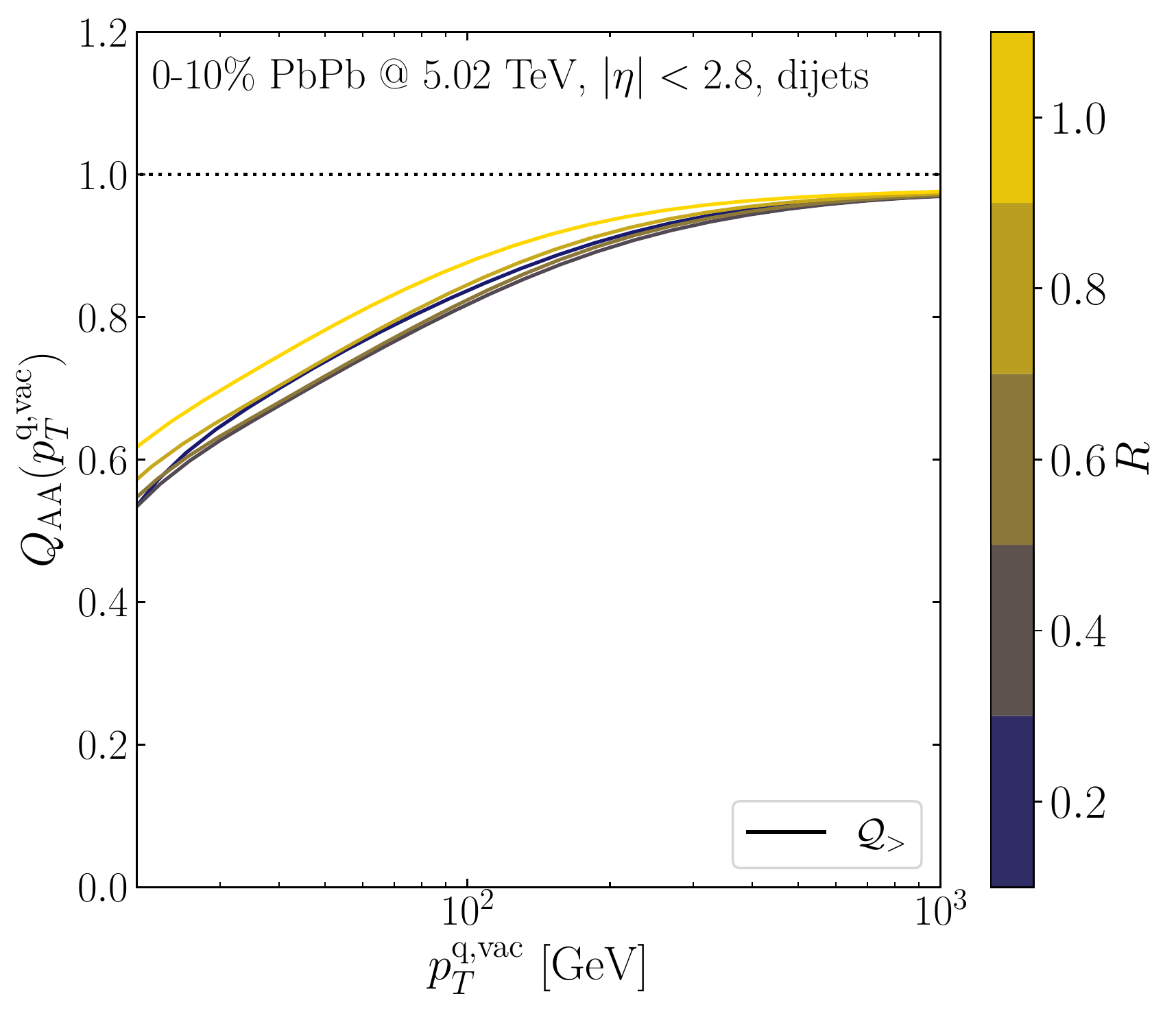}
    \caption{The $R_{\rm AA}$ (left panel) and $Q_{\rm AA}$ (right panel) from Eqs.~\eqref{eq:RAA_PYTHIA} and \eqref{eq:QAA_PYTHIA}, respectively, for single-inclusive jets in dijets events in 0--10\% PbPb collision at $\sqrt{s_{NN}}=$5.02 TeV. The parameters are chosen to reproduce the $R=0.4$ ATLAS~\cite{Aaboud:2018twu} measurement. Data from ALICE~\cite{Acharya:2019jyg} and CMS
    ~\cite{Sirunyan:2021pcp} are also shown.}
    \label{fig:PYTHIA}
\end{figure}

The single-inclusive jet $R_{\rm AA}$, generated from a sample of dijet events, for a set of cone sizes, $0.2< R < 1$, is shown in the left panel of Fig.~\ref{fig:PYTHIA} (solid curves). In the current work, the medium is treated as a static brick with fixed $\hat q_0$ and length $L$, see Tab.~\ref{tab:medium-parameters} for details that is generally a good approximation even for expanding media, see Ref.~\cite{Caucal:2020uic}. There is a notable change of the curves at high $p_T$ due to the inclusion of nPDFs (see also in Fig.~\ref{fig:Spectrum_PYTHIA_fit} in App.~\ref{sec:Generate_spectrum} for only the nPDF effects). The overall $R$ dependence is very modest and will be discussed in more detail shortly. 

The parameters of the calculation are tuned to the measured inclusive jet data from ATLAS~\cite{Aaboud:2018twu} at $p_T \simeq 100$ GeV and $R=0.4$ with $|\eta|<2.8$, cf. Tab.~\ref{tab:medium-parameters}, resulting in good agreement between data and theory for the whole $p_T$ range. The measured inclusive jet $R_{\rm AA}$ from ALICE~\cite{Acharya:2019jyg} for $R=0.2$ and $R=0.4$ are also shown in Fig.~\ref{fig:PYTHIA}, where the rapidity range for the jet selection $|\eta|<0.5$ is slightly different. The recent CMS~\cite{Sirunyan:2021pcp} results are also shown in Fig.~\ref{fig:PYTHIA} for various $R$, where the rapidity is $|\eta|<2$. We would like to note, there is a disagreement between the ATLAS and CMS data that was not pointed out in the CMS' latter publication. We would like to also note that the magnitude of $R_{\rm AA}(p_T)$ in any BDMPS-Z type of calculation is mostly sensitive to the combination $\omega_s\sim \alpha_{\rm med}^2\hat q_0L^2$, as observed in Refs.~\cite{Baier:2001yt,Caucal:2019uvr}. The slope of the $R_{\rm AA}(p_T)$ is quite robust to changes in the parameters.\footnote{By including event-by-event fluctuations in the jet position and path length, the slope becomes flatter in Ref.~\cite{Mehtar-Tani:2021fud}.} Given our simplified modeling of the medium, we do not attempt to reproduce the centrality dependence of the jet $R_{\rm AA}$ at high-$p_T$ which will be left to future work, see also in Ref.~\cite{Mehtar-Tani:2021fud}.

Having constrained the medium parameters with $R_{\rm AA}$, we now turn to the predictions for the quantile momentum ratio which is showed in the right panel of Fig.~\ref{fig:PYTHIA}. There is no drastic change due to the nPDFs in comparison to the $R_{\rm AA}$ at high-$p_T$, demonstrating the robustness of the quantile procedure against modifications in the partonic cross-section. The shape of the curves and even the $R$ dependence is very similar to the $R_{\rm AA}$ and is well captured by the approximate relation $Q_{\rm AA}\sim R_{\rm AA}^{1/(n-1)}$. To date, there are no experimental measurements of the quantile ratio.

Our formalism accounts for the cone size dependence of jet quenching through competing effects related to the early vacuum shower and medium-induced elastic and radiative processes. On the one hand, by opening the cone, one captures more of the particles that are affected by medium interactions and thus recover the lost energy. This is manifested as a suppression of the spectrum $\rmd  I_>/\rmd \omega$ at $\omega > \sqrt{\hat q L}/R$ in Fig.~\ref{fig:dIdwo}, meaning it is less probable to lose energy. For $R\sim R_{\rm rec}$ all quasi-thermalized modes, both in the radiative spectrum ($\omega < \omega_s$) and due to elastic collisions, are recaptured within the jet cone. On the other hand, opening the reconstructed jet cone results in more phase space for vacuum fragmentation at an early stage of the evolution. This leads to a higher multiplicity of vacuum-like emissions and, therefore, more sources for energy loss.

\begin{figure}
    \centering
    \includegraphics[width=0.49\textwidth]{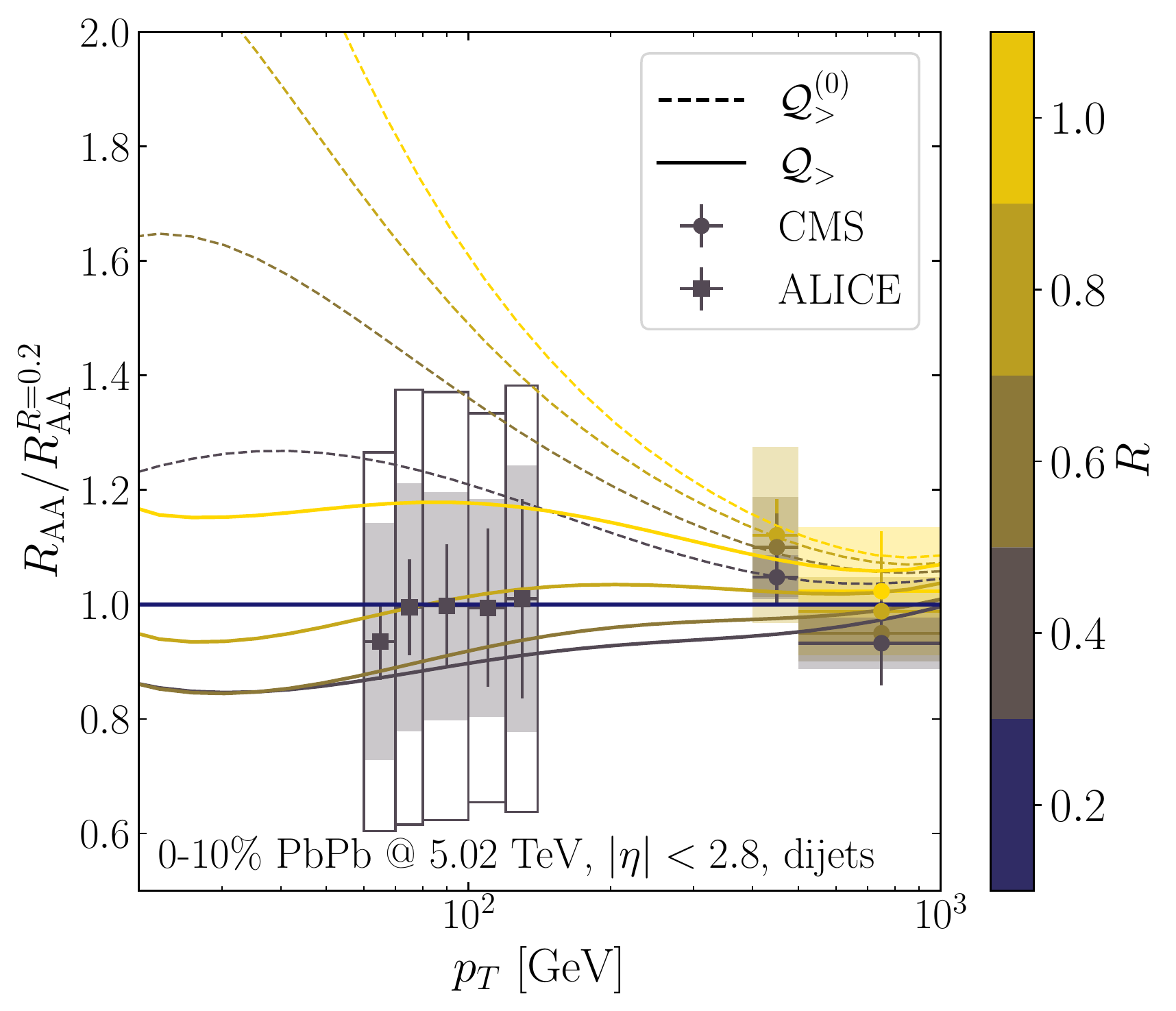}
    \includegraphics[width=0.49\textwidth]{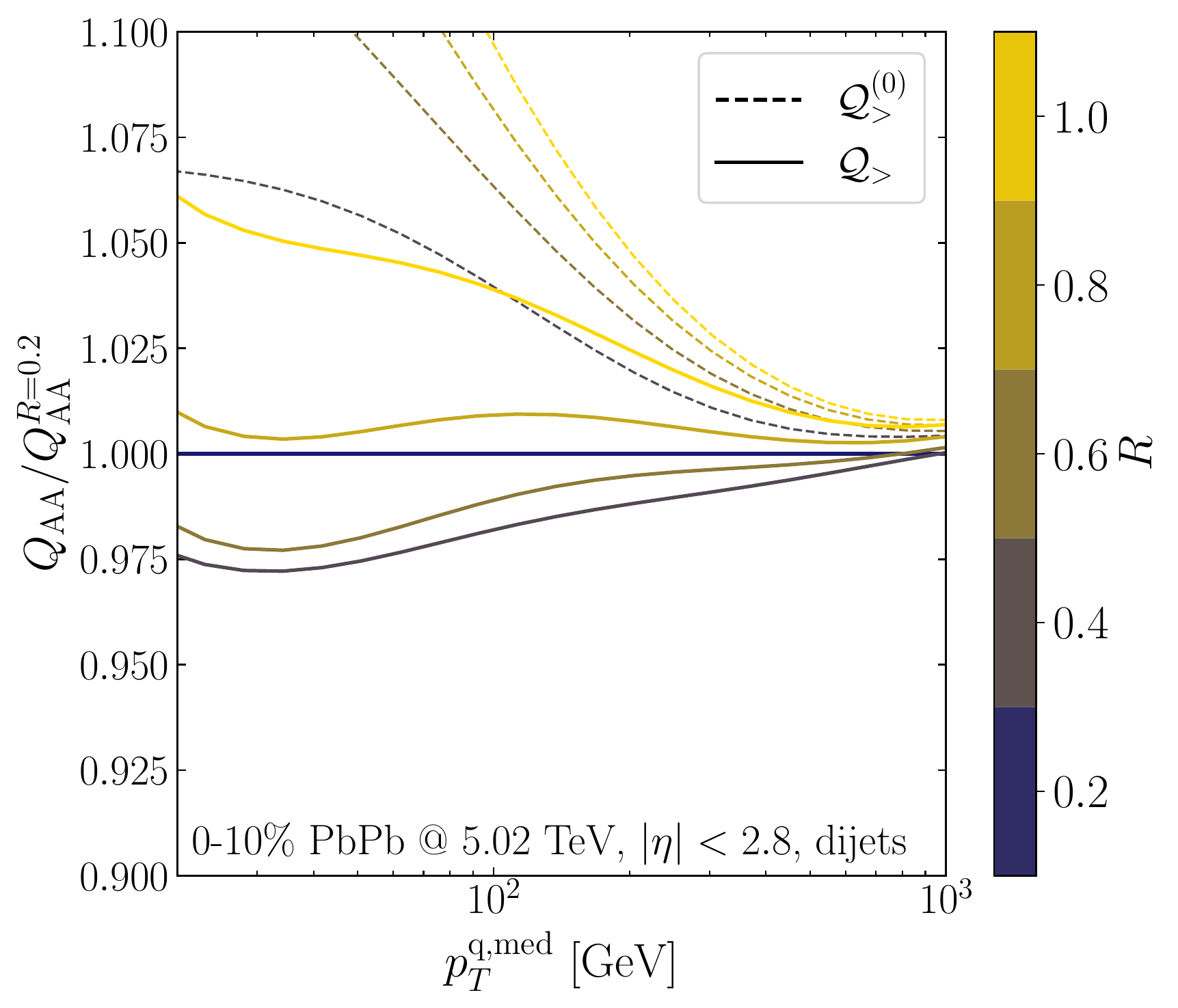}
    \caption{Same plots as in Fig.~\ref{fig:PYTHIA}, but taking the ratio of different cone sizes to enhance the differences between single and multi-parton quenching. On the $x$-axis of the right panel we used $p_T^{\rm q,med}$ instead of $p_T^{\rm q,vac}$ from previous. }
    \label{fig:PYTHIA_ratio}
\end{figure}

The overall effect is a relative cancellation of the $R$ dependence (see Fig.~\ref{fig:PYTHIA}). Figure~\ref{fig:PYTHIA_ratio} shows the $R_{\rm AA}$ (left) and quantile ratio (right) at a given $R$ divided by the $R=0.2$ results. As an illustration, here the dashed curves are the results obtained by using the single-parton quenching factors $Q_>^{(0)}$, which amounts of treating the whole jet as a completely coherent single parton that is not resolved by medium interactions. Their $R$-dependence reflects directly the effect of recapturing energy at large angles without sourcing more energy loss though vacuum fragmentation. The full curves are for the full quenching factors which leads to a more complicated $R$ dependence where, several effects contribute, such as vacuum fragmentation and its recapture, color coherence effects and the thermalization of the medium-induced emissions. This ratio was measured by ALICE~\cite{Acharya:2019jyg} (for $|\eta|<0.5$), and CMS~\cite{Sirunyan:2021pcp} (for $|\eta|<2$), showing great agreement with our model. Reference~\cite{Sirunyan:2021pcp} compares many theory prediction and we can say, currently, our model has the best agreement.

The main differences between the ``bare'', and ``resummed'' quenching factors is at lower $p_T\sim 50-100$ GeV. 
As we mentioned, in our model the $R_{\rm AA}$ is mostly sensitive to the $\omega_s \sim \alpha_{\rm med}^2\hat q_0L^2$ combination of the parameters. A precise measurement on the $R$-dependence would help to constraint more parameters. The right side of Fig.~\ref{fig:PYTHIA_ratio} shows the $R$-dependence of the quantile ratio. It is much less sensitive to the jet cone.

%%%%%%%%%%%%%%%%%%%%%%%%%%%%%%%%%%%%%%%%%%%%%%%%%%%%%%%
\subsection{{Differences between dijets and Z+jets}}
%%%%%%%%%%%%%%%%%%%%%%%%%%%%%%%%%%%%%%%%%%%%%%%%%%%%%%%

At first glance, there should be no difference in the quenching mechanism in dijet and boson+jet events.\footnote{The so-called ``surface bias'' is in our context built in due to the bias from the initial steeply falling spectrum.} 
However, their $R_{\rm AA}$ are not expected to be the same. In this subsection we explain why, and what are the consequences for quark and gluon classification. The Z+jet process is used as an illustration. We also show how can one use the cumulative distribution for quark-gluon discrimination to perform better statistics. 

%%%%%%%%%%%%%%%%%%%%%%%%%%%%%%%%%%%%%%%%%%%%%%%%%%%%%%%
\subsubsection{Difference in the $R_{\rm AA}$}
%%%%%%%%%%%%%%%%%%%%%%%%%%%%%%%%%%%%%%%%%%%%%%%%%%%%%%%
In recent years, a lot of effort has been put to measure and understand the boson+jet processes both in pp and AA collisions. Firstly, it is a favorable process for quark-, and gluon-jet discrimination, because it provides a natural definition of the initial jet flavor. Secondly, measuring the boson momentum, one gets a label on the initial momentum of the recoiling jet. This is especially advantageous in heavy-ion collisions, where the quenching of bosons is suppressed, and one, therefore, gains knowledge about the jet before final-state interactions with the medium.

\begin{figure}[t]
    \centering
    \includegraphics[width=0.49\textwidth]{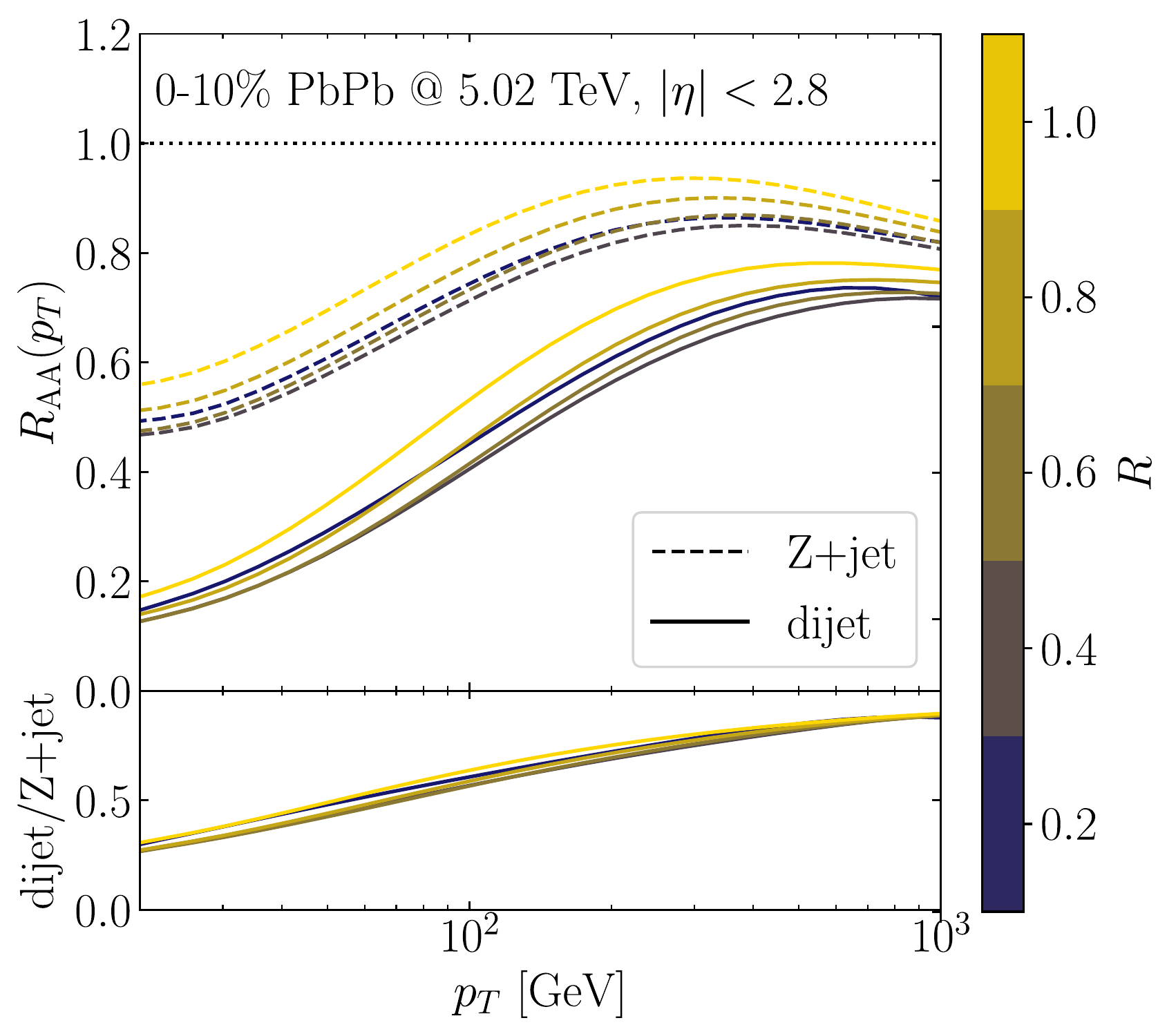}
    \includegraphics[width=0.49\textwidth]{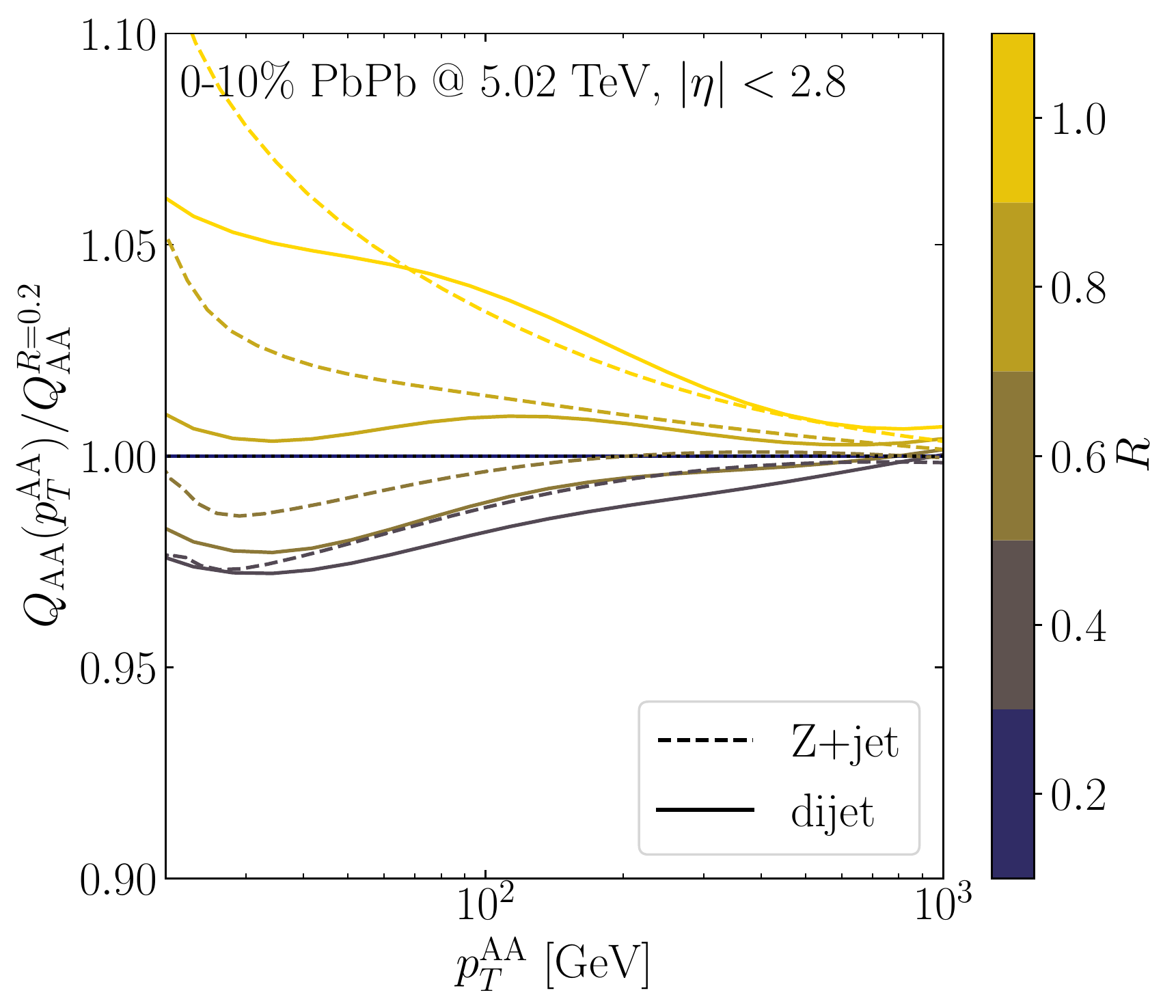}
    \caption{The $R_{\rm AA}$ and $Q_{\rm AA}$ from Eqs.~\eqref{eq:RAA_PYTHIA} and \eqref{eq:QAA_PYTHIA} for single-inclusive jets from dijet and Z+jet events in 0-10\% PbPb collision at 5.02 TeV, keeping the same parameters as in Fig.~\ref{fig:PYTHIA}.}
    \label{fig:Zjet_PYTHIA}
\end{figure}
Here, we focus on Z+jet process, but the arguments are valid for other boson+jet processes. We generated the Z+jet events with Pythia and identified jets using the same cuts as in the dijet study. Further details on the event generation and parameterization are summarized in App.~\ref{sec:Generate_spectrum}. We include the corresponding spectrum using Eq.~\eqref{eq:RAA_PYTHIA} and the result is shown in the right panel of Fig.~\ref{fig:Zjet_PYTHIA}. Dijet and boson+jet processes have different spectra and thus their bias is different on the energy loss distribution. As discussed before, this appears through the spectrum power $n(p_T)$ in the quenching factor. For dijets $n\approx6$, for Z+jets $n\approx4$, therefore based on the approximate $R_{\rm AA}\sim\exp(-2\sqrt{\pi\bar\alpha^2\omega_c n/p_T})$, the bigger the power, the stronger is the quenching. To stress this, the dijet-, and Z+jet-$R_{\rm AA}$ ratio is also shown on the lower panel. The ratio of two exponential remains to be exponential that looks linear on the semi-log scale. Moreover, dijets are gluon dominated, while Z+jets are quark dominated, see in Fig.~\ref{fig:Spectrum_PYTHIA_fit} in App.~\ref{sec:Generate_spectrum}. Therefore we expect Z+jets to have smaller quenching, which is in agreement with Fig.~\ref{fig:Zjet_PYTHIA}. The $p_T$ dependence of the power is also different, resulting in a different slope of $R_{\rm AA}$ as high-$p_T$. Note also that the Z+jet quark contribution is relatively constant in $p_T$ in contrast to the increasing quark contribution in dijets. Since the quenching roughly scales with the color charge (gluons have more quenching), at higher $p_T$, Z+jets has stronger quenching, which is in agreement with the smaller slope in the plot. Although the Z+jet and dijet spectra are different, the relative $R$ dependence is similar.

The quantile ratio is also shown in the right panel of Fig.~\ref{fig:Zjet_PYTHIA}. The difference is smaller between dijets and Z+jets than for the nuclear modification factor, pointing out the reduced sensitivity to the details of the initial spectrum. Using the cumulative spectrum, and the ratio of momenta, one gets much less sensitive to the initial shape of the spectrum. The degree of scaling the initial spectrum of the quantile is nevertheless not as ideal as for the results in Fig.~\ref{fig:R_AA_QAA_n_dependence}. This can be traced to the fact that the single inclusive jets in neither dijet nor Z+jet events are pure samples of quark-, or gluon-initiated jets. The different admixture of parton species, as well as the different level of quenching of the two both, contribute to delaying the onset of scaling effects up to higher transverse momenta. One also has to point out that Fig.~\ref{fig:R_AA_QAA_n_dependence} was obtained by assuming $n=$const, while for realistic calculations $n_q$ and $n_g$ are complicated functions of $p_T$ and differ significantly for dijet and Z+jet events, see Fig.~\ref{fig:Spectrum_PYTHIA_fit}.

All in all, our results are also qualitatively similar to the ones observed in Ref.~\cite{Brewer:2018dfs}. Similarly to dijets, the $R$-dependence is very similar to the Z+jet $R_{\rm AA}$ and is to a great degree captured by the approximate relation $Q_{\rm AA}\sim R_{\rm AA}^{1/(n-1)}$.

%%%%%%%%%%%%%%%%%%%%%%%%%%%%%%%%%%%%%%%%%%%%%%%%%%%%%%%
\subsubsection{Improving quark-gluon discrimination}
%%%%%%%%%%%%%%%%%%%%%%%%%%%%%%%%%%%%%%%%%%%%%%%%%%%%%%%

As mentioned above, jets recoiling from a boson ($\gamma$ or $Z$/$W$) is the preferred process for quark-gluon (q/g) discrimination, because the hard scattering naturally defines the initial flavor of the jet at leading order in perturbation theory. Most q/g discriminators apply cut on jet substructure observables, e.g., jet mass or soft-drop multiplicity, to classify jets, see in Refs.~\cite{Larkoski:2017jix,Larkoski:2019nwj}. While the best performance is achieved by machine learning algorithms, these nonetheless rely on training sets resulting in model dependence. In pp, however, the main description of jets is predominantly perturbative or it can be made perturbative with grooming. Model dependence, therefore, is not as crucial as for heavy-ions, where the description is not yet unique and where non-perturbative effects are more prominent. 

The recently introduced {\it topic modeling} \cite{Metodiev:2018ftz,Komiske:2018vkc,Brewer:2020och} is a data-driven method that is largely model independent, and its outstanding performance was demonstrated for event generator samples both in pp and AA. For this reason, it is also applicable to heavy-ions. There are some caveats of the classifier; (i) it works only with certain observables for which quarks and gluons are {\it mutually irreducible} (usually counting-type observables \cite{Larkoski:2014gra,Frye:2017yrw}), (ii) it is heavily limited by statistical uncertainty, (iii) the performance is limited by the cuts on the phase space, and (iv) it works on statistical samples.

Topic modeling aims to un-mix the sample probability distributions (e.g., dijet and boson+jet samples) to a common basis (quark and gluon distribution). Consider, for example,
\begin{align}\label{eq:Mixture_Zj_2j}
    p^{\rm Zj}(x)&=f_q^{\rm Zj}p_q(x)+(1-f^{\rm Zj}_q)p_g(x)\,, \nonumber \\
    p^{\rm 2j}(x)&=f_q^{\rm 2j}p_q(x)+(1-f^{\rm 2j}_q)p_g(x)\,, 
\end{align}
where $p(x)$ is a probability density of some observable for which Eq.~\eqref{eq:Mixture_Zj_2j} is true (mutual reducibility), and $f_q^{Zj/2j}$ and $f_g^{Zj/2j}$ are the weight factors. To un-mix, one uses the fact that phase-space of $x$ exists, where either the dijet or boson+jet (and thus the quark or gluon) dominates the distribution. This usually happens on the domain border of the observable (e.g., small/big jet multiplicity) \cite{Larkoski:2014gra,Frye:2017yrw}. With this, called anchor, bin one can statistically decouple the basis using $p^{\rm Zj}(x)-\kappa p^{\rm 2j}(x)\geq 0$,
\begin{align}\label{eq:Demix}
    p_q(x)&=\frac{p^{\rm Zj}(x)-\kappa({\rm Zj}|{\rm 2j})p^{\rm 2j}(x)}{1-\kappa({\rm Zj}|{\rm 2j})}\,, \nonumber\\
    p_g(x)&=\frac{p^{\rm 2j}(x)-\kappa({\rm 2j}|{\rm Zj})p^{\rm Zj}(x)}{1-\kappa({\rm 2j}|{\rm Zj})}\,,
\end{align}
where the reducibility factor is
\begin{equation}
    \kappa(i|j)\equiv\inf_{x}\frac{p^{i}(x)}{p^{j}(x)}\,,
\end{equation}
where $i,j={\rm Zj, 2j}$. 

Equation~\eqref{eq:Demix} is true if $\inf_xp_q(x)/p_g(x)=0$, however, for real data it has a finite minimum. The minimum is typically on the edge of the $x$ distribution, and thus the extraction of $\kappa$ is limited by the statistical uncertainty of this corner bin. Because of the linearity, one could integrate both sides of Eq.~\eqref{eq:Mixture_Zj_2j}, and rewrite Eq.~\eqref{eq:Demix} using the cumulative distribution of $p(x)$,
\begin{align}\label{eq:cum-reducibility_factor}
    \tilde \kappa({\rm Zj}|{\rm 2j})\equiv&\inf_x\frac{\int^\infty_x\rmd x\,p^{\rm Zj}(x)}{\int^\infty_x\rmd x\,p^{\rm 2j}(x)}\,,\\
    \tilde \kappa({\rm 2j}|{\rm Zj})\equiv&\inf_x\frac{\int^x_0\rmd x\,p^{\rm 2j}(x)}{\int^x_0\rmd x\,p^{\rm Zj}(x)}\,.
\end{align}
Our cumulative method improves the statistical uncertainty by definition (see Sec.~\ref{sec:tail-cumulative}) that can be trivially tested with arbitrary combined distributions. We would like to note the cumulative in this subsection refers to the $p(x)$ distribution, and has nothing to do with the cumulative of the jet spectrum.

\begin{figure}[t!]
    \centering
    \includegraphics[width=0.49\textwidth]{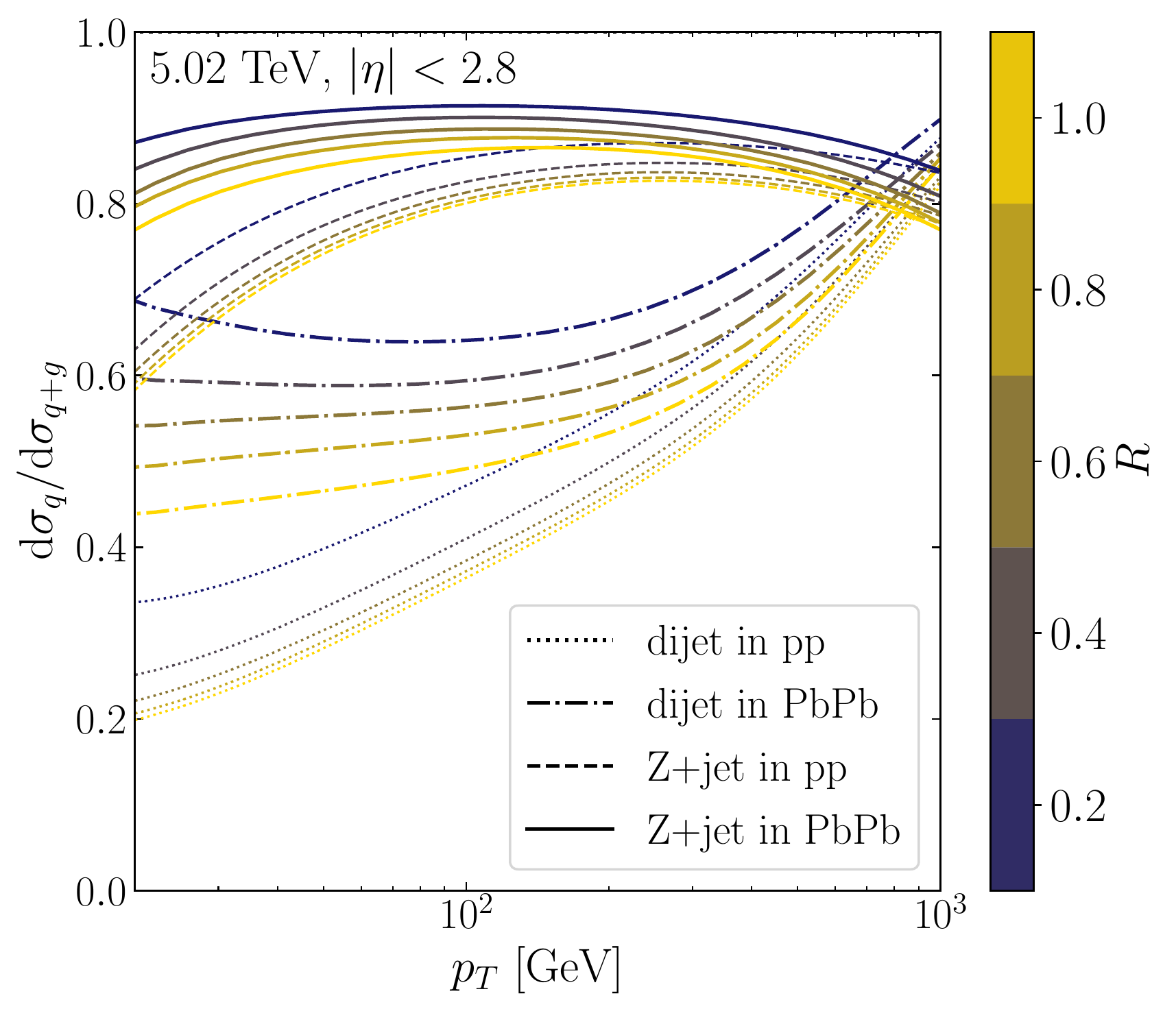}
    \caption{The quark and gluon ratio before and after quenching for dijets and Z+jets in pp and PbPb.} 
    \label{fig:Quenched_quarkandgluon}
\end{figure}
Unfortunately, the jet spectrum is not mutually irreducible. One can see this from Eq.~\eqref{eq:RAA_PYTHIA}, where the quenching factor depends on the quark/gluon spectrum through their $n$ indices. However, the medium modified quark-gluon ratio of the jet spectrum is important for any quark-gluon discriminator, and thus we provide it in Fig.~\ref{fig:Quenched_quarkandgluon}, for both dijets and Z+jets. We observe that at high-$p_T$ the ratio barely changes, however, at lower $p_T$ quarks start to dominate. This is the result of the stronger quenching of gluons, effectively suppressing them in the samples. For future quark-gluon classification, this suppression effect indicates that the Z+jet quark-gluon ratio will be less different from the dijet ratio, making the separation, unfortunately, harder, in line with what was observed in Ref.~\cite{Chien:2018dfn}. 

%%%%%%%%%%%%%%%%%%%%%%%%%%%%%%%%%%%%%%%%%%%%%%%%%%%%%%%
%%%%%%%%%%%%%%%%%%%%%%%%%%%%%%%%%%%%%%%%%%%%%%%%%%%%%%%
\section{Conclusion}
%%%%%%%%%%%%%%%%%%%%%%%%%%%%%%%%%%%%%%%%%%%%%%%%%%%%%%%
%%%%%%%%%%%%%%%%%%%%%%%%%%%%%%%%%%%%%%%%%%%%%%%%%%%%%%%

In heavy-ion collisions, the steeply falling jet spectrum, convolved with the probability for quenching, biases the measured jet observables. In this paper, we studied the origin of this bias and its presence in recently suggested observables, e.g., the quantile ratio and its comparison of single-inclusive jet spectra in dijet and boson+jet events in heavy-ion collisions.

We developed a novel analytic framework based on quenching factors to construct the jet spectrum in heavy-ion collisions. Starting from single parton energy loss, we showed the necessity to include the full medium-induced radiation spectrum, including both multiple-soft (described by the BDMPS-Z spectrum) and single-hard (included in the GLV spectrum) scattering regimes. By keeping track of the jet cone size, the energy is only lost if the emissions propagate out of the cone. We also account for the quenching of multiple jet partons resolved by the medium interactions through the collimator function, see Eq.~\eqref{eq:Quenching_factor_jet}. Therefore, opening the cone, vacuum-like jet fragmentation sources more partons to quench, resulting in a relative cancellation of the cone-size dependence. It is also important to use realistic jet spectrum for predictions by including the $p_T$ dependent spectrum power $n(p_T)$, $p_T$-dependent quark-gluon jet ratio, and nPDFs. We revealed the importance of elastic energy loss, and also included it in our quenching framework for jets.

Different observables have been introduced to study the jet spectrum in heavy-ion collisions, e.g., spectrum shift, cumulative-$R_{\rm AA}$, quantile or pseudo-quantile procedure. We showed their relation to the nuclear modification factor, to each other, and elucidated their dependence on the jet momentum $p_T$, the spectrum power $n$, and the jet cone size $R$. The cumulative-based observables reduce statistical uncertainty, and the momentum-ratio-based ones change the spectrum power dependence. The quantile momentum ratio maximizes both of these advantages, and we provided predictions for its cone-size dependence for the first time. Our approximate formula for the quantile momentum ratio $Q_{\rm AA}\sim R_{\rm AA}^{1/(n-1)}$ captures the rough properties of the observable: it is similar to $R_{\rm AA}$, with reduced spectrum power index $n$.

Finally, we demonstrated the bias effect by comparing quenched dijet and Z+jet spectra. For a pure sample of quark or gluon jets, the quenching factors for single-partons scale like
\begin{equation}\label{eq:qf-behavior-summary}
\renewcommand\arraystretch{2}
    -\ln \mathcal Q_>^{(0)} \approx \left\{ \begin{array}{ll}
        \bar \alpha {\displaystyle\sqrt{\frac{\hat q L^2 n}{p_T}}} & \;\;\text{  for } p_T \ll n\omega_R \\
        \bar \alpha {\displaystyle\frac{\sqrt{\omega_R \hat q  L^2}\, n}{p_T}}
         & \;\;\text{  for } p_T \gg n\omega_R 
    \end{array}\right. \,,
\end{equation}
where $\omega_R = \sqrt{\hat q L}/R$. This qualitative behavior is also numerically confirmed for the resummed quenching factor $\mathcal Q_>$. As a direct manifestation of the bias effect, i.e. the $n$-dependence in Eq.~\eqref{eq:qf-behavior-summary}, the jet spectrum in Z+jet events, which is less steep, results in a smaller nuclear modification factor $R_{\rm AA}$ than for dijet events, see Fig.~\ref{fig:R_AA_QAA_n_dependence} (left).
In other words, the steeper the spectrum the stronger the effect of quenching. 

The quantile ratio $Q_{\rm AA}$ is much more resilient to details of the hard spectrum, both the actual values of $n$ and of the relative admixture of quarks and gluons. For pure samples of quark-, and gluon-jets, the behavior in Eq.~\eqref{eq:qf-behavior-summary} predicts an almost ideal scaling at $p_T \gg n\omega_R$, which was largely confirmed in Fig.~\ref{fig:R_AA_QAA_n_dependence} for $n=$const. For realistic samples of jets in dijet and Z+jet events, however, the universal behavior of the quantile ratio can only be expected to be approximate, see Fig.~\ref{fig:Zjet_PYTHIA} (right).

The dijet and Z+jet events are also useful for quark and gluon discrimination. Following our cumulative spectrum experiences, we improved quark-, and gluon-jet discrimination based on topics modeling in general. However, due to the $n$-dependence of the quenching factors, quarks and gluons in the jet spectrum are not mutually irreducible. We will return to the challenging of quark/gluon discrimination in heavy-ion collisions in future work.

Many of the assumptions underlying the concrete realization behind our numerical results in Sec.~\ref{sec:Quenching} are already implemented in varying degrees in various phenomenological Monte Carlo models. Our framework, therefore, provides theoretical tools to organize the effects of quark/gluon contributions, jet fragmentation, and finally, the details of medium interactions. In the BDMPS-Z framework, these are, to a first approximation, all controlled by a single transport coefficient $\hat q$.

\acknowledgments
We thank T. S. Bir\'o, J. Brewer, P. Caucal, J. Isaksen, G. Milhano, D. Pablos, J. Thaler and D. Perepelitsa for helpful discussions. The work is supported by a Starting Grant from Trond Mohn Foundation (BFS2018REK01) and the University of Bergen. A.T. is also supported by the MCnetITN3 H2020 Marie Curie Initial Training Network, contract 722104, and wishes to thank the Institut de Physique Theorique (IPhT) and Gregory Soyez for the hospitality.

\appendix
 
%%%%%%%%%%%%%%%%%%%%%%%%%%%%%%%%%%%%%%%%%%%%%%%%%%%%%%%
%%%%%%%%%%%%%%%%%%%%%%%%%%%%%%%%%%%%%%%%%%%%%%%%%%%%%%%
\section{Corrections to the Laplace transformation and to the running power}
\label{sec:Corrections_LaplaceAndPower}
%%%%%%%%%%%%%%%%%%%%%%%%%%%%%%%%%%%%%%%%%%%%%%%%%%%%%%%
%%%%%%%%%%%%%%%%%%%%%%%%%%%%%%%%%%%%%%%%%%%%%%%%%%%%%%%

\begin{figure}[t]
    \centering
    \includegraphics[width=0.5\textwidth]{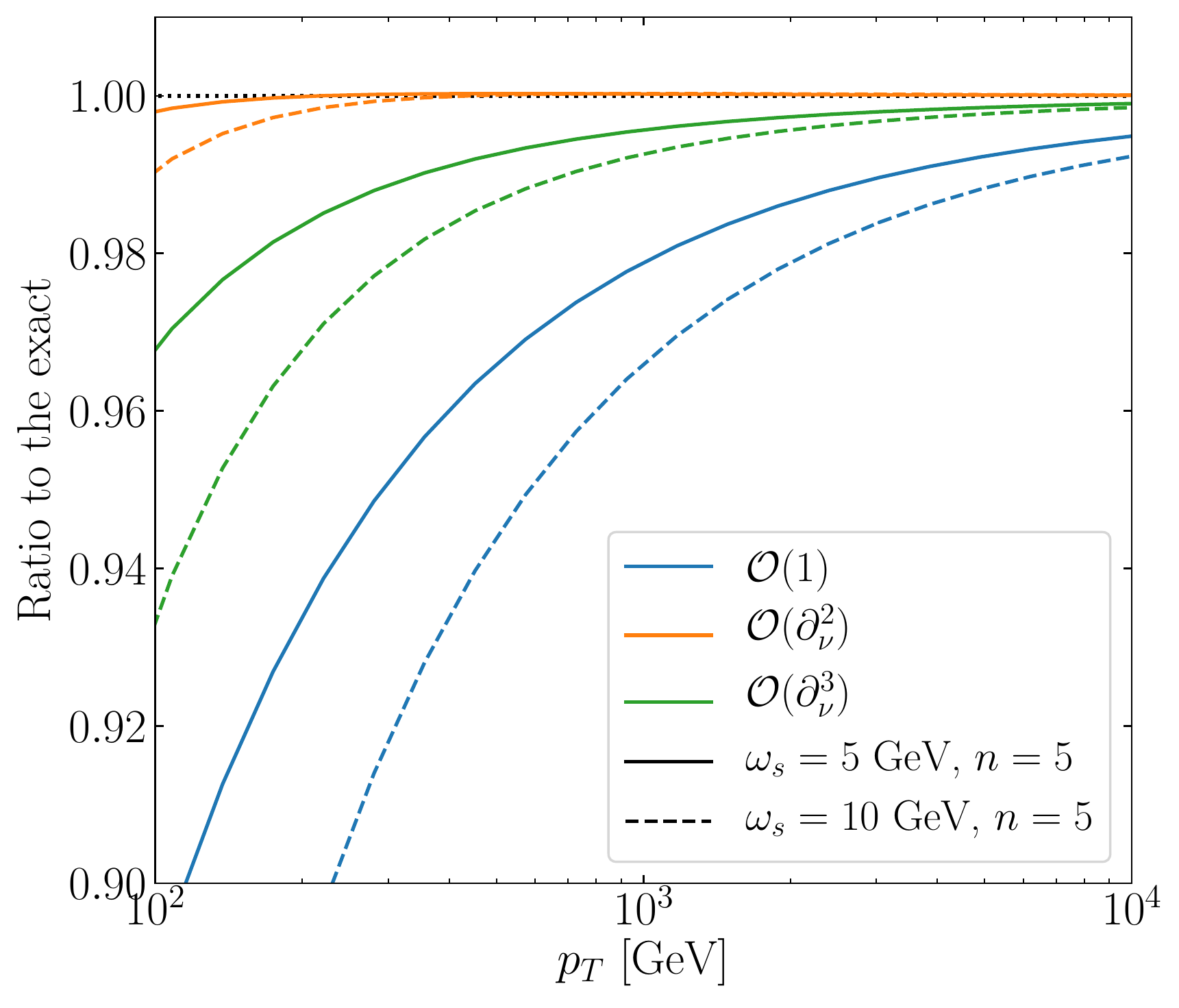}%
    \includegraphics[width=0.5\textwidth]{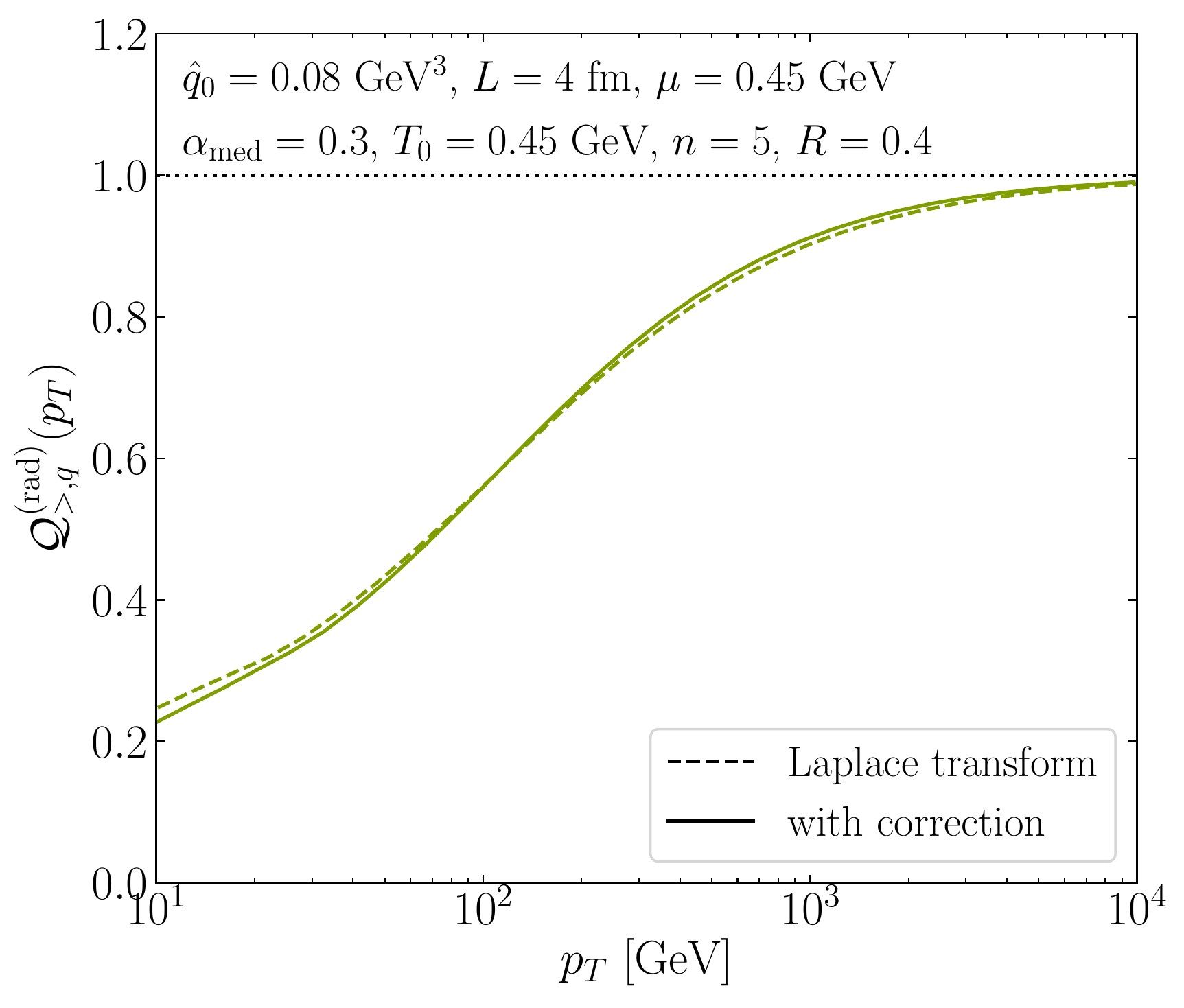}
    \caption{Ratio of the truncated expansion of $R_\text{med}$ from Eq.~\eqref{eq:Laplace_correction} to the exact value for the soft BDMPS-Z model from Eq.~\eqref{eq:Rmed-exact} (left), and the same with the full quenching weight from Eq.~\eqref{eq:Quenching_factor_jet}.}
    \label{fig:laplace-corrections}
\end{figure}

Our objective in this paper, is to compute the ratio of medium to vacuum inclusive jet spectra, which can be written as (see Eqs.~\eqref{eq:medium_spectrum}--\eqref{eq:Q_def})
\begin{equation}\label{eq:Rmed-exact}
    R_\text{med}(p_T) = \int_0^\infty \rmd \varepsilon \, \mathcal{P}(\varepsilon) \left(1 + \frac{\varepsilon}{p_T} \right)^{-n} \,,
\end{equation}
where we omit the $R$ dependence for now, and assume that $n=\text{const}$. In Eq.~\eqref{eq:Q_def2}, we took the $(1+\varepsilon/p_T)^{-n}\approx\exp(-n\varepsilon/p_T)$ approximation. We can, however, easily include corrections to this by noting that
\begin{align}\label{eq:Laplace_correction}
    R_\text{med}(p_T) 
    &=\int_0^\infty \rmd\varepsilon\,\mathcal P(\varepsilon)\left[1+\frac{(\nu\varepsilon)^2}{2n} - \frac{(\nu\varepsilon)^3}{3n^2} + \mathcal O\big( (\nu\varepsilon)^4 \big)\right] {\rm e}^{-\nu\varepsilon} \nonumber\\
    & =\left[1+\frac{\nu^2}{2n}\frac{\partial^2}{\partial\nu^2}  + \frac{\nu^3}{3 n^2} \frac{\partial^3}{\partial \nu^3} + \mathcal O\big(\nu^4 \partial^4_\nu\big)\right]\mathcal Q(\nu)\,,
\end{align}
where $\mathcal{Q}(\nu) \equiv \int_0^\infty \rmd \varepsilon\, \mathcal{P}(\varepsilon) \rme^{-\nu\varepsilon}$ is the Laplace transform of the energy loss distribution and  $\nu=n/p_T$. 
In this appendix, we investigate the impact of these higher-order corrections for a concrete example that can be solved analytically, namely the energy loss distribution obtained in the strictly soft limit of the BDMPS-Z spectrum. It is given by
\begin{equation}
    \mathcal{P}(\varepsilon) = \sqrt{\frac{\omega_s}{\varepsilon^3}}\rme^{- \frac{\pi \omega_s}{\varepsilon}} \,,
\end{equation}
which only depends on the energy scale $\omega_s$ and is properly normalized. In this case, its Laplace transform is $\mathcal Q=\exp(-2\sqrt{\pi\omega_s\nu})$. We can, in fact find any of the terms in Eq.~\eqref{eq:Laplace_correction} by noticing that
\begin{align}
    \mathcal{I}_m &\equiv \frac{\partial^m}{\partial \nu^m} \mathcal{Q}(\nu)=(-1)^m2\left(\frac{\nu}{\pi}\right)^{\frac{1-2m}{4}}\omega_s^\frac{1+2m}{4}K_{m-\frac12}(2\sqrt{\pi\omega_s\nu})\,,
\end{align}
where $K_m(x)$ is the modified Bessel function of the second kind and $\mathcal{I}_0 = \mathcal{Q}(\nu)$. We can therefore write
\begin{equation}
    R_{\rm med}(p_T) = \sum_{m=0}^\infty c_m \mathcal{I}_m \,,
\end{equation}
where, $c_0 = 1$, $c_1=0$, $c_2=\nu^2/(2n)$, $c_3=\nu^3/(3n^2)$ and so forth, by assuming $R_{\rm med}(p_T)$ to be analytic function. On the left of Fig.~\ref{fig:laplace-corrections}, we study the corrections by comparing to the exact value from Eq.~\eqref{eq:Rmed-exact}. The parameters we use are $\omega_s=5$ GeV and $\omega_s = 10$ GeV and $n=5$. For the realistic choice of $\omega_s \lesssim 5$ GeV, the leading behaviour is already of the order of $\mathcal O(10^{-2})$, even at low $p_T\sim100$ GeV.

Finally, the effect of the corrections from Eq.~\eqref{eq:Laplace_correction} for the realistic quenching weight employed in the main body of the paper is showcased in Fig.~\ref{fig:laplace-corrections} (right) up to the leading correction $\mathcal O(\partial^2_\nu)$. The sign of the correction depends on the concavity/convexity of the quenching weight (note the second derivative in $p_T$ in Eq.~\eqref{eq:Laplace_correction}). For our choice of parameters, the correction is tiny $<\mathcal{O}(10^{-3})$, and can safely be neglected.

The conclusions from above also hold if we consider a $p_T$ dependent power $n(p_T)$ of the hard spectrum. One can trivially replace $n\mapsto n(p_T)$ in the quenching weight capturing most of the effects of the running power. The correction to this simple replacement is
\begin{equation}
    R_{\rm med}(p_T)\approx\left[\left(1-\frac{1}{p_T}\frac{\partial n}{\partial p_T}\frac{\partial^2}{\partial\nu^2}\right)\left(1+\ln(p_T)\frac{\partial n}{\partial p_T}\frac{\partial}{\partial\nu}\right)\right]\mathcal Q(\nu)\,,
\end{equation}
resulting in negligible $\ll\mathcal O(10^{-3})$ corrections similar to the corrections to the Laplace transformation.

%%%%%%%%%%%%%%%%%%%%%%%%%%%%%%%%%%%%%%%%%%%%%%%%%%%%%%%
%%%%%%%%%%%%%%%%%%%%%%%%%%%%%%%%%%%%%%%%%%%%%%%%%%%%%%%
\section{\boldmath $p_T$-scaling of the out-of-cone emission quenching factor}
\label{sec:qf-scaling}
%%%%%%%%%%%%%%%%%%%%%%%%%%%%%%%%%%%%%%%%%%%%%%%%%%%%%%%
%%%%%%%%%%%%%%%%%%%%%%%%%%%%%%%%%%%%%%%%%%%%%%%%%%%%%%%

Consider the BDMPS-Z spectrum in the soft limit. In the short formation-time approximation the emission and subsequent broadening of a soft gluon factorizes, and we can write
\begin{equation}
    \omega \frac{\rmd I}{\rmd \omega \rmd k_\perp^2 \rmd t} = \bar \alpha \sqrt{\frac{\hat q}{\omega}} \mathcal P(k_\perp, L-t) \,,
\end{equation}
where $\mathcal P(k_\perp, L-t)$ is the probability for a particle emitted at time $t$ to acquire transverse momentum $k_\perp = \omega \theta$ through elastic scattering up to the end of the medium $L$. In the Gaussian approximation, it reads
\begin{equation}
    \mathcal P(k_\perp, L-t) = \frac{4\pi}{\hat q(L-t)}\rme^{-\frac{k_\perp^2}{\hat q (L-t)}} \,.
\end{equation}
For this simplified ansatz, the out-of-cone spectrum reads
\begin{equation}\label{eq:dIdw_Gauss}
    \frac{\rmd I_>}{\rmd \omega} = \bar \alpha \sqrt{\frac{\hat q L^2}{\omega}} \mathcal B\left(\frac{\omega^2}{\omega^2_R} \right) \,,
\end{equation}
where $\omega_R = \sqrt{\hat q L}/R$ and the broadening factor $\mathcal{B}$ is given by
\begin{equation}
    \mathcal B(y) = \frac{1}{L} \int_0^L \rmd t \int^\infty_{(\omega R^2)}\rmd k_\perp^2\, \mathcal P(k_\perp,L-t) = \rme^{-y} - y \Gamma(0,y)  \,,
\end{equation}
where $y =\omega^2/\omega_R^2$. Therefore, Eq.~\eqref{eq:dIdw_Gauss} falls rapidly for $\omega>\omega_R$, and thus we used $\omega_R$ in the matching definition in Eq.~\eqref{eq:omega_s}. The single-parton quenching factor reads then
\begin{align}
    \ln \mathcal Q^{(0)}_>(\nu) &= -\bar \alpha \sqrt{2\omega_c \nu} \int_{0}^\infty \rmd x \, \frac{1}{x^{3/2}}\left( 1 - \rme^{-x} \right) \mathcal{B}\left(\frac{x^2}{x_R^2} \right) \,, \nonumber\\
    &\approx -\bar \alpha \sqrt{2\omega_c \nu} \int_{0}^{x_R} \rmd x \, \frac{1}{x^{3/2}}\left( 1 - \rme^{-x} \right)
\end{align}
where we changed variables to $x = \omega \nu$ and $x_R \equiv \omega_R \nu$. We will solve this integral in two limiting cases, namely $x_R \ll 1$ and $x_R \gg 1$. In the former case, which corresponds to the high-$p_T$ regime where $p_T \gg \omega_R n$, we can expand the terms in the bracket and find
\begin{equation}\label{eq:out-of-cone-high-pT}
    \left.\ln \mathcal Q^{(0)}_>(\nu)\right|_{x_R \ll 1} = -2\bar \alpha \sqrt{2 \omega_c \omega_R}\nu \,.
\end{equation}
In the opposite limit, for $p_T \ll \omega_R n$, we can extend the upper integration limit to infinity, to obtain
\begin{equation}
    \left. \ln \mathcal{Q}^{(0)}_>(\nu) \right|_{x_R \gg 1} = -2\bar \alpha \sqrt{2\pi \omega_c \nu} \,,
\end{equation}
which is independent of the jet cone.

%%%%%%%%%%%%%%%%%%%%%%%%%%%%%%%%%%%%%%%%%%%%%%%%%%%
\section{Generating and parametrizing the jet spectrum}\label{sec:Generate_spectrum}
%%%%%%%%%%%%%%%%%%%%%%%%%%%%%%%%%%%%%%%%%%%%%%%%%%%

\begin{figure}
    \centering
    \includegraphics[width=0.49\textwidth]{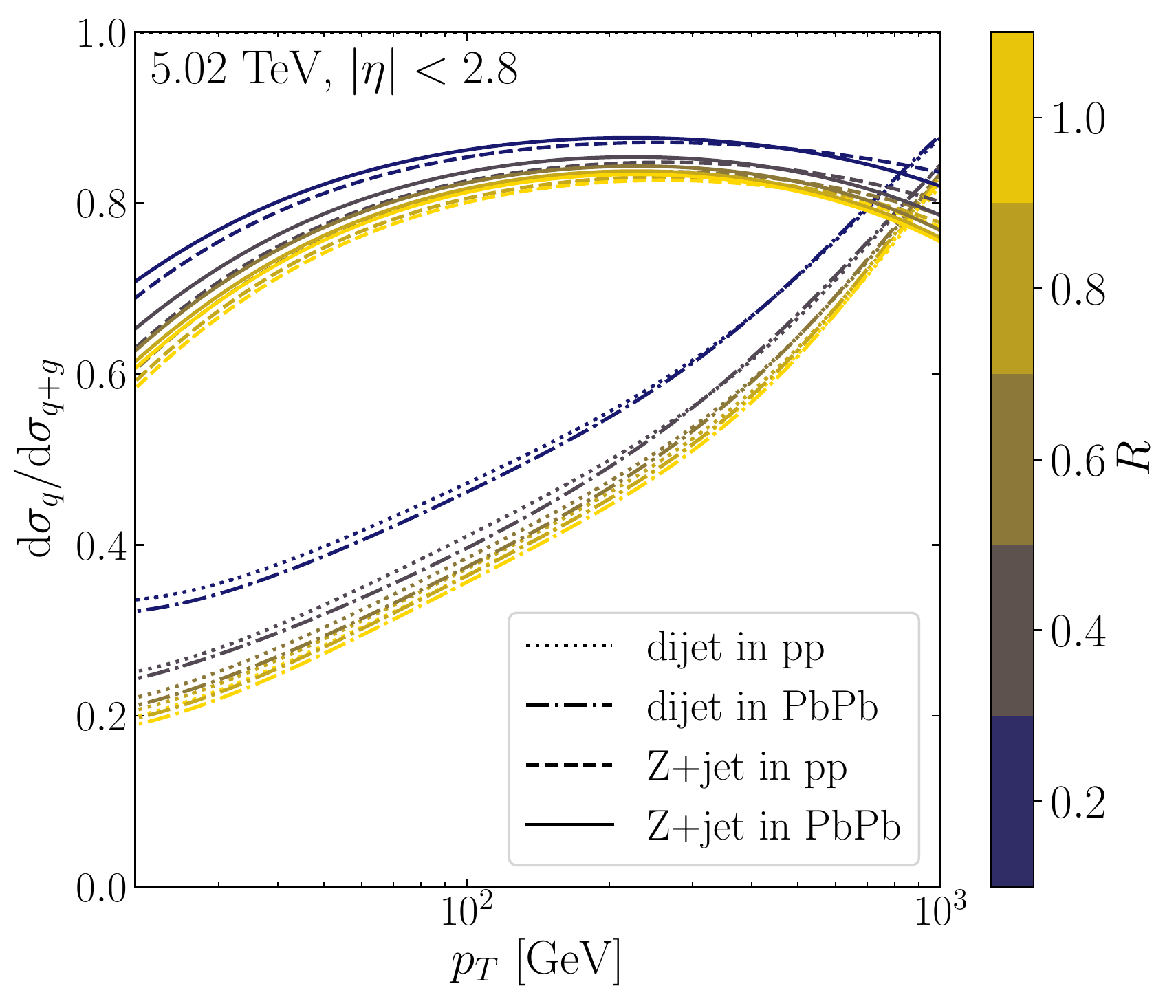}
    \includegraphics[width=0.49\textwidth]{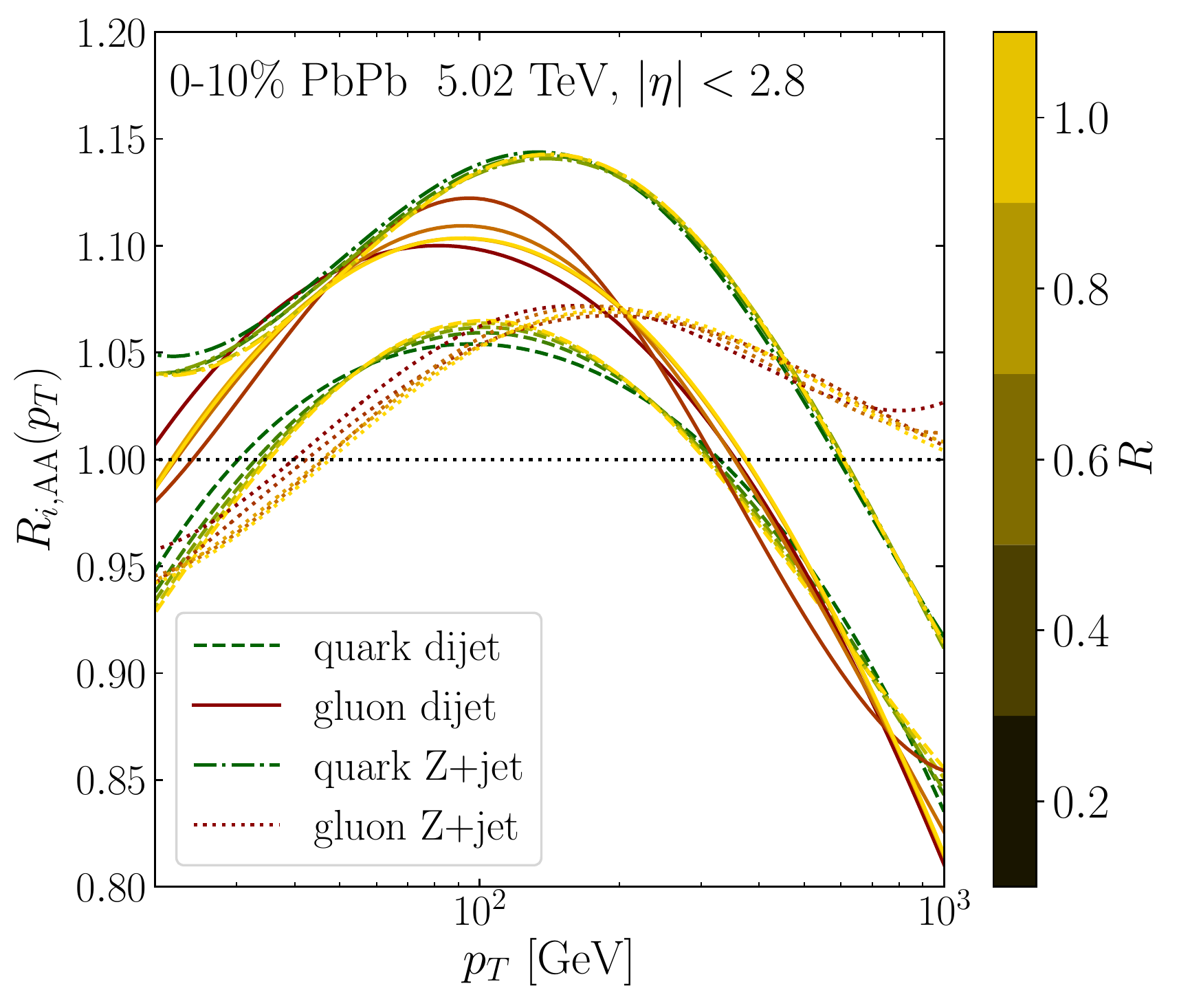}
    \caption{{\it Left}: The parametrized quark jet contribution in pp and in PbPb using the nPDF in Pythia. {\it Right}: The quenching factors for quarks and gluons resulted by the nPDF. }
    \label{fig:Spectrum_PYTHIA_fit}
\end{figure}

As we mentioned in Sec.~\ref{sec:Quenching}, the partonic cross-section to produce a jet with a given $p_T$ and $R$ is perturbatively calculable in the collinear factorization up to high precision~\cite{Dasgupta:2014yra,Dasgupta:2016bnd}. In our work, instead, we extract the spectrum using the Pythia8.235 event generator~\cite{Sjostrand:2014zea}. To generate dijet events we used default settings and tunes with \texttt{HardQCD:All} both in pp and in 0-10\% PbPb collision at 5.02 TeV. This results in LO $2\to2$ matrix elements. The nPDF was EPS09LO which has a relatively important effect on the $R_{\rm AA}$, see in Fig.~\ref{fig:Spectrum_PYTHIA_fit}. The ISR, MPI, and hadronization were turned off to focus on final state radiation only. We reconstructed jets using anti-$k_t$ algorithm with FastJet3~\cite{Cacciari:2011ma} for $R=0-1$, $p_{T,{\rm jet}}=10-1000$ GeV and $|\eta_{\rm jet}|<2.8$, similar to the kinematic cuts of ATLAS.\footnote{At this rapidity selection and jet cones, the ISR and MPI could contribute to jet production, that we address in a future study.} To label the flavors of the jets, we compared them to the outgoing partons from the hard scattering, and we kept the closest in angle if it was less than 2R. We only associated one jet (the hardest) with an initiator, and thus we only kept the two hardest associated jets. This selection detail becomes important for small $R$ jets, where more than one jet can be reconstructed for one initiator. We kept those events in which there no jet passing the criteria, which are important in the proper $R_{\rm AA}$ ratio (before quenching). We parametrized the spectrum following~\cite{Spousta:2015fca},
\begin{align}
    \frac{\rmd\sigma^{\rm pp/AA}_i}{\rmd p_T}(p_T,R)&=c_0\left(\frac{p_T}{p_0}\right)^{-n^{\rm pp/AA}_i(p_T,R)}\,, \\
    n^{\rm pp/AA}_i(p_T,R)&=\sum_{n=1}c_n\log^n\left(\frac{p_T}{p_0}\right)\,,
\end{align}
where $i$ is the flavor of the initiator of the jet and $\{p_0,c_n\}$ are $(p_T,R,i,{\rm pp/AA})$ dependent fitting parameters. We kept terms up to $n=3$, achieving $<3\%$ relative deviation. The resulted parametrization is showed in Fig.~\ref{fig:Spectrum_PYTHIA_fit}. On the left, there is the quark contribution, which increases with $p_T$. The cone size dependence shows, gluons are emitted at larger angles even in the vacuum. On the right the $R_{\rm AA}$ is shown, resulted by the nPDF (no quenching on the plot). The inclusive jet spectrum would be similar to the dijets keeping not only the two hardest jets, however, the flavor assignment would be less trivial especially for smaller cone sizes, therefore we preferred to use the dijet samples.\footnote{For a recent development on jet flavor definition, see Ref.~\cite{Baron:2020xoi}.}

For the Z+jet samples, we followed the logic from previous. We used however the \texttt{WeakBosonAndParton:qg2gmZq} and \texttt{WeakBosonAndParton:qqbar2gmZg} processes and we forced the Z-boson to decay invisibly. In this case, we kept only the hardest recoiling jet if it fulfilled the same criteria as before. In the Z+jet case, the spectrum power is smaller in comparison to dijets $n_{\rm Zj}\approx 4<n_{\rm 2j}\approx 6$. The quark and gluon ratio is different, in comparison to dijets, but the cone size dependence is similar, see on the left in Fig.~\ref{fig:Spectrum_PYTHIA_fit}. In both cases, opening the cone, one captures more gluons meaning, they are radiated at larger angles. The $R_{\rm AA}$ resulted from the nPDF is also shown in Fig.~\ref{fig:Spectrum_PYTHIA_fit}, relatively similar for both dijet and Z+jet processes.

%%%%%%%%%%%%%%%%%%%%%%%%%%%%%%%%%%%%%%%%%%%%%%%%%%%%%%%
%%%%%%%%%%%%%%%%%%%%%%%%%%%%%%%%%%%%%%%%%%%%%%%%%%
\section{Other Models}\label{sec:Other_models}
%%%%%%%%%%%%%%%%%%%%%%%%%%%%%%%%%%%%%%%%%%%%%%%%%%
%%%%%%%%%%%%%%%%%%%%%%%%%%%%%%%%%%%%%%%%%%%%%%%%%%%%%%%
As we mentioned in Sec.~\ref{sec:Quenching} the quenching weight and collimator formalism are independent of the energy loss description. To illustrate this we show how to use this formalism to describe quenching of elastic scattering and within the strong coupling approximation.

%%%%%%%%%%%%%%%%%%%%%%%%%%%%%%%%%%%%%%%%%%%%%%%%%%%%%%%
\subsection{Momentum shift and elastic energy loss}\label{sec:Corrections_elastic}
%%%%%%%%%%%%%%%%%%%%%%%%%%%%%%%%%%%%%%%%%%%%%%%%%%%%%%%
The simplest example for the jet energy loss probability $\mathcal P(\varepsilon)$ is to take a momentum shift of vacuum jets $\mathcal P(\varepsilon)=\delta(\varepsilon-S(p_T))$, resulting equivalent definition to the old fashioned momentum shift parameter $\rmd\sigma_{\rm med}(p_T)/\rmd p_T\equiv\rmd\sigma_{\rm vac}(p_T+S(p_T))/\rmd p_T$~\cite{Baier:2001yt,Adler:2006bw}. By expanding the steeply falling spectrum in $S(p_T)\ll p_T$, 
\begin{equation}
    \frac{\rmd\sigma_{\rm med}}{\rmd p_T}(p_T)=\sum_n^\infty\langle\varepsilon^n\rangle\frac{\rmd^n}{\rmd p_T^n}\left(\frac{\rmd\sigma_{\rm vac}}{\rmd p_T}\right)\approx\frac{\rmd\sigma_{\rm vac}}{\rmd p_T}(p_T+\langle\varepsilon\rangle)\,,
\end{equation}
where we used $\langle\varepsilon^i\rangle\approx\langle\varepsilon\rangle^i$. The shift parameter is roughly the mean energy loss
\begin{equation}\label{eq:momentumshift_averageenergy}
    S(p_T)\approx\langle\varepsilon\rangle=\int\rmd\varepsilon\,\varepsilon\,\mathcal{P}(\varepsilon)\equiv\Delta E\,.
\end{equation}
The quenching factor in the simple power-law case is
\begin{equation}\label{eq:RAA_momentumshift}
    R_{\rm med}(p_T)=p_T^{n(p_T)-n(p_T+S(p_T))}\left[1+\frac{S(p_T)}{p_T}\right]^{-n(p_T+S(p_T))}\,,
\end{equation}
which is well approximated by $R_{\rm med}(p_T)\approx [1+S(p_T)/p_T]^{-n(p_T)}$ (see in App.~\ref{sec:Corrections_LaplaceAndPower}). The quantile ratio is straightforward by using Eq.~\eqref{eq:Quantile_def} with $\mathcal Q=[1+S(p_T)/p_T]^{-n(p_T)}$. In the constant power approximation it is
\begin{equation}
    Q_{\rm med}(p_T)\approx p_T\left[(n-1)\int_{p_T}^\infty\rmd p\,(p_T+S(p_T))^{-n}\right]^{-\frac{1}{1-n}}\,.
\end{equation}

\begin{figure}
    \centering
    \includegraphics[width=0.59\textwidth]{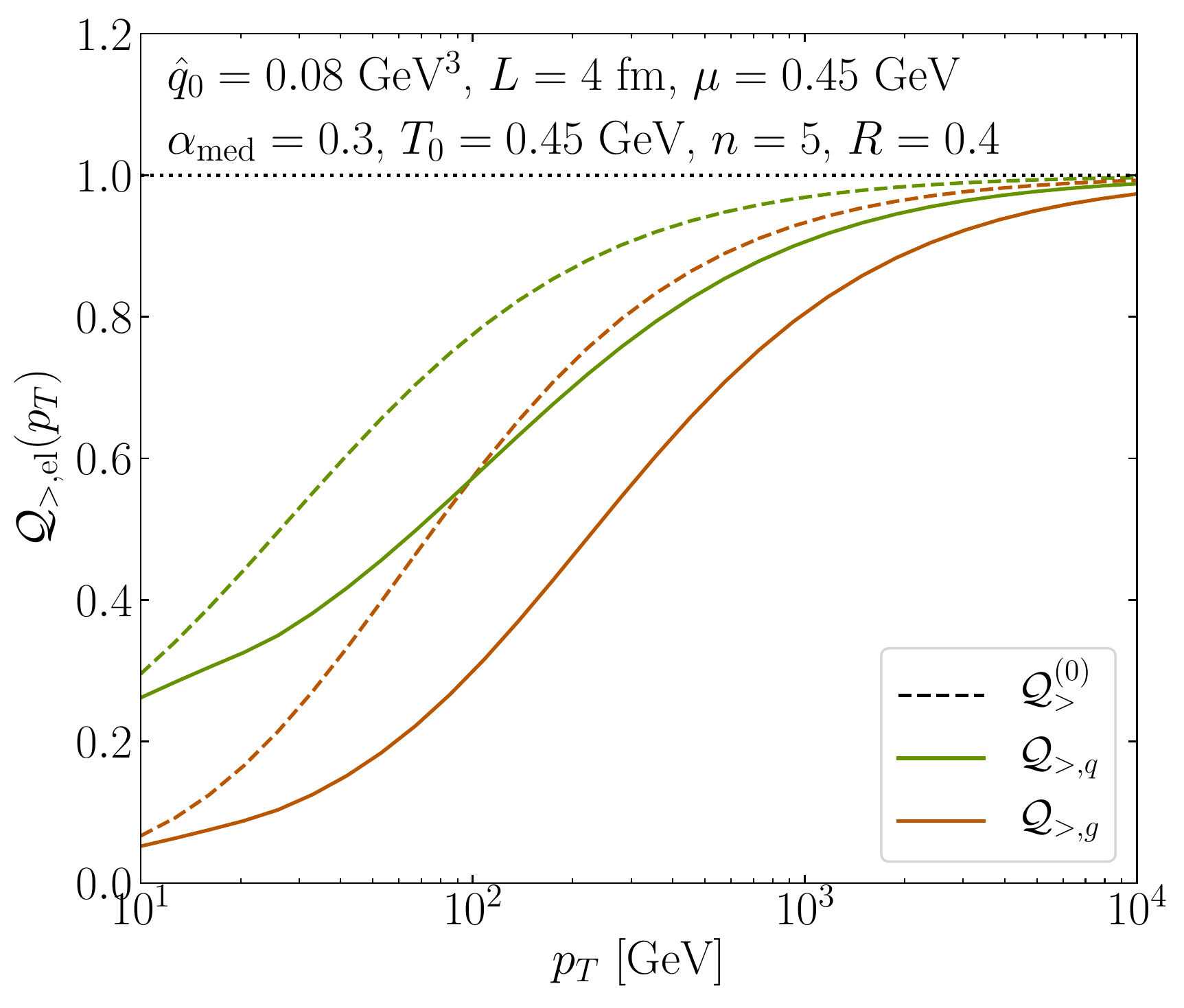}
    \caption{The quenching factor of elastic scatterings from Eq.~\eqref{eq:Q0el} and with the collimator from Eq.~\eqref{eq:Collimator_lin}. }
    \label{fig:Corrections}
\end{figure}

The elastic scattering is approximated by a constant momentum shift $\mathcal{P}_{{\rm el},i}(\varepsilon)=\delta(\varepsilon-\hat e_iL)$, where $\hat e=-\rmd \langle E\rangle/\rmd t\approx C_i\hat q_0/(4N_cT_0)$, and $T_0\approx0.45$ GeV~\cite{Qin:2015srf,Tachibana:2017syd}. This can be translated to an additional quenching weight compared to the form Eq.~\eqref{eq:Quenching_factor_singleparton},
\begin{equation}\label{eq:Q0el}
    \mathcal{Q}^{(0)}_{{\rm >,el}}(p_T,R)=\exp\left[-\frac{\hat eL\,n}{p_T}\left(1-\frac{R^2}{R^2_{\rm rec}}\right)\right],
\end{equation}
where we included some energy recapture through the second term with $R_{\rm rec}=\pi/2$. Therefore in total $\mathcal Q^{(0)}_{\rm >,tot}=\mathcal Q^{(0)}_{\rm >,rad}\mathcal Q^{(0)}_{\rm >,el}$. This factor runs slower with $p_T$ than the BDMPS-Z, similar to the ``$N=1$'' (or GLV) spectrum. We can also include the elastic energy loss of each jet constituent, by using the collimator function form Eq.~\eqref{eq:Collimator_lin}, $\mathcal Q_{\rm >,tot}=\mathcal Q_{\rm >,rad}\mathcal Q_{\rm >,el}$. Fig.~\ref{fig:Corrections} shows Eq.~\eqref{eq:Q0el} with dashed lines and with the collimator with full lines, indicating the importance of elastic scattering in the overall quenching. Therefore we included this effect in Sec.~\ref{sec:Numerical_results}. 

%%%%%%%%%%%%%%%%%%%%%%%%%%%%%%%%%%%%%%%%%%%%%%%%%%%%%%%
\subsection{Hybrid weak-, strong-coupling}
%%%%%%%%%%%%%%%%%%%%%%%%%%%%%%%%%%%%%%%%%%%%%%%%%%%%%%%

\begin{figure}
    \centering
    \includegraphics[width=0.49\textwidth]{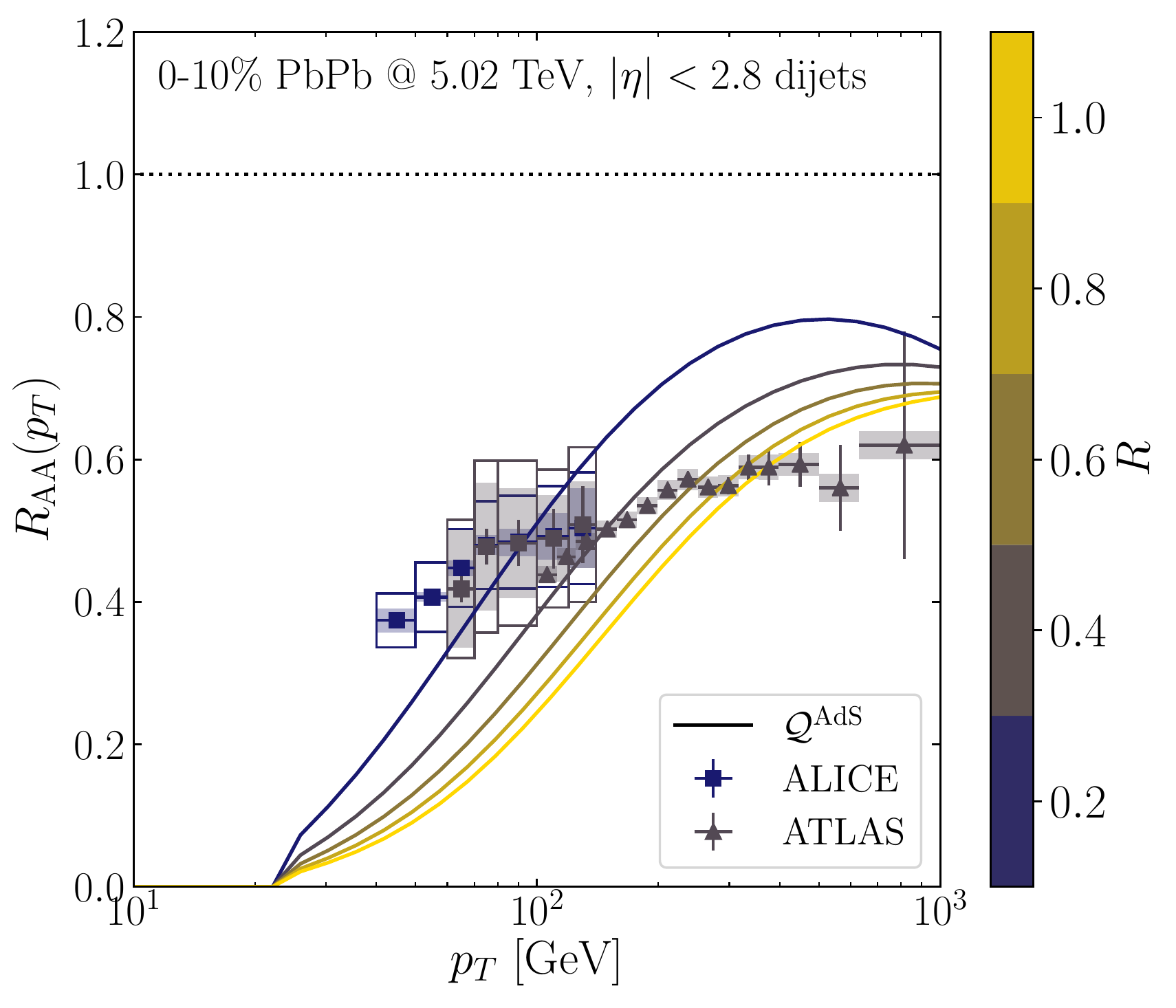}
    \includegraphics[width=0.49\textwidth]{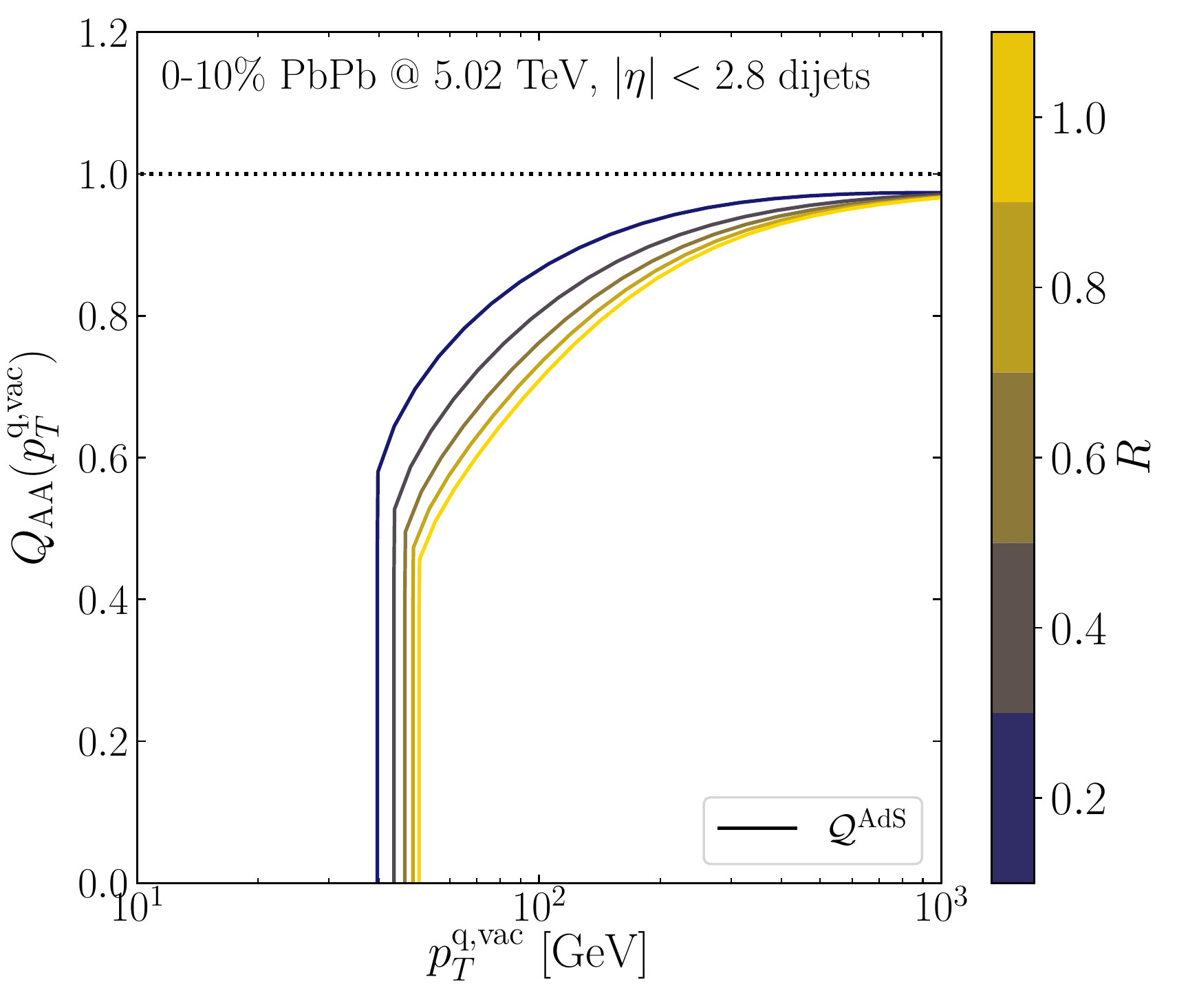}
    \caption{The quenching and the quantile ratio in the strong coupling assumption with the proper quark and gluon mixture using Eq.~\eqref{eq:RAA_AdS} and including the collimator to consider multi-parton effects from Eq.~\eqref{eq:Collimator_lin}.}
    \label{fig:RAA_AdS}
\end{figure}

In the strong coupling approximation, the energy loss is assumed to be described by a semi-classical string falling inside a black hole horizon~\cite{Chesler:2014jva}. This model does not use $\rmd I/\rmd\omega$ because there are no emitted gluons, and thus the energy loss is directly connected with $\mathcal{P}^{(0)}(\varepsilon)$. The average lost energy of a single parton traversing through an $L$ sized, $T$ temperature strongly coupled medium is
\begin{equation}
    \frac{\Delta E}{E}=1-\frac2\pi\left[\frac{L}{x_s}\sqrt{1-\left(\frac{L}{x_s}\right)^2}+\cos^{-1}\left(\frac{L}{x_s}\right)\right],
\end{equation}
where $E$ is the initial energy, $x_s=E^{1/3}/(2\kappa_{\rm sc}T^{4/3})$ is the stopping length and $\kappa_{\rm sc}=1.05\,g^{1/3}N_c^{1/6}$. Using the definition of momentum shift from Eq.~\eqref{eq:momentumshift_averageenergy}, the single parton quenching is estimated by
\begin{equation}\label{eq:RAA_AdS}
    \mathcal Q^{(0)}(p_T)=\left[2-\frac2\pi\left(\frac{L}{x_s}\sqrt{1-\left(\frac{L}{x_s}\right)^2}+\cos^{-1}\left(\frac{L}{x_s}\right)\right)\right]^{-n}\,,
\end{equation}
where the $p_T$ dependence is presented in $x_s$. For high $p_T$, $\mathcal Q^{(0)}(p_T)\approx1-32n\kappa_{\rm sc}^3L^3T^4/(3\pi p_T)$ the same $p_T$ dependence obtained from GLV (see Eq.~\eqref{eq:qf-single-hard} and below). We did not include here the broadening in and out of the cone due to the lack of particles, however, one could include the linearized hydro response through $R_{\rm rec}(\eta,\phi)$. We include multi parton quenching and thus the $R$-dependence by using the collimator from Eq.~\eqref{eq:Collimator_lin}. We used Eq.~\eqref{eq:RAA_AdS} as $\mathcal Q^{(0)}(p_T)$, and for $\Theta_{\rm res}$ we used the resolution condition used in the hybrid model~\cite{Hulcher:2017cpt}, $t_{\rm f}<t_{\rm d}=(\theta\mu)^{-1}<L$, where $\mu=\pi T/2$ is the IR screening scale used in the hybrid model (the Debye mass would be $gT$). Eq.~\eqref{eq:RAA_AdS} is shown in Fig.~\ref{fig:RAA_AdS} with the Pythia spectrum, resulting similar quenching to the GLV assumption. The curves in Fig.~\ref{fig:RAA_AdS} similar to the results in~\cite{Pablos:2019ngg} without the medium response. The $R$-dependence is simplified in our case, because we neglected the medium response. We also estimate the quantile ratio using Eq.~\eqref{eq:RAA_momentumshift}. The parameters are $T=0.27$ GeV, $L=4$ fm and $\kappa_{sc}=0.4$.

\bibliography{ref} 

\providecommand{\href}[2]{#2}\begingroup\raggedright\begin{thebibliography}{10}

\bibitem{Larkoski:2017jix}
A.~J. Larkoski, I.~Moult, and B.~Nachman, {\it {Jet Substructure at the Large
  Hadron Collider: A Review of Recent Advances in Theory and Machine
  Learning}},  {\em Phys. Rept.} {\bf 841} (2020) 1--63,
  [\href{http://arxiv.org/abs/1709.04464}{{\tt arXiv:1709.04464}}].

\bibitem{Marzani:2019hun}
S.~Marzani, G.~Soyez, and M.~Spannowsky, {\em {Looking inside jets: an
  introduction to jet substructure and boosted-object phenomenology}},
  vol.~958.
\newblock Springer, 2019.

\bibitem{Dasgupta:2020fwr}
M.~Dasgupta, F.~A. Dreyer, K.~Hamilton, P.~F. Monni, G.~P. Salam, and G.~Soyez,
  {\it {Parton showers beyond leading logarithmic accuracy}},  {\em Phys. Rev.
  Lett.} {\bf 125} (2020), no.~5 052002,
  [\href{http://arxiv.org/abs/2002.11114}{{\tt arXiv:2002.11114}}].

\bibitem{Asquith:2018igt}
R.~Kogler et~al., {\it {Jet Substructure at the Large Hadron Collider:
  Experimental Review}},  {\em Rev. Mod. Phys.} {\bf 91} (2019), no.~4 045003,
  [\href{http://arxiv.org/abs/1803.06991}{{\tt arXiv:1803.06991}}].

\bibitem{Dasgupta:2014yra}
M.~Dasgupta, F.~Dreyer, G.~P. Salam, and G.~Soyez, {\it {Small-radius jets to
  all orders in QCD}},  {\em JHEP} {\bf 04} (2015) 039,
  [\href{http://arxiv.org/abs/1411.5182}{{\tt arXiv:1411.5182}}].

\bibitem{Dasgupta:2016bnd}
M.~Dasgupta, F.~A. Dreyer, G.~P. Salam, and G.~Soyez, {\it {Inclusive jet
  spectrum for small-radius jets}},  {\em JHEP} {\bf 06} (2016) 057,
  [\href{http://arxiv.org/abs/1602.01110}{{\tt arXiv:1602.01110}}].

\bibitem{Rajagopal:2016uip}
K.~Rajagopal, A.~V. Sadofyev, and W.~van~der Schee, {\it {Evolution of the jet
  opening angle distribution in holographic plasma}},  {\em Phys. Rev. Lett.}
  {\bf 116} (2016), no.~21 211603, [\href{http://arxiv.org/abs/1602.04187}{{\tt
  arXiv:1602.04187}}].

\bibitem{Casalderrey-Solana:2018wrw}
J.~Casalderrey-Solana, Z.~Hulcher, G.~Milhano, D.~Pablos, and K.~Rajagopal,
  {\it {Simultaneous description of hadron and jet suppression in heavy-ion
  collisions}},  {\em Phys. Rev. C} {\bf 99} (2019), no.~5 051901,
  [\href{http://arxiv.org/abs/1808.07386}{{\tt arXiv:1808.07386}}].

\bibitem{Du:2020pmp}
Y.-L. Du, D.~Pablos, and K.~Tywoniuk, {\it {Deep learning jet modifications in
  heavy-ion collisions}},  \href{http://arxiv.org/abs/2012.07797}{{\tt
  arXiv:2012.07797}}.

\bibitem{Brewer:2018dfs}
J.~Brewer, J.~G. Milhano, and J.~Thaler, {\it {Sorting out quenched jets}},
  {\em Phys. Rev. Lett.} {\bf 122} (2019), no.~22 222301,
  [\href{http://arxiv.org/abs/1812.05111}{{\tt arXiv:1812.05111}}].

\bibitem{Chatrchyan:2012gt}
{\bf CMS} Collaboration, S.~Chatrchyan et~al., {\it {Studies of jet quenching
  using isolated-photon+jet correlations in PbPb and $pp$ collisions at
  $\sqrt{s_{NN}}=2.76$ TeV}},  {\em Phys. Lett. B} {\bf 718} (2013) 773--794,
  [\href{http://arxiv.org/abs/1205.0206}{{\tt arXiv:1205.0206}}].

\bibitem{Sirunyan:2017jic}
{\bf CMS} Collaboration, A.~M. Sirunyan et~al., {\it {Study of Jet Quenching
  with $Z+\text{jet}$ Correlations in Pb-Pb and $pp$ Collisions at
  ${\sqrt{s}}_{NN}=5.02\text{ }\text{ }\mathrm{TeV}$}},  {\em Phys. Rev. Lett.}
  {\bf 119} (2017), no.~8 082301, [\href{http://arxiv.org/abs/1702.01060}{{\tt
  arXiv:1702.01060}}].

\bibitem{Aaboud:2018anc}
{\bf ATLAS} Collaboration, M.~Aaboud et~al., {\it {Measurement of
  photon\textendash{}jet transverse momentum correlations in 5.02 TeV Pb + Pb
  and $pp$ collisions with ATLAS}},  {\em Phys. Lett. B} {\bf 789} (2019)
  167--190, [\href{http://arxiv.org/abs/1809.07280}{{\tt arXiv:1809.07280}}].

\bibitem{Aaboud:2019oac}
{\bf ATLAS} Collaboration, M.~Aaboud et~al., {\it {Comparison of Fragmentation
  Functions for Jets Dominated by Light Quarks and Gluons from $pp$ and Pb+Pb
  Collisions in ATLAS}},  {\em Phys. Rev. Lett.} {\bf 123} (2019), no.~4
  042001, [\href{http://arxiv.org/abs/1902.10007}{{\tt arXiv:1902.10007}}].

\bibitem{Brewer:2020och}
J.~Brewer, J.~Thaler, and A.~P. Turner, {\it {Data-driven quark and gluon jet
  modification in heavy-ion collisions}},
  \href{http://arxiv.org/abs/2008.08596}{{\tt arXiv:2008.08596}}.

\bibitem{He:2018gks}
Y.~He, L.-G. Pang, and X.-N. Wang, {\it {Bayesian extraction of jet energy loss
  distributions in heavy-ion collisions}},  {\em Phys. Rev. Lett.} {\bf 122}
  (2019), no.~25 252302, [\href{http://arxiv.org/abs/1808.05310}{{\tt
  arXiv:1808.05310}}].

\bibitem{Baier:1996sk}
R.~Baier, Y.~L. Dokshitzer, A.~H. Mueller, S.~Peigne, and D.~Schiff, {\it
  {Radiative energy loss and p(T) broadening of high-energy partons in
  nuclei}},  {\em Nucl. Phys. B} {\bf 484} (1997) 265--282,
  [\href{http://arxiv.org/abs/hep-ph/9608322}{{\tt hep-ph/9608322}}].

\bibitem{Zakharov:1996fv}
B.~Zakharov, {\it {Fully quantum treatment of the Landau-Pomeranchuk-Migdal
  effect in QED and QCD}},  {\em JETP Lett.} {\bf 63} (1996) 952--957,
  [\href{http://arxiv.org/abs/hep-ph/9607440}{{\tt hep-ph/9607440}}].

\bibitem{Baier:1998kq}
R.~Baier, Y.~L. Dokshitzer, A.~H. Mueller, and D.~Schiff, {\it {Medium induced
  radiative energy loss: Equivalence between the BDMPS and Zakharov
  formalisms}},  {\em Nucl. Phys. B} {\bf 531} (1998) 403--425,
  [\href{http://arxiv.org/abs/hep-ph/9804212}{{\tt hep-ph/9804212}}].

\bibitem{Gyulassy:2000er}
M.~Gyulassy, P.~Levai, and I.~Vitev, {\it {Reaction operator approach to
  nonAbelian energy loss}},  {\em Nucl. Phys. B} {\bf 594} (2001) 371--419,
  [\href{http://arxiv.org/abs/nucl-th/0006010}{{\tt nucl-th/0006010}}].

\bibitem{CaronHuot:2010bp}
S.~Caron-Huot and C.~Gale, {\it {Finite-size effects on the radiative energy
  loss of a fast parton in hot and dense strongly interacting matter}},  {\em
  Phys. Rev. C} {\bf 82} (2010) 064902,
  [\href{http://arxiv.org/abs/1006.2379}{{\tt arXiv:1006.2379}}].

\bibitem{Feal:2018sml}
X.~Feal and R.~Vazquez, {\it {Intensity of gluon bremsstrahlung in a finite
  plasma}},  {\em Phys. Rev. D} {\bf 98} (2018), no.~7 074029,
  [\href{http://arxiv.org/abs/1811.01591}{{\tt arXiv:1811.01591}}].

\bibitem{Andres:2020vxs}
C.~Andres, L.~Apolin\'ario, and F.~Dominguez, {\it {Medium-induced gluon
  radiation with full resummation of multiple scatterings for realistic
  parton-medium interactions}},  {\em JHEP} {\bf 07} (2020) 114,
  [\href{http://arxiv.org/abs/2002.01517}{{\tt arXiv:2002.01517}}].

\bibitem{Mehtar-Tani:2019tvy}
Y.~Mehtar-Tani, {\it {Gluon bremsstrahlung in finite media beyond multiple soft
  scattering approximation}},  {\em JHEP} {\bf 07} (2019) 057,
  [\href{http://arxiv.org/abs/1903.00506}{{\tt arXiv:1903.00506}}].

\bibitem{Mehtar-Tani:2019ygg}
Y.~Mehtar-Tani and K.~Tywoniuk, {\it {Improved opacity expansion for
  medium-induced parton splitting}},  {\em JHEP} {\bf 06} (2020) 187,
  [\href{http://arxiv.org/abs/1910.02032}{{\tt arXiv:1910.02032}}].

\bibitem{Barata:2020sav}
J.~a. Barata and Y.~Mehtar-Tani, {\it {Improved opacity expansion at NNLO for
  medium induced gluon radiation}},  {\em JHEP} {\bf 10} (2020) 176,
  [\href{http://arxiv.org/abs/2004.02323}{{\tt arXiv:2004.02323}}].

\bibitem{Baier:2001yt}
R.~Baier, Y.~L. Dokshitzer, A.~H. Mueller, and D.~Schiff, {\it {Quenching of
  hadron spectra in media}},  {\em JHEP} {\bf 09} (2001) 033,
  [\href{http://arxiv.org/abs/hep-ph/0106347}{{\tt hep-ph/0106347}}].

\bibitem{Salgado:2003gb}
C.~A. Salgado and U.~A. Wiedemann, {\it {Calculating quenching weights}},  {\em
  Phys. Rev. D} {\bf 68} (2003) 014008,
  [\href{http://arxiv.org/abs/hep-ph/0302184}{{\tt hep-ph/0302184}}].

\bibitem{Mehtar-Tani:2017web}
Y.~Mehtar-Tani and K.~Tywoniuk, {\it {Sudakov suppression of jets in QCD
  media}},  {\em Phys. Rev. D} {\bf 98} (2018), no.~5 051501,
  [\href{http://arxiv.org/abs/1707.07361}{{\tt arXiv:1707.07361}}].

\bibitem{Miller:2007ri}
M.~L. Miller, K.~Reygers, S.~J. Sanders, and P.~Steinberg, {\it {Glauber
  modeling in high energy nuclear collisions}},  {\em Ann. Rev. Nucl. Part.
  Sci.} {\bf 57} (2007) 205--243,
  [\href{http://arxiv.org/abs/nucl-ex/0701025}{{\tt nucl-ex/0701025}}].

\bibitem{Arleo:2017ntr}
F.~Arleo, {\it {Quenching of Hadron Spectra in Heavy Ion Collisions at the
  LHC}},  {\em Phys. Rev. Lett.} {\bf 119} (2017), no.~6 062302,
  [\href{http://arxiv.org/abs/1703.10852}{{\tt arXiv:1703.10852}}].

\bibitem{Spousta:2015fca}
M.~Spousta and B.~Cole, {\it {Interpreting single jet measurements in Pb $+$ Pb
  collisions at the LHC}},  {\em Eur. Phys. J. C} {\bf 76} (2016), no.~2 50,
  [\href{http://arxiv.org/abs/1504.05169}{{\tt arXiv:1504.05169}}].

\bibitem{Qiu:2019sfj}
J.-W. Qiu, F.~Ringer, N.~Sato, and P.~Zurita, {\it {Factorization of jet cross
  sections in heavy-ion collisions}},  {\em Phys. Rev. Lett.} {\bf 122} (2019),
  no.~25 252301, [\href{http://arxiv.org/abs/1903.01993}{{\tt
  arXiv:1903.01993}}].

\bibitem{Mehtar-Tani:2021fud}
Y.~Mehtar-Tani, D.~Pablos, and K.~Tywoniuk, {\it {Cone size dependence of jet
  suppression in heavy-ion collisions}},
  \href{http://arxiv.org/abs/2101.01742}{{\tt arXiv:2101.01742}}.

\bibitem{Zakharov:1997uu}
B.~Zakharov, {\it {Radiative energy loss of high-energy quarks in finite size
  nuclear matter and quark - gluon plasma}},  {\em JETP Lett.} {\bf 65} (1997)
  615--620, [\href{http://arxiv.org/abs/hep-ph/9704255}{{\tt hep-ph/9704255}}].

\bibitem{Baier:1996kr}
R.~Baier, Y.~L. Dokshitzer, A.~H. Mueller, S.~Peigne, and D.~Schiff, {\it
  {Radiative energy loss of high-energy quarks and gluons in a finite volume
  quark - gluon plasma}},  {\em Nucl. Phys. B} {\bf 483} (1997) 291--320,
  [\href{http://arxiv.org/abs/hep-ph/9607355}{{\tt hep-ph/9607355}}].

\bibitem{Blaizot:2012fh}
J.-P. Blaizot, F.~Dominguez, E.~Iancu, and Y.~Mehtar-Tani, {\it {Medium-induced
  gluon branching}},  {\em JHEP} {\bf 01} (2013) 143,
  [\href{http://arxiv.org/abs/1209.4585}{{\tt arXiv:1209.4585}}].

\bibitem{Apolinario:2014csa}
L.~Apolin\'ario, N.~Armesto, J.~G. Milhano, and C.~A. Salgado, {\it
  {Medium-induced gluon radiation and colour decoherence beyond the soft
  approximation}},  {\em JHEP} {\bf 02} (2015) 119,
  [\href{http://arxiv.org/abs/1407.0599}{{\tt arXiv:1407.0599}}].

\bibitem{Gyulassy:1993hr}
M.~Gyulassy and X.-n. Wang, {\it {Multiple collisions and induced gluon
  Bremsstrahlung in QCD}},  {\em Nucl. Phys. B} {\bf 420} (1994) 583--614,
  [\href{http://arxiv.org/abs/nucl-th/9306003}{{\tt nucl-th/9306003}}].

\bibitem{Wiedemann:2000za}
U.~A. Wiedemann, {\it {Gluon radiation off hard quarks in a nuclear
  environment: Opacity expansion}},  {\em Nucl. Phys. B} {\bf 588} (2000)
  303--344, [\href{http://arxiv.org/abs/hep-ph/0005129}{{\tt hep-ph/0005129}}].

\bibitem{Andres:2020kfg}
C.~Andres, F.~Dominguez, and M.~G. Martinez, {\it {From soft to hard radiation:
  the role of multiple scatterings in medium-induced gluon emissions}},
  \href{http://arxiv.org/abs/2011.06522}{{\tt arXiv:2011.06522}}.

\bibitem{Feal:2019xfl}
X.~Feal, C.~A. Salgado, and R.~A. Vazquez, {\it {Jet quenching tests of the QCD
  Equation of State}},  \href{http://arxiv.org/abs/1911.01309}{{\tt
  arXiv:1911.01309}}.

\bibitem{Barata:2020rdn}
J.~a. Barata, Y.~Mehtar-Tani, A.~Soto-Ontoso, and K.~Tywoniuk, {\it {Revisiting
  transverse momentum broadening in dense QCD media}},
  \href{http://arxiv.org/abs/2009.13667}{{\tt arXiv:2009.13667}}.

\bibitem{MehtarTani:2012cy}
Y.~Mehtar-Tani, C.~A. Salgado, and K.~Tywoniuk, {\it {The Radiation pattern of
  a QCD antenna in a dense medium}},  {\em JHEP} {\bf 10} (2012) 197,
  [\href{http://arxiv.org/abs/1205.5739}{{\tt arXiv:1205.5739}}].

\bibitem{Mehtar-Tani:2017ypq}
Y.~Mehtar-Tani and K.~Tywoniuk, {\it {Radiative energy loss of neighboring
  subjets}},  {\em Nucl. Phys. A} {\bf 979} (2018) 165--203,
  [\href{http://arxiv.org/abs/1706.06047}{{\tt arXiv:1706.06047}}].

\bibitem{Blok:2019uny}
B.~Blok and K.~Tywoniuk, {\it {Higher-order corrections to heavy-quark jet
  quenching}},  {\em Eur. Phys. J. C} {\bf 79} (2019), no.~7 560,
  [\href{http://arxiv.org/abs/1901.07864}{{\tt arXiv:1901.07864}}].

\bibitem{Casalderrey-Solana:2014bpa}
J.~Casalderrey-Solana, D.~C. Gulhan, J.~G. Milhano, D.~Pablos, and
  K.~Rajagopal, {\it {A Hybrid Strong/Weak Coupling Approach to Jet
  Quenching}},  {\em JHEP} {\bf 10} (2014) 019,
  [\href{http://arxiv.org/abs/1405.3864}{{\tt arXiv:1405.3864}}]. [Erratum:
  JHEP 09, 175 (2015)].

\bibitem{Adler:2006bw}
{\bf PHENIX} Collaboration, S.~Adler et~al., {\it {A Detailed Study of
  High-p(T) Neutral Pion Suppression and Azimuthal Anisotropy in Au+Au
  Collisions at s(NN)**(1/2) = 200-GeV}},  {\em Phys. Rev. C} {\bf 76} (2007)
  034904, [\href{http://arxiv.org/abs/nucl-ex/0611007}{{\tt nucl-ex/0611007}}].

\bibitem{Kang:2016mcy}
Z.-B. Kang, F.~Ringer, and I.~Vitev, {\it {The semi-inclusive jet function in
  SCET and small radius resummation for inclusive jet production}},  {\em JHEP}
  {\bf 10} (2016) 125, [\href{http://arxiv.org/abs/1606.06732}{{\tt
  arXiv:1606.06732}}].

\bibitem{Dai:2016hzf}
L.~Dai, C.~Kim, and A.~K. Leibovich, {\it {Fragmentation of a Jet with Small
  Radius}},  {\em Phys. Rev. D} {\bf 94} (2016), no.~11 114023,
  [\href{http://arxiv.org/abs/1606.07411}{{\tt arXiv:1606.07411}}].

\bibitem{Sjostrand:2014zea}
T.~Sjöstrand, S.~Ask, J.~R. Christiansen, R.~Corke, N.~Desai, P.~Ilten,
  S.~Mrenna, S.~Prestel, C.~O. Rasmussen, and P.~Z. Skands, {\it {An
  introduction to PYTHIA 8.2}},  {\em Comput. Phys. Commun.} {\bf 191} (2015)
  159--177, [\href{http://arxiv.org/abs/1410.3012}{{\tt arXiv:1410.3012}}].

\bibitem{Dasgupta:2007wa}
M.~Dasgupta, L.~Magnea, and G.~P. Salam, {\it {Non-perturbative QCD effects in
  jets at hadron colliders}},  {\em JHEP} {\bf 02} (2008) 055,
  [\href{http://arxiv.org/abs/0712.3014}{{\tt arXiv:0712.3014}}].

\bibitem{Aaboud:2018twu}
{\bf ATLAS} Collaboration, M.~Aaboud et~al., {\it {Measurement of the nuclear
  modification factor for inclusive jets in Pb+Pb collisions at
  $\sqrt{s_\mathrm{NN}}=5.02$ TeV with the ATLAS detector}},  {\em Phys. Lett.
  B} {\bf 790} (2019) 108--128, [\href{http://arxiv.org/abs/1805.05635}{{\tt
  arXiv:1805.05635}}].

\bibitem{Moore:2004tg}
G.~D. Moore and D.~Teaney, {\it {How much do heavy quarks thermalize in a heavy
  ion collision?}},  {\em Phys. Rev. C} {\bf 71} (2005) 064904,
  [\href{http://arxiv.org/abs/hep-ph/0412346}{{\tt hep-ph/0412346}}].

\bibitem{Tachibana:2017syd}
Y.~Tachibana, N.-B. Chang, and G.-Y. Qin, {\it {Full jet in quark-gluon plasma
  with hydrodynamic medium response}},  {\em Phys. Rev. C} {\bf 95} (2017),
  no.~4 044909, [\href{http://arxiv.org/abs/1701.07951}{{\tt
  arXiv:1701.07951}}].

\bibitem{Caucal:2020uic}
P.~Caucal, E.~Iancu, and G.~Soyez, {\it {Jet radiation in a longitudinally
  expanding medium}},  \href{http://arxiv.org/abs/2012.01457}{{\tt
  arXiv:2012.01457}}.

\bibitem{Acharya:2019jyg}
{\bf ALICE} Collaboration, S.~Acharya et~al., {\it {Measurements of inclusive
  jet spectra in pp and central Pb-Pb collisions at $\sqrt{s_{\rm{NN}}}$ = 5.02
  TeV}},  {\em Phys. Rev. C} {\bf 101} (2020), no.~3 034911,
  [\href{http://arxiv.org/abs/1909.09718}{{\tt arXiv:1909.09718}}].

\bibitem{Sirunyan:2021pcp}
{\bf CMS} Collaboration, A.~M. Sirunyan et~al., {\it {First measurement of
  large area jet transverse momentum spectra in heavy-ion collisions}},
  \href{http://arxiv.org/abs/2102.13080}{{\tt arXiv:2102.13080}}.

\bibitem{Caucal:2019uvr}
P.~Caucal, E.~Iancu, and G.~Soyez, {\it {Deciphering the $z_g$ distribution in
  ultrarelativistic heavy ion collisions}},  {\em JHEP} {\bf 10} (2019) 273,
  [\href{http://arxiv.org/abs/1907.04866}{{\tt arXiv:1907.04866}}].

\bibitem{Larkoski:2019nwj}
A.~J. Larkoski and E.~M. Metodiev, {\it {A Theory of Quark vs. Gluon
  Discrimination}},  {\em JHEP} {\bf 10} (2019) 014,
  [\href{http://arxiv.org/abs/1906.01639}{{\tt arXiv:1906.01639}}].

\bibitem{Metodiev:2018ftz}
E.~M. Metodiev and J.~Thaler, {\it {Jet Topics: Disentangling Quarks and Gluons
  at Colliders}},  {\em Phys. Rev. Lett.} {\bf 120} (2018), no.~24 241602,
  [\href{http://arxiv.org/abs/1802.00008}{{\tt arXiv:1802.00008}}].

\bibitem{Komiske:2018vkc}
P.~T. Komiske, E.~M. Metodiev, and J.~Thaler, {\it {An operational definition
  of quark and gluon jets}},  {\em JHEP} {\bf 11} (2018) 059,
  [\href{http://arxiv.org/abs/1809.01140}{{\tt arXiv:1809.01140}}].

\bibitem{Larkoski:2014gra}
A.~J. Larkoski, I.~Moult, and D.~Neill, {\it {Power Counting to Better Jet
  Observables}},  {\em JHEP} {\bf 12} (2014) 009,
  [\href{http://arxiv.org/abs/1409.6298}{{\tt arXiv:1409.6298}}].

\bibitem{Frye:2017yrw}
C.~Frye, A.~J. Larkoski, J.~Thaler, and K.~Zhou, {\it {Casimir Meets Poisson:
  Improved Quark/Gluon Discrimination with Counting Observables}},  {\em JHEP}
  {\bf 09} (2017) 083, [\href{http://arxiv.org/abs/1704.06266}{{\tt
  arXiv:1704.06266}}].

\bibitem{Chien:2018dfn}
Y.-T. Chien and R.~Kunnawalkam~Elayavalli, {\it {Probing heavy ion collisions
  using quark and gluon jet substructure}},
  \href{http://arxiv.org/abs/1803.03589}{{\tt arXiv:1803.03589}}.

\bibitem{Cacciari:2011ma}
M.~Cacciari, G.~P. Salam, and G.~Soyez, {\it {FastJet User Manual}},  {\em Eur.
  Phys. J. C} {\bf 72} (2012) 1896, [\href{http://arxiv.org/abs/1111.6097}{{\tt
  arXiv:1111.6097}}].

\bibitem{Baron:2020xoi}
J.~Baron, D.~Reichelt, S.~Schumann, N.~Schwanemann, and V.~Theeuwes, {\it
  {Soft-drop grooming for hadronic event shapes}},
  \href{http://arxiv.org/abs/2012.09574}{{\tt arXiv:2012.09574}}.

\bibitem{Qin:2015srf}
G.-Y. Qin and X.-N. Wang, {\it {Jet quenching in high-energy heavy-ion
  collisions}},  {\em Int. J. Mod. Phys. E} {\bf 24} (2015), no.~11 1530014,
  [\href{http://arxiv.org/abs/1511.00790}{{\tt arXiv:1511.00790}}].

\bibitem{Chesler:2014jva}
P.~M. Chesler and K.~Rajagopal, {\it {Jet quenching in strongly coupled
  plasma}},  {\em Phys. Rev. D} {\bf 90} (2014), no.~2 025033,
  [\href{http://arxiv.org/abs/1402.6756}{{\tt arXiv:1402.6756}}].

\bibitem{Hulcher:2017cpt}
Z.~Hulcher, D.~Pablos, and K.~Rajagopal, {\it {Resolution Effects in the Hybrid
  Strong/Weak Coupling Model}},  {\em JHEP} {\bf 03} (2018) 010,
  [\href{http://arxiv.org/abs/1707.05245}{{\tt arXiv:1707.05245}}].

\bibitem{Pablos:2019ngg}
D.~Pablos, {\it {Jet Suppression From a Small to Intermediate to Large
  Radius}},  {\em Phys. Rev. Lett.} {\bf 124} (2020), no.~5 052301,
  [\href{http://arxiv.org/abs/1907.12301}{{\tt arXiv:1907.12301}}].

\end{thebibliography}\endgroup

\end{document}